\documentclass[12pt]{JHEP3}
\pdfoutput=1

\usepackage{amsmath,epsfig}
\usepackage{amssymb,amsfonts}
\usepackage{latexsym}
\usepackage[latin1]{inputenc}
\usepackage{tocvsec2}
\usepackage{subeqnarray}
\usepackage{xcolor}
\let\normalcolor\relax
\usepackage{graphicx}
\usepackage{longtable}

\relax
\renewcommand{\theequation}{\arabic{section}.\arabic{equation}}
\def\be{\begin{equation}}
\def\ee{\end{equation}}
\def\bea{\begin{eqnarray}}
\def\eea{\end{eqnarray}}

\newcommand\fverb{\setbox\pippobox=\hbox\bgroup\verb}
\newcommand\fverbdo{\egroup\medskip\noindent%
                        \fbox{\unhbox\pippobox}\ }
\newcommand\fverbit{\egroup\item[\fbox{\unhbox\pippobox}]}

\newcommand{\bear}{\begin{eqnarray}}

\newcommand{\eear}{\end{eqnarray}}

\newcommand{\bsea}{\begin{subeqnarray}}
\newcommand{\esea}{\end{subeqnarray}}
\newbox\pippobox

\def\6{\partial}

\def\a{\alpha}

\def\nn{\nonumber}

\def\sq
\def\a{\alpha}

\def\nn{\nonumber}

\newcommand{\comments}[1]{}
\allowdisplaybreaks[3]

\title{\Large{Intertwined Orders in Holography: Pair and Charge Density Waves}}
\author{ \large Sera Cremonini$^{a}$,
  Li Li$^{a}$, Jie Ren$^b$\\
~\\
$^a$ Department of Physics, Lehigh University, Bethlehem, PA, 18018, USA.\\
~\\
$^b$Racah Institute of Physics, The Hebrew University of Jerusalem, 91904, Israel.
~\\
~\\
E-mail: cremonini@lehigh.edu, lil416@lehigh.edu, jie.ren@mail.huji.ac.il
}

\abstract{Building on \cite{Cremonini:2016rbd}, we examine a holographic model
in which a U(1) symmetry and translational invariance are broken spontaneously at the same time.
The symmetry breaking is realized through the St\"{u}ckelberg mechanism, and
leads to a scalar condensate and a charge density which are spatially modulated and exhibit unidirectional stripe order.
Depending on the choice of parameters, the oscillations of the scalar condensate can average out to zero, with a frequency which is half of that of the charge density.
In this case the system realizes some of the key features of pair density wave order.
The model also admits a phase with co-existing superconducting and charge density wave orders, in which the scalar condensate has a uniform component.
In our construction the various orders are intertwined with each other and have a common origin.
The fully backreacted geometry is computed numerically, including for the case in which the theory contains axions. The latter can be added to
explicitly break translational symmetry and mimic lattice-type effects.}


\begin{document}

\newpage
\section{Introduction}

Behind the unconventional behavior of many strongly interacting quantum systems is an
intrinsically complex phase diagram exhibiting a variety of orders, some of which exist at comparable temperatures.
A prime example is that of the high $T_c$ superconductors, whose peculiar properties seem to be linked to the interplay between different phases,
which may not only compete but also cooperate with each other (see \emph{e.g.}~\cite{EmeryKivelsonTranquada,Berg:2009dga,Colloquium}).
Indeed, in certain regions of the phase diagram many of these orders are believed to
have a common origin and to be \emph{intertwined}.

Our interest here is in a very particular quantum phase of matter \cite{HimedaPRL,BergPRL}, the so-called pair density wave (PDW),
which describes a state in which charge density wave (CDW) and
superconducting (SC) orders -- and possibly spin density wave (SDW) order -- are intertwined in a very specific manner.
Thus, these are peculiar types of striped superconductors.
The most compelling experimental evidence of a PDW is found in the cuprate superconductor $\text{La}_{2-x} \text{Ba}_{x} \text{CuO}_{4}$ (LBCO), but
there is also computational evidence suggesting that it might be a robust feature of strongly correlated electron systems~\cite{Colloquium}.
The defining properties of PDW order that we will focus on are the following:
\begin{itemize}
\item
The superconducting order itself is spatially modulated in such a way that its uniform
component is zero.
\item
The charge density oscillates at twice the frequency of the condensate.
\end{itemize}
Thus, a PDW differs from a state with co-existing SC and CDW orders, in which the superconducting condensate does have a homogeneous component.

As argued in~\cite{Colloquium}, one way to understand highly complex phase diagrams such as those of the cuprates
may be to view the many observed phases as having originated from a
``parent phase'' which spontaneously breaks a number of symmetries. This is also tied
to the idea that certain orders may not compete, but rather cooperate with each other.
In this paper we explore these ideas within the framework of holography, by providing an example of intertwined orders which realizes the two PDW features described above.
In particular, we extend the work we started in~\cite{Cremonini:2016rbd} and examine a four dimensional holographic model
in which translational invariance and a U(1) symmetry are broken spontaneously and at the same time.
As a result, in the dual theory one finds a scalar condensate that is spatially modulated, and similarly an oscillating charge density,
exhibiting unidirectional stripe order. Thus, this is a model of a striped superconductor\,\footnote{We will loosely refer to this
as a superconductor, as typically done in the literature, although technically the U(1) symmetry of the dual field theory is global, as in a superfluid, and not local.}.

Depending on the choice of parameters, our model exhibits either PDW order or co-existing SC+CDW orders.
In the former case the oscillations of the condensate average out to zero,
\begin{equation}
\langle O_\chi\rangle \propto \cos(k\,x) \, ,
\end{equation}
and those of the charge density are induced (in a manner which will be described in Section \ref{PDWvsCDW}) and oscillate at twice the frequency,
\begin{equation}
\rho(x) = \rho_0 + \rho_1 \cos(2k\,x) \, .
\end{equation}
On the other hand, in the phase with co-existing SC+CDW orders the scalar condensate has a uniform component, and now oscillates at the same frequency as the CDW.
As we will discuss shortly, the model also contains a second ``charge'' density (which could in principle mimic a spin density) which also oscillates at the frequency of the condensate.

The setup we consider is not of the form of the standard holographic superconductor~\cite{Gubser:2008px,Hartnoll:2008vx}, but instead
falls within the~\emph{generalized} class of theories examined in~\cite{Franco:2009yz}, where a complex scalar is replaced by two real scalar degrees of freedom, which have the advantage of admitting less restrictive scalar couplings.
Indeed, in our construction the spontaneous symmetry breaking of the global U(1) symmetry
of the dual field theory occurs via the St\"{u}ckelberg mechanism\,\footnote{We will discuss  the nature of the U(1) symmetry in more detail later in the paper. Studies of such kinds of generalized holographic superconductors can be found \emph{e.g.} in~\cite{Gubser:2009qm,Aprile:2009ai,Aprile:2010yb,Liu:2010ka,Peng:2011gh,Cai:2012es,Gouteraux:2012yr,Kiritsis:2015oxa}.}, and  the two real scalar fields $\chi$ and $\theta$ are not necessarily associated with the magnitude and phase of a complex scalar.
While it should be possible to construct models in which the PDW phase is associated with the condensation of a complex scalar -- more appropriate
for applications to \emph{e.g.} cuprate high $T_c$ superconductors -- this would likely entail using a more involved matter sector, and in this paper as a first step we restrict ourselves to this particular framework. We will come back to this issue in Section \ref{FinalRemarks}.

Our theory involves gravity coupled to two real scalars, $\chi$ and $\theta$, and two vector fields $A_\mu$ and $B_\mu$, with $A_\mu$ playing the role of the gauge
field which provides the charge density of the dual field theory.
The high temperature phase is described on the gravity side by the
standard electrically charged AdS-Reissner Nordstr\"{o}m black brane supported by $A_\mu$, with
the scalars and the second vector $B_\mu$ in the vacuum.
At sufficiently low temperatures, on the other hand, $\chi$ and $B_\mu$ spontaneously develop non-trivial profiles and become dynamical\,\footnote{The second scalar $\theta$ can be set to zero by gauge fixing.}.
When this happens the U(1) symmetry of the high temperature phase is broken spontaneously and a scalar condensate develops in the dual field theory.
By allowing for perturbations which are spatially modulated, the instability can be seen to lead to a condensate and charge density which are striped.
In particular, depending on whether the scalar $\chi$ is charged under the second vector field $B_\mu$ or not,
we find PDW vs. SC+CDW order.
The physical motivation of the second vector field in this setup is not obvious. It can be thought of as playing a somewhat passive role,
imprinting its own oscillations onto the charge density associated with $A_\mu$, \emph{i.e.} giving rise to the properties of the CDW.
On the other hand, the oscillations of its density can in principle be associated with those of a spin density, \emph{i.e.} generating SDW order in the system.
This interpretation, however, requires an explanation for the specific couplings we have chosen in the setup, which we discuss in the next section.
At this stage we simply point out that it is possible to obtain some of the key oscillatory features of intertwined SC, CDW and SDW orders,
but more work is clearly needed to realize a model which would capture the detailed phenomenology of these highly complex systems.

Holographic superconductors with spatially modulated orders have been studied in a number of setups, starting with \cite{Flauger:2010tv}, in which striped modulations were sourced by a periodic chemical potential, with many generalizations in the literature since then, \emph{e.g.} \cite{Hutasoit:2011rd,Hutasoit:2012ib,Erdmenger:2013zaa,Kuang:2013jma,Arean:2013mta}. A study of backreaction by adding a lattice in the form of a periodic potential was reported in~\cite{Horowitz:2013jaa}.
Note that in these setups the translational invariance was broken explicitly by introducing some kind of spatially modulated source.
The competition between superfluid and striped phases has also been examined in holography, see
 \cite{Donos:2012yu,Cremonini:2014gia} for top down models.
The spontaneous formation of striped order in a holographic model with a scalar coupled to two U(1) gauge fields
was first studied in~\cite{Donos:2011bh} and more recently in~\cite{Donos:2013gda,Ling:2014saa,Kiritsis:2015hoa,Donos:2016hsd}
(but the U(1) symmetry was preserved in these models).
Here we have extended such constructions by spontaneously breaking both symmetries at the same time, and focusing on the differences between PDW order and the better studied CDW order in holographic superconductors\,\footnote{A holographic superconductor in which the metric is coupled to a U(1) gauge field $A_\mu$ and a complex two-form $C_{\mu\nu}$ has been shown to generate helical orders spontaneously~\cite{Donos:2011ff,Donos:2012gg}. }.
What has recently become apparent in holographic models of strongly correlated electrons
is the advantage of using multiple vector fields, as they typically lead to richer physics. Our construction provides a further example of this idea.
Note that while in our analysis the mass of the vector $B_\mu$ is not expected to affect the phase structure in a qualitative way, it would play a role for applications to transport.

In this paper we have extended~\cite{Cremonini:2016rbd} in two main ways.
First of all, we have performed an analysis of the fully backreacted geometry, done numerically using the DeTurck method~\cite{Headrick:2009pv}.
We have also included the perturbation equations for the system, at next-to-leading order in fluctuations, which provide intuition for the numerical results.
Second, we have examined the effect of including axions into the model,
introduced to mimic the effects of an underlying lattice structure and to provide a mechanism for momentum relaxation.

The outline of the paper is as follows.
We introduce our model in Section \ref{SetupSection} and examine the critical temperature for the onset of spatially modulated instabilities in Section \ref{section:spatial}.
The difference between PDW and SC+CDW orders is explained in some detail in Section \ref{PDWvsCDW}, and the numerical analysis of the backreacted striped solutions
is presented in Section \ref{backreaction} (no axions) and Section \ref{AxionsSection} (axions). We conclude with open questions and remarks in Section \ref{FinalRemarks}.
The full perturbation analysis up to next-to-leading order is included in Appendix \ref{app:eoms}, while Appendices \ref{therm}
and \ref{app:test} contain, respectively, a discussion of the free energy and a check on the quality of our numerics.

\section{Holographic Setup}
\label{SetupSection}

We work with a four-dimensional gravitational model which describes two real scalar fields $\chi$ and $\theta$
coupled to two abelian vector fields $A_\mu$ and $B_\mu$,
\begin{eqnarray}
S&=&\frac{1}{2\kappa_N^2}\int d^{4}x \sqrt{-g} \left[\mathcal{R}-2\Lambda+\mathcal{L}_{m}\right], \nonumber \\
\mathcal{L}_{m} &=& -\frac{1}{2}\partial_{\mu}\chi \partial^{\mu}\chi-\frac{Z_A(\chi)}{4}F_{\mu\nu}F^{\mu\nu}-\frac{Z_B(\chi)}{4}\tilde{F}_{\mu\nu}\tilde{F}^{\mu\nu}-\frac{Z_{AB}(\chi)}{2}F_{\mu\nu}\tilde{F}^{\mu\nu}\nonumber\\
&&-\mathcal{K}(\chi)(\partial_\mu\theta-q_A A_\mu-q_B B_\mu)^2-\frac{m_v^2}{2}B_\mu B^\mu-V(\chi)\,,
\label{actions}
\end{eqnarray}
where $F_{\mu\nu}=\partial_\mu A_\nu-\partial_\nu A_\mu$ and $\tilde{F}_{\mu\nu}=\partial_\mu B_\nu-\partial_\nu B_\mu$ are
their respective field strengths.
This model falls within the generalized class of holographic \emph{St\"{u}ckelberg superconductors\,\footnote{Technically, these are superfluids, since the symmetry
in the boundary theory is global and not local, but we will refer
to them as being superconductors, as is typically done in the literature.}} studied in~\cite{Franco:2009yz},
in which the spontaneous breaking of the global $U(1)$ symmetry happens via the St\"{u}ckelberg mechanism
and the local gauge invariance is encoded in the transformations
\begin{eqnarray}
\label{localgaugeinvariance}
\theta \rightarrow \theta + \alpha(x^\mu) \, , \quad \quad  A_\mu \rightarrow  A_\mu + \frac{1}{q_A} \partial_\mu \alpha(x^\mu) \, ,
\end{eqnarray}
with an analogous transformation for the second vector field $B_\mu$ when it is massless.
Within the framework of the St\"{u}ckelberg mechanism $\chi$ and $\theta$ are not necessarily the magnitude and phase of a complex scalar field, and in particular $\chi$ does not have to be positive.
As we will see, this will be important in order to realize the main features of PDW order in this particular model.
It has implications for the nature of the U(1) symmetry, which in our case turns out to be non-compact, as there are no restrictions on the range of $\theta$.
We will come back to this point in Section \ref{FinalRemarks}.
For our purposes the advantage of the St\"{u}ckelberg model (\ref{actions}) is that it is less
constraining than the standard holographic superconductor~\cite{Gubser:2008px,Hartnoll:2008vx}, and allows for the type of couplings that are needed to realize the striped orders we are after.
One should keep in mind that there may be other ways to engineer similar phases, using a different matter content.

For completeness here we take the scalar $\chi$ to be generically charged under both vectors.
However, as we will see in detail the case with $q_B=0$ and $q_A\neq 0$ will enable us to describe PDW order.
Our focus will be on breaking the global U(1) symmetry associated with $A_\mu$, the gauge field responsible for
the finite charge density in the boundary theory. Thus, here we will always keep $q_A \neq 0$.
Moreover, note that while the current dual to $A_\mu$ is always conserved, this is not the case for
the second vector field $B_\mu$ when $m_v^2\neq 0$.
We will consider separately the cases in which $B_\mu$ is massless or massive.
We allow for a mass term because it is expected to play a key role in the transport properties of the system, which we plan to examine
in future work, even though it does not affect in any qualitative way the analysis done in this paper.

We have taken all the gauge fields' couplings to depend on the real scalar $\chi$, and have allowed for an interaction
$\sim Z_{AB}(\chi)$ between the two fluxes, which, as we will see, will be responsible for
the breaking of translational invariance.
Finally, the cosmological constant is given by $\Lambda=-\frac{3}{L^2}$, with $L$ denoting the AdS radius.
Einstein's equations resulting from the action above read
\begin{equation}\label{eommetric}
\begin{split}
\mathcal{R}_{\mu\nu} =&\frac{1}{2}\partial_\mu\chi\partial_\nu\chi+ \frac{Z_A}{2} (F_{\mu\rho}{F_\nu}^\rho-\frac{1}{4}g_{\mu\nu} F_{\rho\lambda}F^{\rho\lambda})
+\frac{Z_B}{2} (\tilde{F}_{\mu\rho}{\tilde{F}_\nu}\,^\rho-\frac{1}{4}g_{\mu\nu} \tilde{F}_{\rho\lambda}\tilde{F}^{\rho\lambda})\\
+&\frac{Z_{AB}}{2}(F_{\mu\rho}{\tilde{F}_\nu}\,^\rho+F_{\nu\rho}{\tilde{F}_\mu}\,^\rho-\frac{1}{2}g_{\mu\nu} F_{\rho\lambda}\tilde{F}^{\rho\lambda})+\frac{1}{2}(V-\frac{6}{L^2})g_{\mu\nu}\\
+&\mathcal{K}(\nabla_\mu\theta-q_A A_\mu-q_B B_\mu)(\nabla_\nu\theta-q_A A_\nu-q_B B_\nu)+\frac{m_v^2}{2}B_\mu B_\nu\,.
\end{split}
\end{equation}
The equations of motion for the matter fields on the other hand are
\begin{equation}
\nabla_\mu[\mathcal{K}(\nabla^\mu\theta-q_A A^\mu-q_B B^\mu)]= 0\,,
\end{equation}
\begin{equation}
\begin{split}
 \Box \chi-\mathcal{K}'(\partial_\mu\theta-q_A A_\mu-q_B B_\mu)^2-\left(\frac{Z_A^\prime}{4}F^2+\frac{Z_B^\prime}{4}\tilde{F}^2+\frac{Z_{AB}^\prime}{2}F \tilde{F}\right)-V^\prime=0\,,
 \end{split}
\end{equation}
\begin{equation}
\nabla_\nu(Z_A F^{\nu\mu}+Z_{AB}\tilde{F}^{\nu\mu})+2\mathcal{K}q_A(\nabla^\mu\theta-q_A A^\mu-q_B B^\mu)=0\,,
\end{equation}
\begin{equation}
\nabla_\nu(Z_B \tilde{F}^{\nu\mu}+Z_{AB} F^{\nu\mu})+2\mathcal{K}q_B(\nabla^\mu\theta-q_A A^\mu-q_B B^\mu)-m_v^2 B^\mu=0\,,
\end{equation}
where we used $^\prime = \partial_\chi$, $F^2=F_{\mu\nu}F^{\mu\nu}$, $\tilde{F}^2=\tilde{F}_{\mu\nu}\tilde{F}^{\mu\nu}$ and $F\tilde{F} =F_{\mu\nu}\tilde{F}^{\mu\nu}$ for short.
The scalar couplings and the potential will be chosen so that in the limit $\chi\rightarrow 0$ they can be expanded in the following way,
\begin{equation}\label{coupling}
\begin{split}
&Z_A(\chi)=1+\frac{a}{2}\chi^2+\cdots,\quad Z_B(\chi)=1+ \frac{b}{2} \chi^2 + \cdots\,,\\
&Z_{AB}(\chi)=c\, \chi + \cdots,\quad V( \chi)=\frac{1}{2}m^2\chi^2+\cdots,\quad \mathcal{K}(\chi)=\frac{\kappa}{2}\chi^2+\cdots\,,
\end{split}
\end{equation}
with $(a,b,c,m,\kappa)$ constants. The motivation behind these particular choices will become apparent when we discuss the symmetry breaking mechanism.

We want to work with a theory whose dual has a finite charge density with respect to the global U(1) symmetry associated with the gauge field $A_\mu$. One consequence one shall see immediately is that the expansion~\eqref{coupling} allows for the standard electrically charged AdS Reissner-Nordstr\"{o}m (AdS-RN) black brane supported by $A_\mu$,
in which the two scalars and the second gauge field are in the vacuum,
\begin{equation}\label{RNads}
\begin{split}
&ds^2=\frac{1}{f(r)}dr^2- f(r)dt^2+\frac{r^2}{L^2}(dx^2+dy^2),\\
&f(r)=\frac{r^2}{L^2}\left(1-\frac{r_h^3}{r^3}\right)+\frac{\mu^2 r_h^2}{4r^2}\left(1-\frac{r}{r_h}\right),\quad A_t=\mu\left(1-\frac{r_h}{r}\right)=\mu-2\kappa_N^2 L^2\frac{\rho}{r} \, .\\
\end{split}
\end{equation}
Here $r_h$ denotes the horizon, $\mu$ the chemical potential and $\rho$ the charge density.
The associated temperature is
\be
\label{temp}
T=\frac{r_h}{4\pi}\left[\frac{3}{L^2}-\frac{\mu^2}{4r_h^2}\right]  ,
\ee
and in the extremal limit $T=0$ the horizon radius can be easily seen to be $r_h=\frac{L\mu}{2\sqrt{3}}$.
It is well-known that the extremal, near horizon geometry is $AdS_2\times R^2$,
\begin{equation}\label{ads2}
\begin{split}
ds^2=\frac{L^2}{6 \tilde{r}^2 }d\tilde{r}^2-\frac{6 \tilde{r}^2}{L^2}dt^2+\frac{r_h^2}{L^2}(dx^2+dy^2),\quad A_t=\frac{2\sqrt{3}}{L}\tilde{r},
\end{split}
\end{equation}
with $\tilde{r}=r-r_h$ and the horizon radius given by $L_{(2)}=L/ \sqrt{6}$.

The AdS-RN solution~\eqref{RNads} describes the high temperature phase in which the dual theory has a global U(1) symmetry.
However, we are interested in backgrounds in which $\chi$ and $B_\mu$ are dynamical 
and $\mathcal{K}(\chi)\neq 0$, which will describe phases with a broken U(1) symmetry ($\theta$ can be gauge-fixed to zero).

Indeed, in our model at sufficiently low temperatures the scalar $\chi$ and the second vector field $B_\mu$ spontaneously develop non-trivial profiles.
When this happens, and in particular when $\mathcal{K}(\chi)$ no longer vanishes, the U(1) symmetry of the high temperature phase (associated with the $A_\mu$ vector field) is broken spontaneously, and
a scalar condensate forms in the dual field theory.
Moreover, by considering perturbations which source spatial modulations, one can trigger instabilities to striped superconducting phases.
Indeed, we showed in \cite{Cremonini:2016rbd} that at finite charge density our model can exhibit
spatially modulated phases which spontaneously break translational invariance
at the same time as they break the U(1) symmetry.
The detailed features of the model will be particularly sensitive to the value of $q_B$, as we discuss next.

\section{Spatially Modulated Instabilities}
\label{section:spatial}

Here we come back to the analysis done in~\cite{Cremonini:2016rbd}, where we examined
the spatially modulated static mode in the spectrum of fluctuations around the unbroken phase~\eqref{RNads}.
We are going to first consider instabilities of the zero temperature domain wall solutions which interpolate between $AdS_4$ and
an electrically charged $AdS_2\times R^2$ solution in the far IR.
In particular, we are going to construct PDW or CDW-type modes which violate the IR $AdS_2$ BF bound.
As is well known, the presence of such zero temperature modes
is a strong indication that analogous instabilities should exist at finite temperature, and that
there should be a region in which one has spatially modulated superconducting order.
After examining analytically the perturbations of the zero temperature IR background, we repeat the analysis at finite temperature and focus on the
behavior of the critical temperature for the onset of the phase transition, as a function of wave number $k$.

\subsection{Instabilities of the $AdS_2\times R^2$ geometry}

To investigate striped instabilities of the electrically charged AdS-RN black blane~\eqref{RNads},
we start by considering
linearized perturbations in the $AdS_2\times R^2$ background \eqref{ads2} describing its extremal IR limit,
\begin{equation}\label{ads2bk2}
\begin{split}
ds^2=\frac{L^2}{6 r^2 }dr^2-\frac{6 r^2}{L^2} dt^2+\frac{r_h^2}{L^2}(dx^2+dy^2)\,,\quad A_t=\frac{2\sqrt{3}}{L}r,\quad r_h=\frac{\mu L}{2\sqrt{3}}\,,
\end{split}
\end{equation}
where we have relabeled $\tilde{r} \rightarrow r$ for convenience.
We turn on the following spatially modulated perturbations,
\begin{eqnarray}
\label{firstorder}
\delta \chi =  \varepsilon\, w(r)\cos(k\,x)\,,\quad \delta B_{t} =  \varepsilon\,  b_{t}(r)\cos(k\,x)\,,
\end{eqnarray}
where we introduce $\varepsilon$ as a formal expansion parameter which we will use to
make sure that the perturbations are indeed small when compared to the background~\eqref{ads2bk2}.

Working at linear level $\mathcal{O}(\varepsilon)$ in perturbations, we then obtain the two coupled equations
\begin{eqnarray}
\frac{6}{L^2}\partial_r(r^2 \partial_r w)-\frac{2\sqrt{3}\, c}{L}b_t'-\left(m^2-\frac{6 a}{L^2}-2\kappa q_A^2+\frac{k^2 L^2}{r_h^2}\right)w& = & 0\,, \\
\frac{6 r^2}{L^2}b_t''+\frac{12\sqrt{3}\, c\, r^2}{L^3}w'-\left(m_v^2+\frac{k^2 L^2}{r_h^2}\right)b_t &=&0\,,
\end{eqnarray}
where $(a,c,\kappa,m)$ have been defined in equation~\eqref{coupling}.
Note that they are only coupled to each other when $c \neq 0$ (recall that $c$ controls the coupling $Z_{AB}$ between the scalar and the two vectors)
and that $b$ does not appear in the linearized equations.
We consider the following mode
\begin{equation}
w(r)=u_1\, r^\lambda,\quad b_t(r)=u_2\, r^{\lambda+1}\,,
\end{equation}
where $u_1,u_2$ are constants and $\lambda$ is related to the scaling dimensions of the operators
in the one-dimensional CFT dual to the $AdS_2$ geometry.
The linearized  equations can then be written in matrix form
\begin{equation}\label{matrixeom}
\begin{pmatrix}
\frac{6 \lambda(\lambda+1)}{L^2}-M_{(2)}^2-\frac{k^2L^2}{r_h^2} & \frac{2\sqrt{3}c(\lambda+1)}{L}  \\
    \frac{12\sqrt{3}c \lambda}{L^3} & \frac{6 \lambda(\lambda+1)}{L^2}-m_v^2-\frac{k^2L^2}{r_h^2}
\end{pmatrix}
\begin{pmatrix}
      u_1    \\
      u_2
\end{pmatrix}
=0\,,
\end{equation}
where we have defined
\begin{equation}
M_{(2)}^2=m^2-2 \kappa \, q_A^2-\frac{6a}{L^2}\,.
\end{equation}
To have a non-trivial solution for $u_1$ and $u_2$, the determinant of the matrix on the left hand of~\eqref{matrixeom}
should be vanishing, from which we can solve  for $\lambda$,
\begin{equation}
\lambda_+^\pm=\frac{-1+\sqrt{1+4 m_\pm^2}}{2}\,,\quad \lambda_-^\pm=\frac{-1-\sqrt{1+4 m_\pm^2}}{2}\,,
\end{equation}
with
\begin{equation}
\begin{split}
m_\pm^2&=\frac{1}{12}\left(M_{(2)}^2L^2+m_v^2L^2+12 c^2+24 k^2L^2\right) \\
&\pm\frac{1}{12}\sqrt{144 c^4+(M_{(2)}^2L^2-m_v^2L^2)^2+24 c^2(M_{(2)}^2L^2+m_v^2L^2+24 k^2L^2)}\,.
\end{split}
\end{equation}
Notice that we have fixed the chemical potential and taken it to be $\mu=1$.

The onset of the instability associated with the violation of the $AdS_2$ BF bound occurs when $\lambda$ becomes imaginary, \emph{i.e.} when
\begin{equation}
m_-^2<-\frac{1}{4}\,.
\end{equation}
Thus, in order to have a striped instability, one must identify
a non-zero wave number $k$ at which the value of $\lambda$ is imaginary,
for a fixed choice of Lagrangian parameters.
Such BF-bound violating modes are then associated with spatially modulated phases.
It is easy to check that by choosing parameters appropriately, the BF bound can be violated.
For an explicit example, note that when $m_v=0, m^2=-8, a=4, \kappa= q_A= \mu=1, L=1/2$ and $c=3$, we find
a mass which is clearly below the $AdS_2$ BF bound for a certain range of $k$.

\subsection{Finite temperature instabilities}

The zero temperature instability analysis of the IR $AdS_2$ region indicates that similar spatially modulated unstable modes should be present at finite temperature,
in the black brane background~\eqref{RNads}.
Indeed, we are now ready to calculate the critical temperature below which the AdS-RN geometry becomes unstable, as a function of wave number $k$.
When the scalar field instabilities are associated with a non-zero value of $k$, the condensate is spatially modulated.

In analogy with the $AdS_2$ analysis, we turn on the fluctuations given in~\eqref{firstorder}.
To identify the critical temperature for the instability, it is again sufficient to work to linear order in perturbations.
After expanding around the AdS-RN background~\eqref{RNads} we obtain the two coupled ODEs,
\begin{eqnarray}
&&w''+\left(\frac{2}{r}+\frac{f'}{f}\right)w'+\frac{c \mu r_h}{r^2 f}b_t' -\frac{1}{f}\left(m^2-\frac{\kappa\, q^2\mu^2 (r-r_h)^2}{r^2 f}+\frac{k^2 L^2}{r^2}-\frac{a \mu^2 r_h^2}{2r^4}\right)w= 0\,,
\nonumber \\
&&\label{linearbt}b_t''+\frac{2}{r} b_t'+\frac{c \mu r_h}{r^2}w'-\frac{1}{f}\left(m_v^2+\frac{k^2 L^2}{r^2}\right)b_t=0 \, ,
\end{eqnarray}
with the coupling between the two modes once again proportional to $c$.
We require the fluctuations to be regular at the horizon $r=r_h$, with the expansion
\begin{equation}
w(r) = w^h+\mathcal{O}(r-r_h)\,,\quad\quad b_t(r)=b_t^h(r-r_h)+\mathcal{O}(r-r_h)^2 \, .
\end{equation}
The UV expansion (as $r\rightarrow \infty$) is instead given by
\begin{equation}
\begin{split}
&w(r)=\frac{w_s}{r^{3-\Delta_\chi}}+\cdots+\frac{w_v}{r^{\Delta_\chi}}+\cdots,\quad \quad \Delta_\chi=\frac{1}{2}(3+\sqrt{9+4m^2L^2})\,,  \\
&b_{t}(r) =\frac{b_s}{r^{2-\Delta_B}}+\cdots-\frac{b_v}{r^{\Delta_B-1}}+\cdots,\quad \Delta_B=\frac{1}{2}(3+\sqrt{1+4m_v^2L^2})\,,
\end{split}
\end{equation}
where $\Delta_\chi$ and $\Delta_B$ are the scaling dimensions of the operators dual to $\chi$ and $B_\mu$,
respectively\,\footnote{Note that the vector field $B_\mu$ is dual to a relevant current when $\Delta_B<3$.}.
The sources for the operators, encoded by the parameters $w_s$ and $b_s$, are going to be turned off,
since we want to break the U(1) symmetry and translational invariance spontaneously and not explicitly.

After fixing the parameters in the model, for a given wave number $k$ there will be a normalizable zero mode appearing at a particular temperature,
which is therefore the critical temperature associated with the phase transition.
We are going to take the mass of the scalar field $\chi$ to be $m^2 L^2=-2$ throughout this section, so that the scaling dimension of the dual operator is $\Delta_\chi=2$.
We have considered two different cases for the vector field $B_\mu$.
First, we have taken it to be massless, corresponding to a conserved current in the dual field theory,
with a UV expansion given by
\begin{equation}
b_{t}(r) =b_s-\frac{b_v}{r}+\cdots \; .
\end{equation}
Next, we have allowed for a mass term
$m_v^2L^2=0.11$, in which case the current dual to $B_\mu$ is no longer conserved and the expansion is instead given by
\begin{equation}
b_{t}(r) =\frac{b_s}{r^{-0.1}}-\frac{b_v}{r^{1.1}}+\cdots \; .
\end{equation}

\begin{figure}[ht!]
\begin{center}
\includegraphics[width=.75\textwidth]{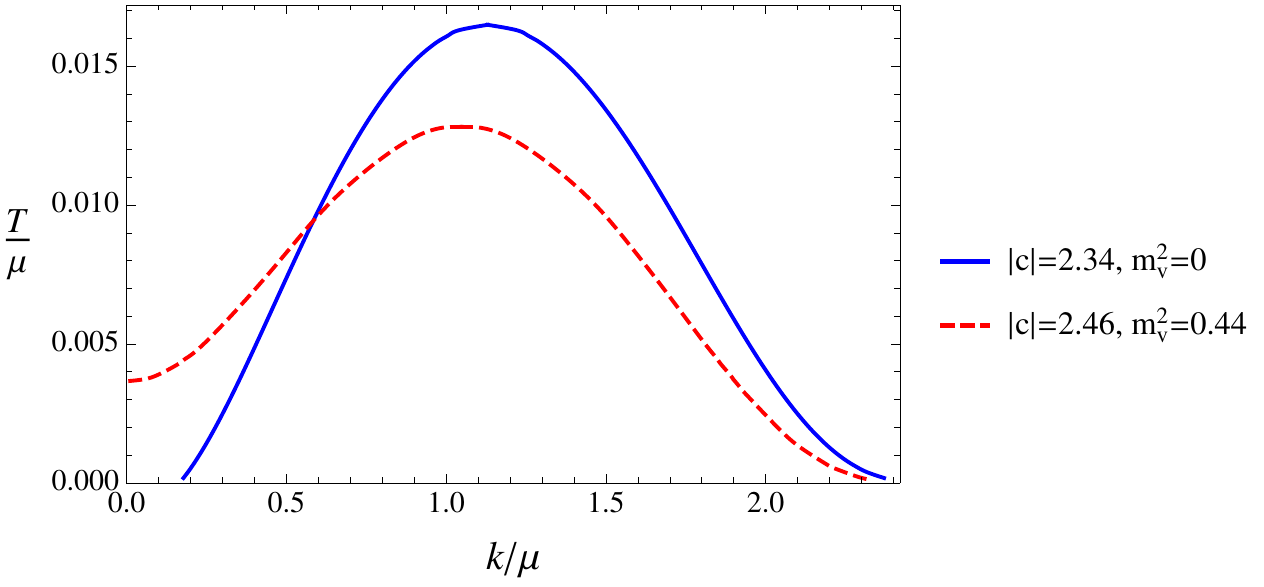}
\caption{Critical temperature versus wave number for the onset of striped instabilities for the massless case
(solid line) and massive case (dashed line).
The peaks are located at $(k\approx 1.130\mu, T\approx 0.01649\mu)$ for the solid curve and at $(k\approx 1.048\mu, T\approx 0.01281\mu)$ for the dashed one.
The remaining parameters are $m^2=-8, L=1/2, a=4, \kappa=q_A=1$.}
\label{fig:mvtckc}
\end{center}
\end{figure}

\begin{figure}[ht!]
\begin{center}
\includegraphics[width=.7\textwidth]{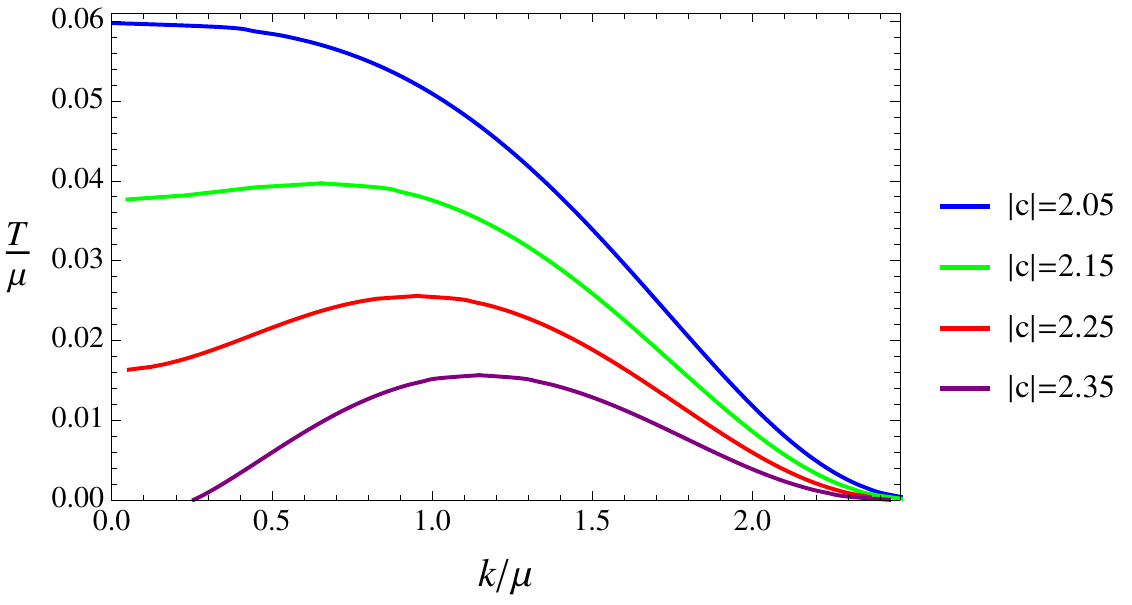}
\caption{Critical temperature as a function of wave number, for different choices of coupling $c$.
From top to bottom the coupling is increased.
We have $m^2=-8, m_v^2=0, L=1/2, a=4, \kappa=q_A=1$.}
\label{fig:tckccs}
\end{center}
\end{figure}

The associated critical temperatures are shown in Figure~\ref{fig:mvtckc}, and correspond to having
set $w_s=b_s=0$.
The figure exhibits the typical bell curve behavior as a function of wave number -- the curves are peaked at non-zero values of $k$, indicating that the condensate is indeed driven by the momentum-dependence of the spatially modulated instabilities.
We emphasize that the mass of the vector does not play any significant role in the behavior of $T_c$ -- by examining more explicit examples one finds that the two cases do not exhibit any qualitative differences.
A more important role is played by the strength $c$ of the coupling $Z_{AB}\sim c \chi $ between the two vector fields, as expected, since the latter is responsible for couplings the modes to each other.
As seen in Figure~\ref{fig:tckccs}, the finite-momentum peak becomes less and less pronounced (and shifts towards zero) as $|c|$ decreases.
What this is indicating is that, while for small $|c|$ one may still have a superconducting instability, it is not going to be striped.
As a result, in this model the coupling must be non-zero and sufficiently large, in order for the phase transition to occur at a finite value of $k$.
This singles out the coupling $c$ between the scalar and the two vector fields as the leading parameter controlling the onset of the striped phase transition.
However, note that when $|c|$
becomes too large the instability once again disappears, because the BF bound can no longer be violated.

%

\section{PDW vs. SC+CDW}
\label{PDWvsCDW}
%

By performing a perturbative analysis and working to leading order
in perturbation theory, we have just found the static, normalizable zero modes
in the electrically charged AdS-RN background, and obtained the critical temperature
at which the  geometry becomes unstable to the formation of a spatially modulated phase.
From the bell-type behavior of Figures~\ref{fig:mvtckc} and~\ref{fig:tckccs} we see that there is a particular wave number
$k_c$ associated with the critical temperature $T_c$ at which a new branch of black branes appears spontaneously.
The wave number $k_c$ thus characterizes the breaking of translational symmetry.

As we have shown numerically and discuss in detail in the next section, for temperatures just below $T_c$
the scalar operator $O_\chi$ dual to $\chi$ acquires spontaneously a spatially modulated expectation value,
\begin{equation}
\label{scalarcond}
\left<O_\chi\right> \sim \cos(k_c\, x) \, ,
\end{equation}
breaking the (non-compact) global U(1) symmetry of the boundary theory.
Therefore,  this spatially modulated phase is always associated with
a non-vanishing condensate, \emph{i.e} the new phase is a type of striped superconductor.
Moreover, the ``charge'' density $\rho_B$ dual to $B_\mu$ becomes striped,
and this, in conjunction with $\left<O_\chi\right>$, induces a modulation in the charge density $\rho_A$ associated with $A_\mu$.

The crucial feature in this model is that the frequency of oscillations of $\rho_A$ as compared to that of $\left<O_\chi\right>$, which is one of the determining factors for whether the state is a PDW or a SC+CDW, depends on whether the charge $q_B$ vanishes or not\,\footnote{More precisely,
it depends on whether $q_A q_B=0 $ or not. However, since our interest is to spontaneously break the U(1) symmetry associated with
$A_\mu$, we are always going to keep $q_A \neq 0$.}.
In particular, when $q_B=0$ we find that
the time component of the vector operator $J_A^t$ corresponding to $A_t$ becomes spatially modulated with
\begin{equation}
\rho_A=\left<J_A^t\right> \sim \cos(2 k_c\, x),
\end{equation}
which implies that the associated period of oscillations is precisely one half of that of the scalar condensate given in (\ref{scalarcond}).
Meanwhile, the time component $J^t_B$ of the operator corresponding to $B_t$ oscillates as in
\begin{equation}
\rho_B=\left<J_B^t\right> \sim \cos(k_c\, x),
\end{equation}
with the same frequency as the scalar condensate.
In addition, when $q_B=0$ the scalar condensate does not have a homogeneous component, so that its oscillations average precisely to zero.
The behavior of $\rho_A$ and $\left<O_\chi\right>$ match precisely with those of a PDW phase, as discussed in the introduction.

The situation is different when $q_B \neq 0$. In this case the charge density $\rho_A$ associated with the $A_\mu$ vector field oscillates at the same frequency as the scalar condensate, and the latter develops a homogeneous component.
Thus, the $q_B \neq 0$ case describes a phase with coexisting SC and CDW orders.
Notably, while the second vector field $B_\mu$ does not determine
the type of order (PDW or SC+CDW) developed in the system, it can in principle be associated with spin degrees of freedom,
with its modulated density $\rho_B$ describing SDW order.
Recall that if we set $m_v^2=0$, the dual theory has a second conserved current corresponding to $B_\mu$.
In the numerical analysis of the next section, for numerical convenience we will focus precisely on this massless case, and leave the massive case to future work (we anticipate that it may be
useful when discussing the transport behavior).

Our numerical results can be explained in part by inspecting the structure of the perturbations at
next-to-leading order.
In particular, to quadratic order in $\varepsilon$, the perturbative parameter which is proportional to $\sqrt{1-T/T_c}$, we have
\begin{equation}\label{secondorders}
\begin{split}
&\delta \chi =\varepsilon\, w(r)\cos(k\,x)+\varepsilon^2[\chi^{(1)}(r)+ \chi^{(2)}(r)\cos(2k\,x)]\, , \\
&\delta B_t =\varepsilon\, b_t(r)\cos(k\,x)+\varepsilon^2[b_{t}^{(1)}(r)+ b_{t}^{(2)}(r)\cos(2k\,x)]\,, \\
&\delta A_t = \varepsilon^2[a_{t}^{(1)}(r)+ a_{t}^{(2)}(r)\cos(2k\,x)]\,,
\end{split}
\end{equation}
where we are singling out the scalar and vector fields for the sake of space.
We refer the reader to Appendix~\ref{app:eoms} for the full perturbation analysis.
We find that the order ${\cal O} (\varepsilon^2)$ components of $\delta \chi$ and $ \delta B_t $ are sourced by ${\cal O} (\varepsilon)$ terms with a prefactor proportional to $q_A \, q_B$,
and therefore vanish when $q_B=0$.
In particular, note that this implies that the homogenous perturbations $\chi^{(1)}(r) $  and $ b_{t}^{(1)}(r)$, which fall under Set (I) of Appendix~\ref{app:eoms},
can both consistently be chosen to vanish when $q_B=0$.
This causes the scalar condensate $\langle O_\chi\rangle$ modulations, which to leading order are $\propto \cos (k\,x)$,  to average out to zero.
By the same argument the oscillations of the charge density $\rho_B $ also average to zero.
Also, their period agrees with that of the scalar condensate, which is consistent with SDW order in a PDW.
On the other hand, when $q_A q_B \neq 0$ the two uniform components $\chi^{(1)}(r) $  and $ b_{t}^{(1)}(r)$ must be turned on, and we thus lose PDW order -- the SC condensate now has a homogeneous component.

\begin{figure}[ht!]
\begin{center}
\includegraphics[width=.82\textwidth]{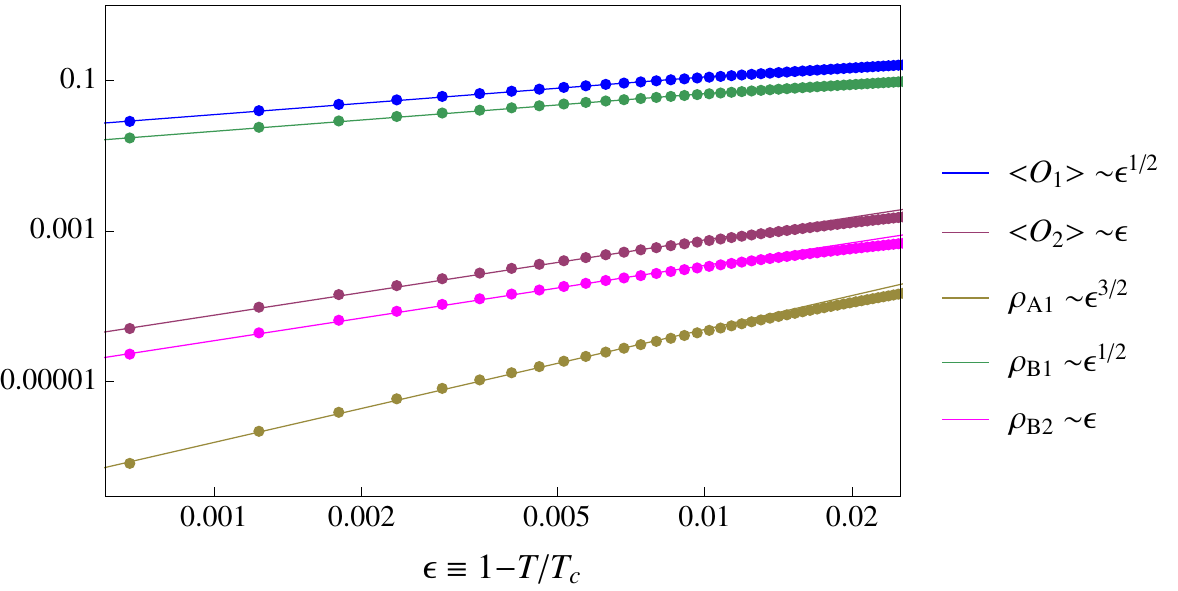}
\caption{The log-log plot of the VEVs  near $T_c$.
Straight lines denote the expected scaling behaviors shown in (\ref{VEVscalings}), and dotted lines denote numerical values.
We consider the model~(\ref{couplingnumerics}) with $q_B=0.5$ and $k/\mu=1$. The data at the top corresponds to $\langle{O}_1\rangle$ and ${\rho}_{B1}$, the data in the middle to $\langle O_{2} \rangle$ and $\rho_{B2}$, and
at the bottom to $\rho_{A1}$.}
\label{fig:fit}
\end{center}
\end{figure}

To understand the behavior of the charge density $\rho_A$ we must inspect the perturbations in the two channels (III) and (IV) of Appendix ~\ref{app:eoms}.
First of all, we emphasize that the modes $a_{t}^{(1)}(r)$ and $ a_{t}^{(2)}(r)$ are sourced by the ${\cal O}(\varepsilon)$ (leading order) perturbations in $\delta\chi$ and $\delta B_t$, and are therefore non-vanishing.
Indeed, the $\sim \cos(2k\,x)$ oscillation of $\rho_A$ is a next-to-leading order effect, which is sourced by
the leading order oscillations $\propto \cos(k\,x)$  of $\chi$ and $B_t$. In this sense, it is \emph{induced}, as needed in a phase with PDW order.
Precisely because when $q_B=0$ the  $\sim \cos(2k\,x)$ oscillations in the scalar condensate are absent, the frequency of the $\rho_A$ oscillations is twice that of the scalar field.
On the other hand, when $q_B \neq 0$ the frequency of the oscillations of $\rho_A$ is the same as that of the scalar condensate\,\footnote{This can be explained by noting that there is a $\cos(k\,x)$ mode appearing at ${\cal O}(\varepsilon^3)$ when $q_B \neq 0$, which is confirmed by our numerics (see Figure~\ref{fig:fit}).}, which now has a uniform component.
Thus, what we have is a co-existing SC+CDW state, and not a PDW.

Note that from the perturbation expressions~\eqref{secondorders} and the fact that $ \varepsilon \sim \sqrt{1-T/T_c}$, when $q_B\neq 0$ we expect the following scaling behavior near $T_c$,
\begin{equation}
\label{VEVscalings}
\langle{O}_{1}\rangle, \; {\rho}_{B1}\sim (1-T/T_c)^{1/2},\quad \langle O_{2} \rangle, \; \rho_{B2}\sim (1-T/T_c),\quad \rho_{A1}\sim (1-T/T_c)^{3/2}.
\end{equation}
Here $\langle{O}_{1}\rangle$ and ${\rho}_{B1}$ are the $\cos(k\,x)$ components of $\langle{O}_{\chi}\rangle$ and $\rho_{B}$,
$\langle{O}_{2}\rangle$ and ${\rho}_{B2}$ are their $\cos(2k\,x)$ components, and finally $\rho_{A1}$ is the $\cos(k\,x)$ component of $\rho_A$.
This is precisely observed in our numerics, as we show in Figure~\ref{fig:fit}.

In summary, the perturbative analysis suggests that the system behaves like a PDW when $q_B=0$ (no uniform component to $\langle O_\chi\rangle$, and
doubling of oscillation frequency), while when $q_B \neq 0$ it describes a SC+CDW state (the uniform contribution~$\propto \chi^{(1)}(r) $ is sourced and the frequencies match), as confirmed explicitly by the numerics. Table \ref{table1} summarizes the key differences between two cases.
\begin{table}[h]
\centering
\begin{tabular}{c|c|c}
\hline
VEVs& $q_B=0$ & $q_B\neq 0$\\
\hline
$\langle\bar{O}_\chi\rangle\equiv\frac{k}{2\pi}\int_0^{2\pi/k}\langle O_\chi(x)\rangle dx$ & $\langle\bar{O}_\chi\rangle=0$ & $\langle\bar{O}_\chi\rangle\neq 0$\\
\hline
$\rho_{A1}\equiv \frac{k}{\pi}\int_0^{2\pi/k}\rho_A(x)\cos(x)dx$ & $\rho_{A1}=0$ & $\rho_{A1}\neq 0$\\
\hline
$\bar{\rho}_B\equiv\frac{k}{2\pi}\int_0^{2\pi/k} \rho_B(x) dx$ & $\bar{\rho}_B=0$ & $\bar{\rho}_B\neq 0$\\
\hline
$\rho_{B2}\equiv\frac{k}{\pi}\int_0^{2\pi/k}\rho_B(x)\cos(2x)dx$ & $\rho_{B2}=0$ & $\rho_{B2}\neq 0$\\
\hline
\end{tabular}
\caption{Comparison of various VEVs between $q_B=0$ and $q_B\neq 0$. The former case corresponds to PDW order, while the latter to a state with co-existing SC+CDW orders. }
\label{table1}
\end{table}
%

\section{The Striped Black Brane Solutions}
\label{backreaction}

In this section we construct the non-linear solutions corresponding to the striped black branes. To be more explicit, we take the couplings (\ref{coupling}) to be given by
\begin{equation}\label{couplingnumerics}
\begin{split}
& Z_A(\chi)=1+2\chi^2,\quad Z_B(\chi)=1,\quad Z_{AB}(\chi)=-2.34\, \chi\,,     \\
& V( \chi)=-4\chi^2,\quad \mathcal{K}(\chi)=\frac{1}{2}\chi^2\,,
\end{split}
\end{equation}
which correspond to $m^2=-8, m_v^2=0, L=1/2, c=-2.34, a=4, \kappa=q_A=1$.
The critical temperature for the striped instability as a function of wave number for this choice of parameters
is shown by the solid blue curve of Figure~\ref{fig:mvtckc}.
For our numerical analysis it is convenient to introduce a new radial coordinate,
\begin{equation}\label{rtoz}
1-z^2=\frac{r_h}{r} \, ,
\end{equation}
in terms of which the AdS-RN black brane geometry~\eqref{RNads} reads

\begin{equation}\label{RNadsz}
\begin{split}
&ds^2=\frac{r_h^2}{L^2 (1-z^2)^2}\left[-F(z)dt^2+\frac{4 z^2 L^4}{r_h^2 F(z)}dz^2+(dx^2+dy^2)\right]\,,\\
&F(z)=z^2\left[2-z^2+(1-z^2)^2-\frac{L^2 \mu^2}{4 r_h^2}(1-z^2)^3\right],\quad A_t=\mu\, z^2\,.
\end{split}
\end{equation}
Note that the horizon is now located at $z=0$ and the AdS boundary is at $z=1$.
The black hole temperature, expressed in terms of $r_h$, is given in~\eqref{temp}.

\subsection{The ansatz and boundary conditions}
We adopt the following ansatz for the striped  black brane geometry,
\bea\label{ansatzbh}
&&ds^2=\frac{r_h^2}{L^2 (1-z^2)^2}\left[-F(z)Q_{tt} \, dt^2+\frac{4 z^2 L^4 Q_{zz}}{r_h^2 F(z)}\, dz^2+Q_{xx}(dx-2 z(1-z^2)^2Q_{xz}dz)^2+Q_{yy}\, dy^2\right], \nn \\
&&\chi=(1-z^2)\phi \, , \qquad A_t=\mu\, z^2 \alpha, \qquad B_t=z^2 \beta,
\eea
where the eight functions $\mathcal{Q}=(\phi,\alpha,\beta,Q_{tt},Q_{zz},Q_{xx},Q_{yy}, Q_{xz})$ depend on both $z$ and $x$.
Notice that we recover the AdS-RN solution~\eqref{RNadsz} by choosing their background values to be $\mathcal{Q}^{(0)}=(\phi=\beta=Q_{xz}=0$, $\alpha=Q_{tt}=Q_{zz}=Q_{xx}=Q_{yy}=1$).

We will consider two different cases, corresponding to $q_B=0$ and $q_B\neq 0$.
Since we are looking for solutions with a regular horizon at $z=0$, all functions depend smoothly on $z^2$.
Thus, at the horizon one can easily impose Neumann boundary conditions of the type $\partial_z \phi(0,x)=0$, and similarly for the remaining components.
Additionally, we impose the Dirichlet boundary condition $Q_{tt}(0,x)=Q_{zz}(0,x)$, so that the temperature of the black brane~\eqref{ansatzbh} is still given by~\eqref{temp}.

We employ the DeTurck method~\cite{Headrick:2009pv} to solve the resulting PDEs.
The key point behind the method is to introduce a new term into the Ricci tensor,
\begin{equation}
\mathcal{R}_{\mu\nu}\rightarrow \mathcal{R}_{\mu\nu}^H=\mathcal{R}_{\mu\nu}-\nabla_{(\mu}\xi_{\nu)}\,,
\end{equation}
where the DeTurck vector field is given by
\begin{equation}\label{eqxi}
\xi^\mu=g^{\lambda\sigma}[\Gamma_{\lambda\sigma}^\mu(g)-\bar{\Gamma}_{\lambda\sigma}^\mu(\bar{g})]\,,
\end{equation}
and $\bar{\Gamma}_{\lambda\sigma}^\mu(\bar{g})$ is the Levi-Civita connection associated with a reference metric $\bar{g}_{\mu\nu}$.
We choose the reference metric to be the metric of the AdS-RN solution~\eqref{RNadsz}.
To avoid Ricci solitons~\cite{Headrick:2009pv}, we should ensure that the quantity $\xi^\mu\xi_\mu$ is zero to the required numerical accuracy (see Appendix~\ref{app:test}).

To know the asymptotic behavior near the AdS boundary, we start from the scaling dimensions of the operators in the dual field theory.
Labeling the eight unknown functions by $\mathcal{Q}_i=\mathcal{Q}_i^{(0)}+\delta \mathcal{Q}_i$, we are going to assume
that the perturbations are of the form
\begin{equation}
\delta \mathcal{Q}_i=v_i(1-z^2)^{\delta_i},
\end{equation}
where $v_i$ denote constants.
After substituting into the equations of motion, we obtain a matrix equation of the form
$$ M_{ij}\,v_j=0 \, ,$$ where $M_{ij}$ is an $8\times 8$ matrix that depends on the exponents $\delta_i$.
In order to have non-trivial solutions we demand the matrix determinant to vanish, which in turn allows us to solve for
the possible values of $\delta_i$,
\begin{gather}
\delta_{1,2,3}=\{0,1\},\qquad \delta_{4,5}=\{0,3\},\qquad \delta_6=\{-3,2\},\qquad
\delta_{7,8}^\pm=\frac{1}{2}(3\pm\sqrt{33}) \, ,
\end{gather}
where $\delta_{1,2,3,4,5,6}$ correspond to, respectively, the scalar, the two vector fields and the metric.
The last two pairs $  \delta_{7,8}^\pm $ are due to the dynamical gauge fixing procedure~\cite{Donos:2016hsd}.

 After imposing the source free condition for the scalar $\chi$ and vector field $B_t$ and  fixing the boundary metric to be Minkoswski,
the asymptotic expansion in the UV behaves as
\begin{equation}\label{uvexpand}
\begin{split}
\phi(z,x)&=\phi_v(x) \,(1-z^2)+\mathcal{O}((1-z^2)^2)\,,\\
\alpha(z,x)&=1+\rho_a(x) (1-z^2)+\mathcal{O}((1-z^2)^2)\,,\\
\beta(z,x)&=\rho_b(x) (1-z^2)r_h+\mathcal{O}((1-z^2)^2)\,,\\
Q_{tt}(z,x)&=1+q_{tt}(x)\,(1-z^2)^3+\mathcal{O}((1-z^2)^4)\,,\\
Q_{xx}(z,x)&=1+q_{xx}(x)\, (1-z^2)^3+\mathcal{O}((1-z^2)^4)\,,\\
Q_{yy}(z,x)&=1+q_{yy}(x)\, (1-z^2)^3+\mathcal{O}((1-z^2)^4)\,,\\
Q_{zz}(z,x)&=1+\mathcal{O}((1-z^2)^4)\,,\\
Q_{xz}(z,x)&=\mathcal{O}((1-z^2)^2)\,,
\end{split}
\end{equation}
where we have imposed the following boundary conditions at $z=1$,
\begin{equation}\label{uvcondition}
\begin{split}
&Q_{tt}(1,x)=Q_{zz}(1,x)=Q_{xx}(1,x)=Q_{yy}(1,x)=\alpha(1,x)=1\,,\\
&\phi(1,x)=\beta(1,z)=Q_{xz}(1,x)=0\,.
\end{split}
\end{equation}
The coefficients $(\phi_v,\rho_a,\rho_b,q_{tt}, q_{xx}, q_{yy})$ appearing in the boundary expansion are all dimensionless.
Finally, the equations of motion lead to the following conditions
\begin{equation}\label{uvrelation}
q_{tt}(x)+q_{xx}(x)+q_{yy}(x)=0\,,\quad \quad \partial_x \, q_{xx}(x)=0\,,
\end{equation}
which means that the energy momentum tensor of the dual field theory is traceless and conserved (see the discussion in Appendix \ref{therm}).

In our numerics we first use the standard pseudospectral collocation approximation to change the PDEs into non-linear algebraic
equations. More precisely, we adopt Fourier discretization in the $x$ direction and Chebyshev polynomials in the $z$ direction. The resulted system is then solved using a Newton-Raphson method.

\subsection{Numerical solutions}

\begin{figure}[ht!]
\begin{center}
\includegraphics[width=.46\textwidth]{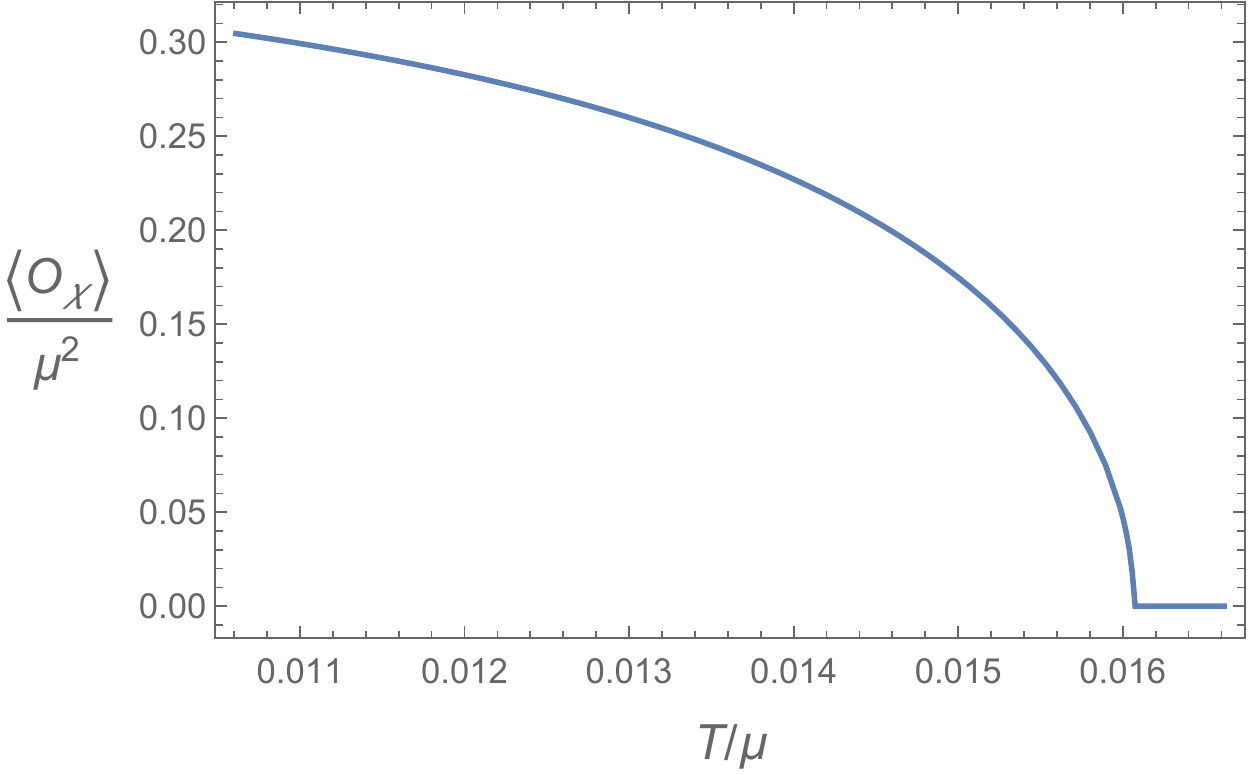}\quad
\includegraphics[width=.48\textwidth]{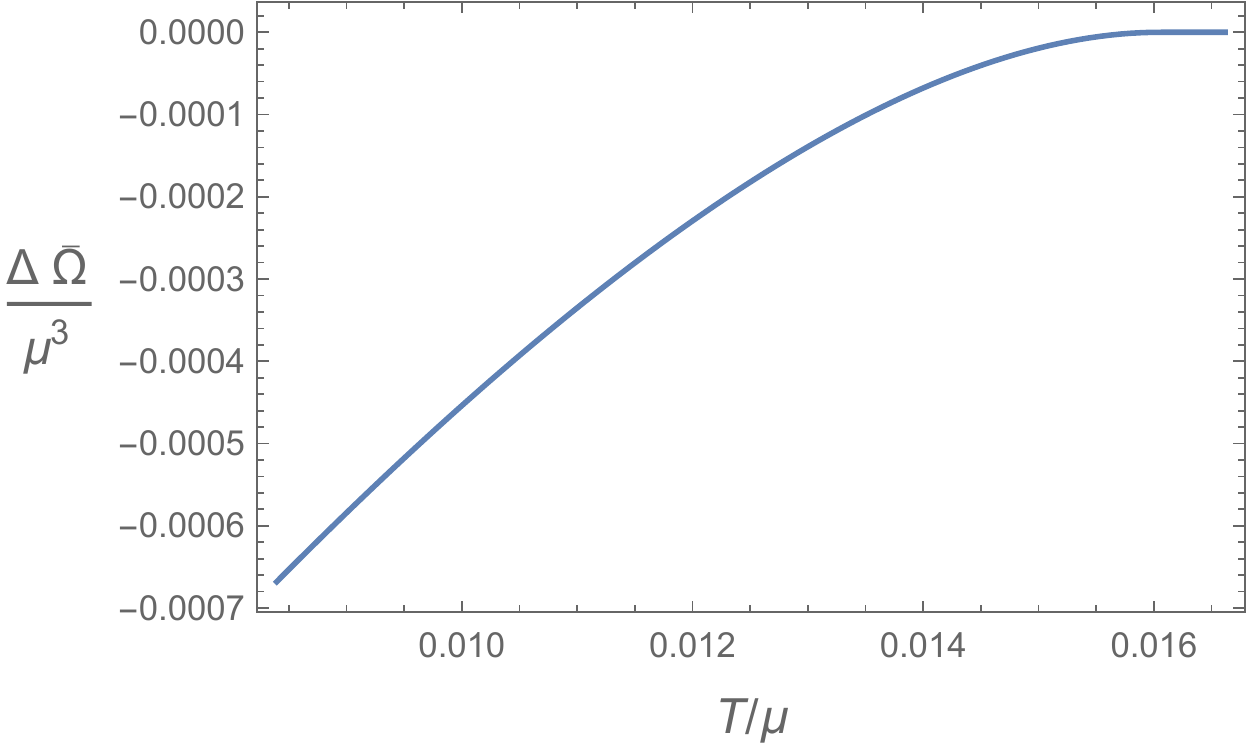}
\caption{The left plot shows the VEV of the scalar condensate at $x=0$ as a function of temperature.
The right plot shows the difference in the average value of the free energy $\Omega$ between the striped phase and the normal
phase. The parameters chosen here are  $k=\mu=1, q_B=0$ and  the critical temperature is $T_c=0.0161$.}
\label{fig:Conden-FreeEn}
\end{center}
\end{figure}

As expected from the analytical analysis, at sufficiently low temperatures the expectation value of the operator dual to $\chi$ is non-zero,
and we have a phase with a striped condensate.
As a typical example, throughout we are going to focus on the $k/\mu=1$ branch, with the corresponding critical temperature being $T_c=0.016\mu$.
We will work in the grand canonical ensemble with 
 $\mu=1$ and choose the convention $2\kappa_N^2=1$.
We plot the temperature dependence of the VEV of the scalar condensate in the left panel of Figure~\ref{fig:Conden-FreeEn}, for $q_B=0$.
The right panel on the other hand shows the temperature dependence of the averaged free energy, which indicates that the broken phase is indeed thermodynamically preferred over the normal phase.

\begin{figure}[ht!]
\begin{center}
\includegraphics[width=.49\textwidth]{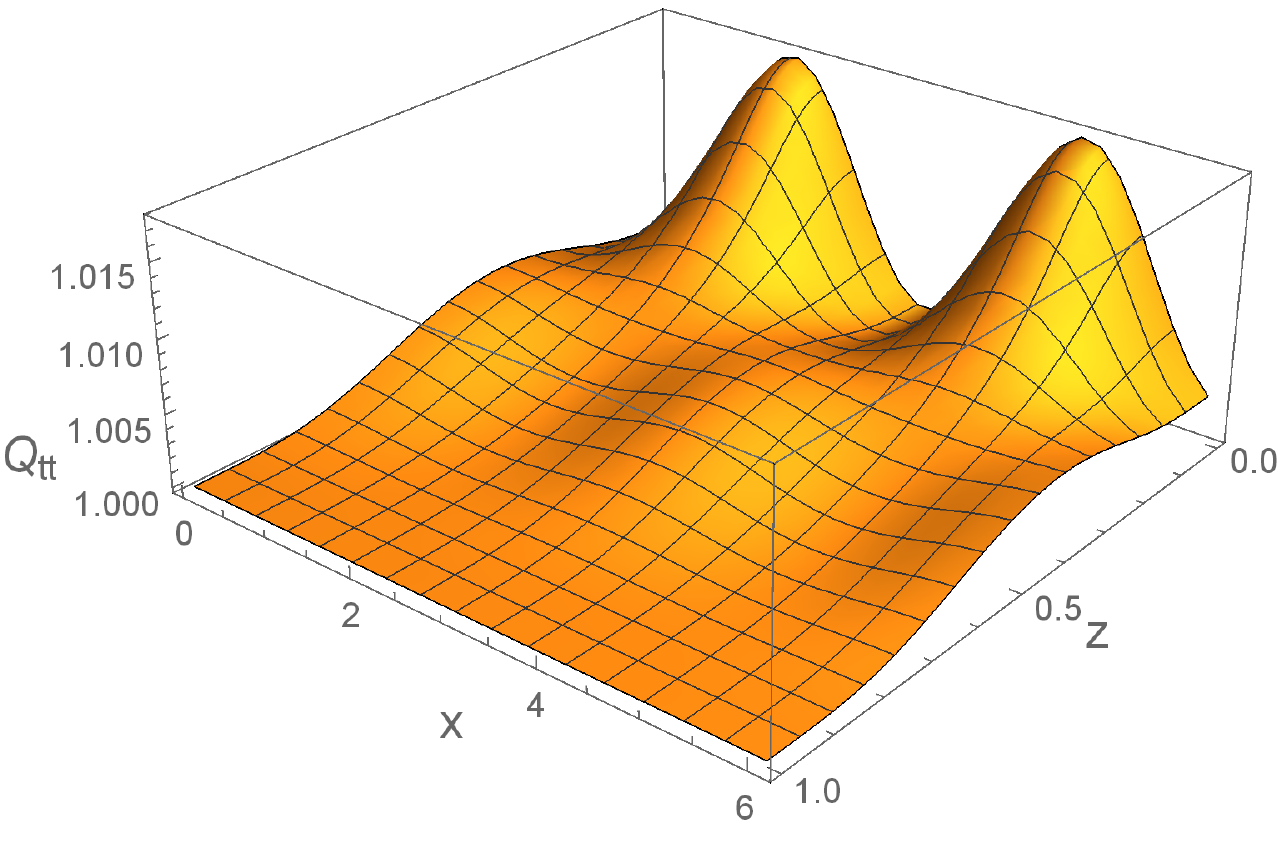}
\includegraphics[width=.49\textwidth]{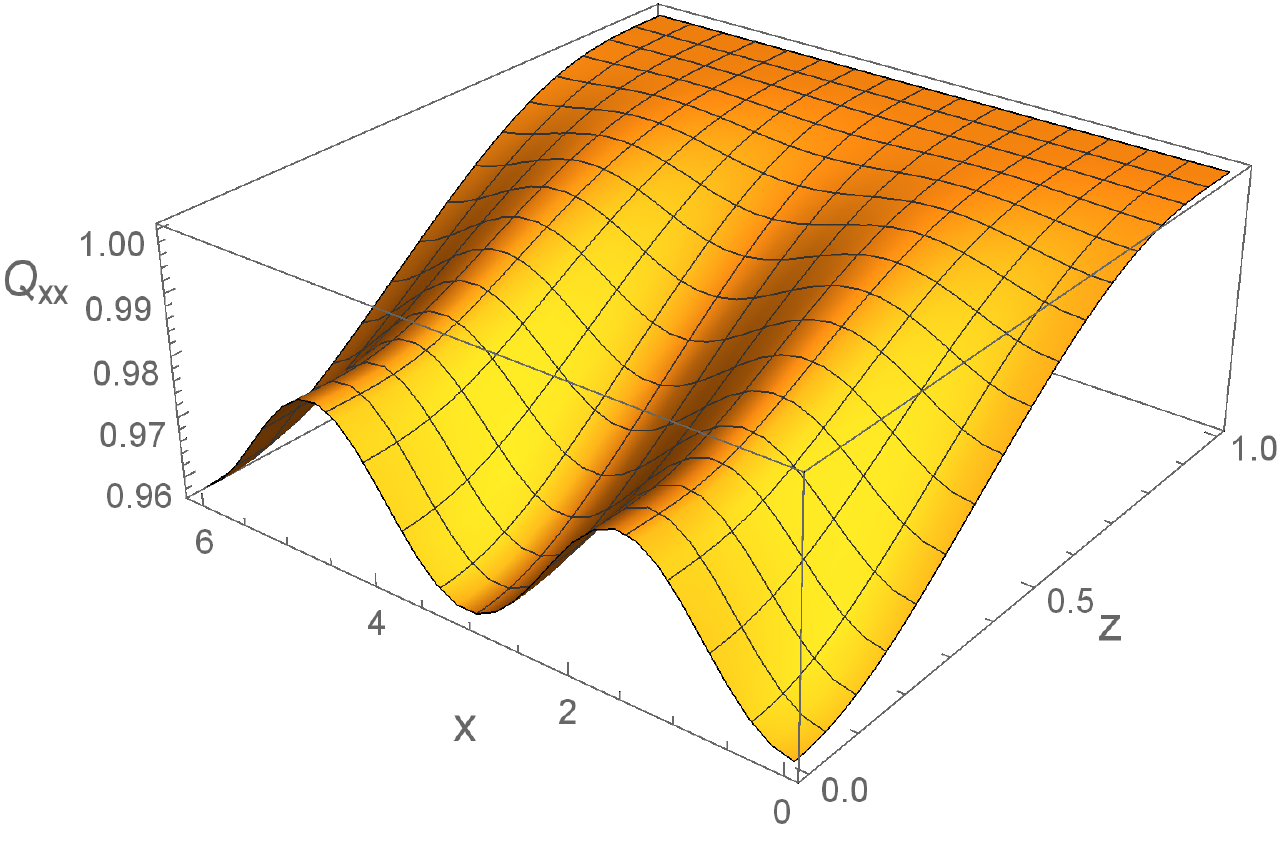}
\includegraphics[width=.49\textwidth]{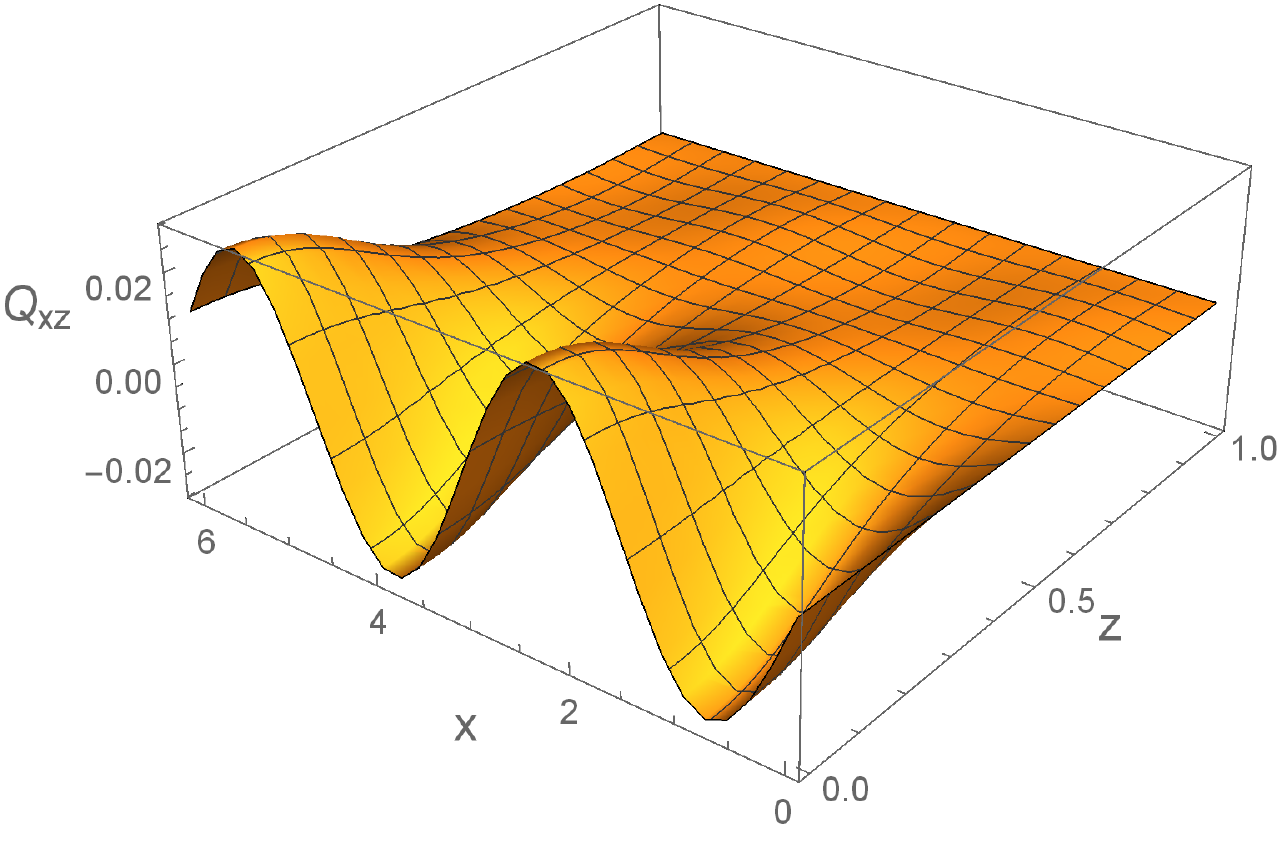}
\includegraphics[width=.49\textwidth]{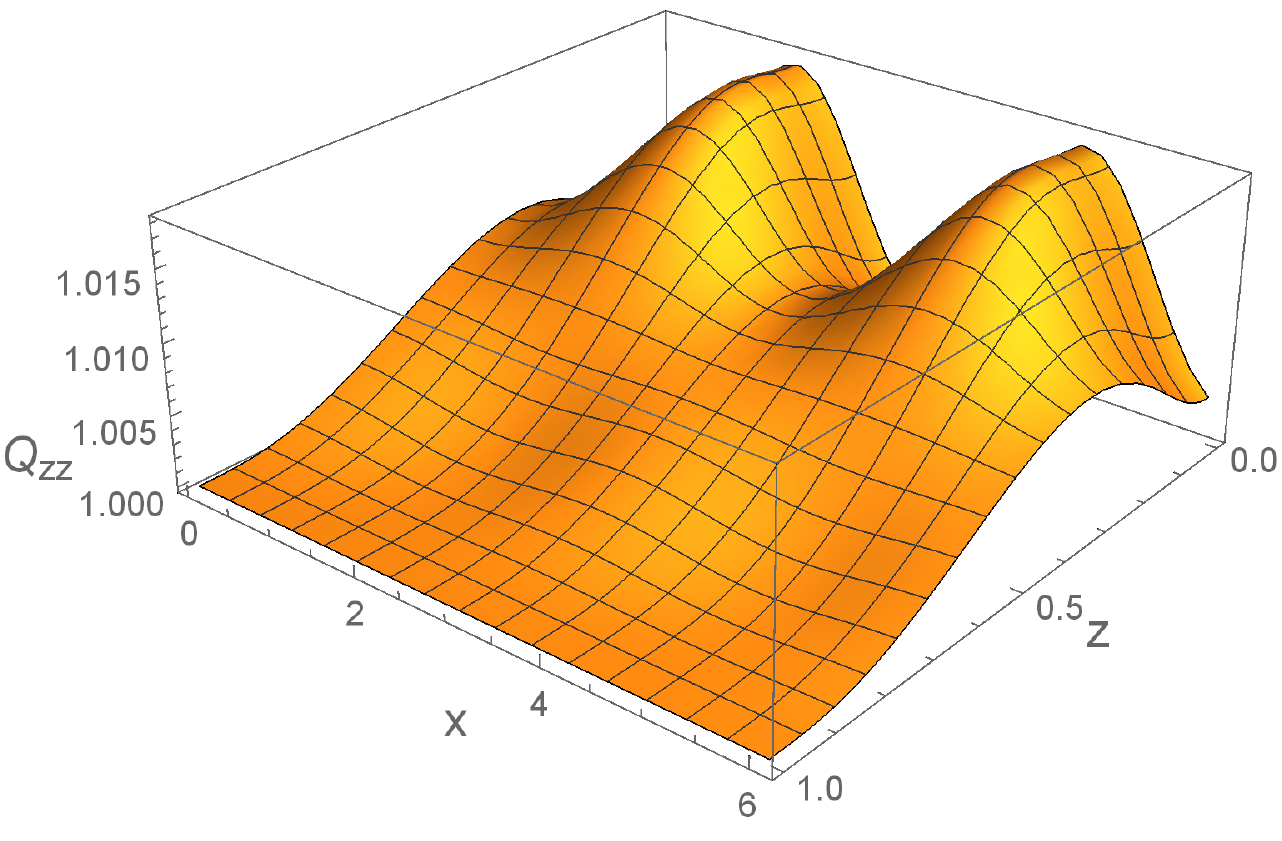}
\caption{Bulk solutions for the metric fields $Q_{tt}, Q_{xx},Q_{xz}, Q_{zz}$ for $T=0.01426$ and $k=1$, for the parameters chosen in (\ref{couplingnumerics}) and $q_B=0$. The horizon is at $z=0$ and the AdS boundary at $z=1$.}
\label{fig:solutionmetric}
\end{center}
\end{figure}

\begin{figure}[ht!]
\begin{center}
\includegraphics[width=.45\textwidth]{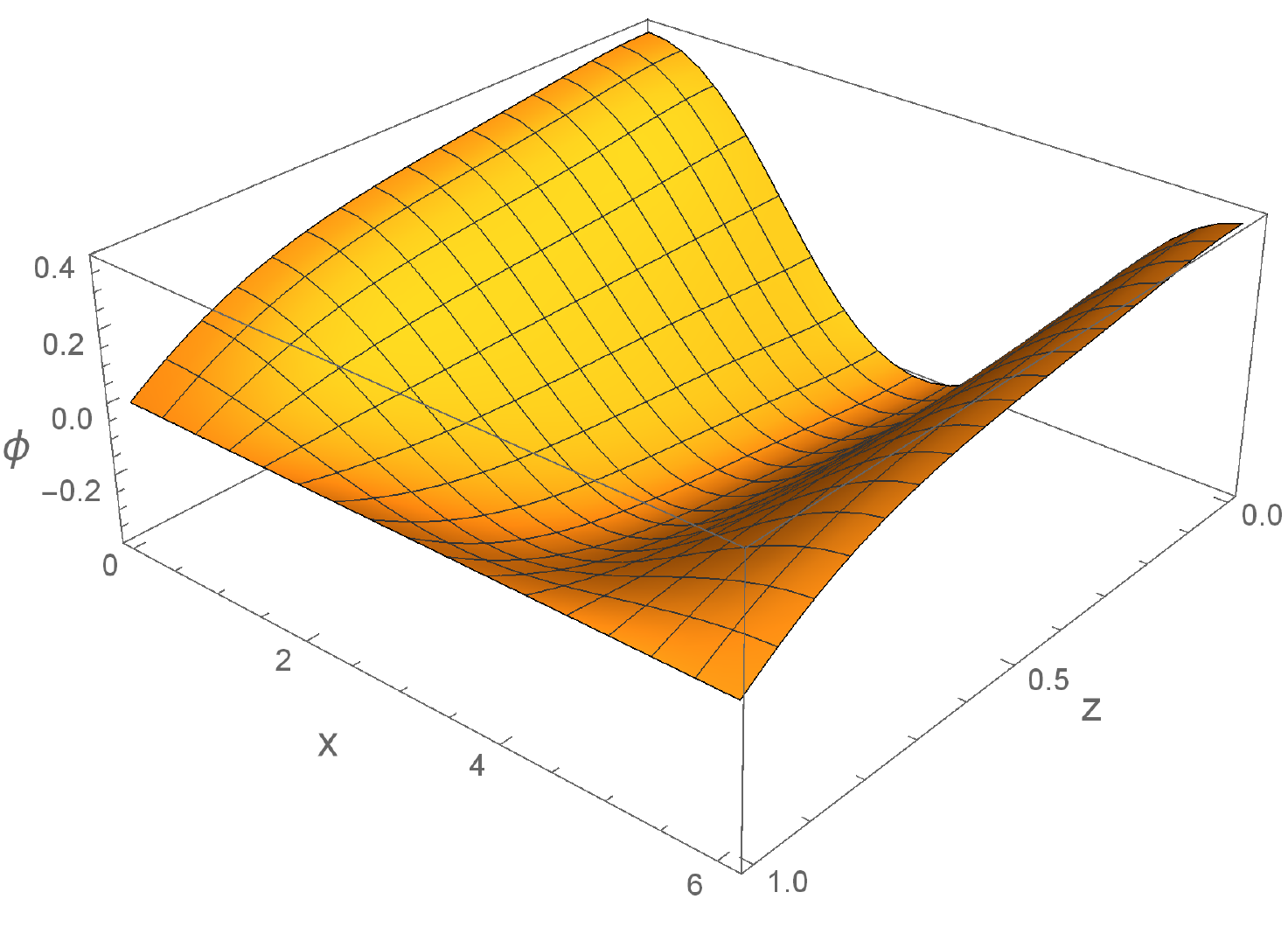}\qquad
\includegraphics[width=.45\textwidth]{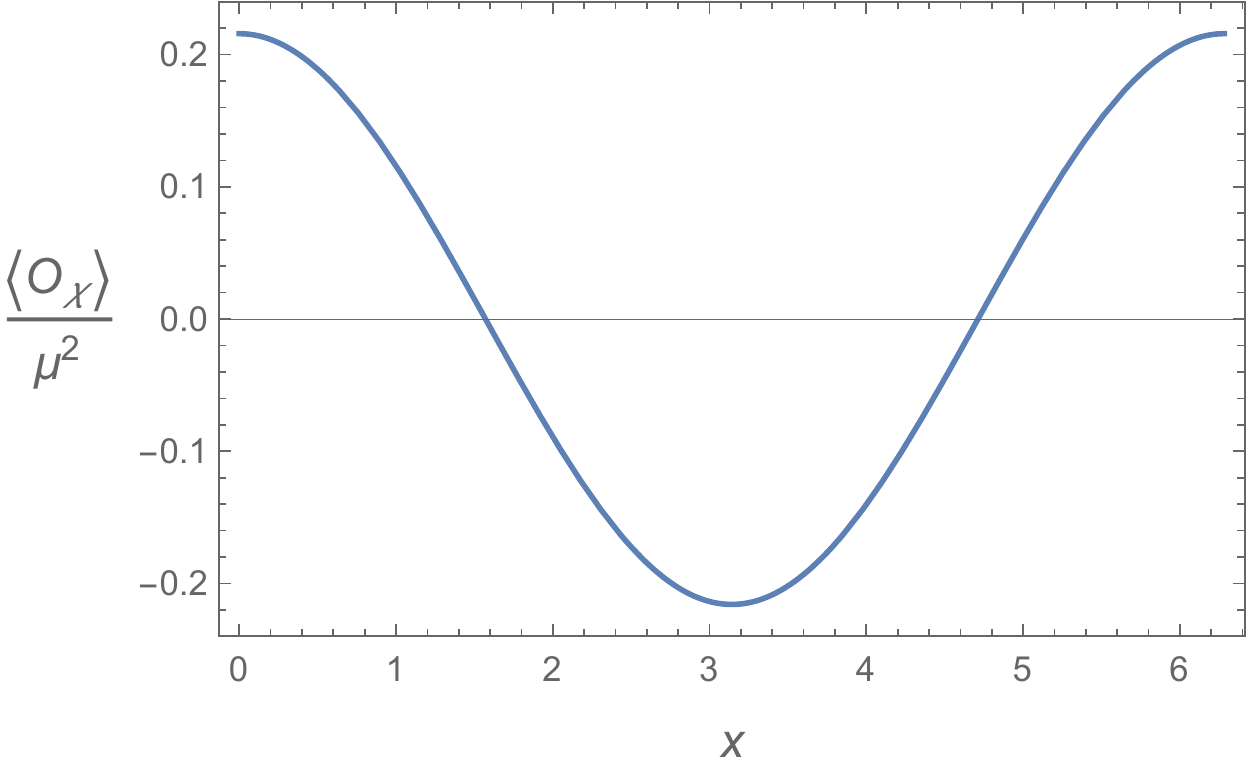}
\caption{Profile for the scalar field $\phi$ (left panel) and scalar VEV (right panel) for $T=0.01426$ and $k=1$,
for the parameters chosen in (\ref{couplingnumerics}) and $q_B=0$. }
\label{fig:solutionphi}
\end{center}
\end{figure}

\begin{figure}[ht!]
\begin{center}
\includegraphics[width=.45\textwidth]{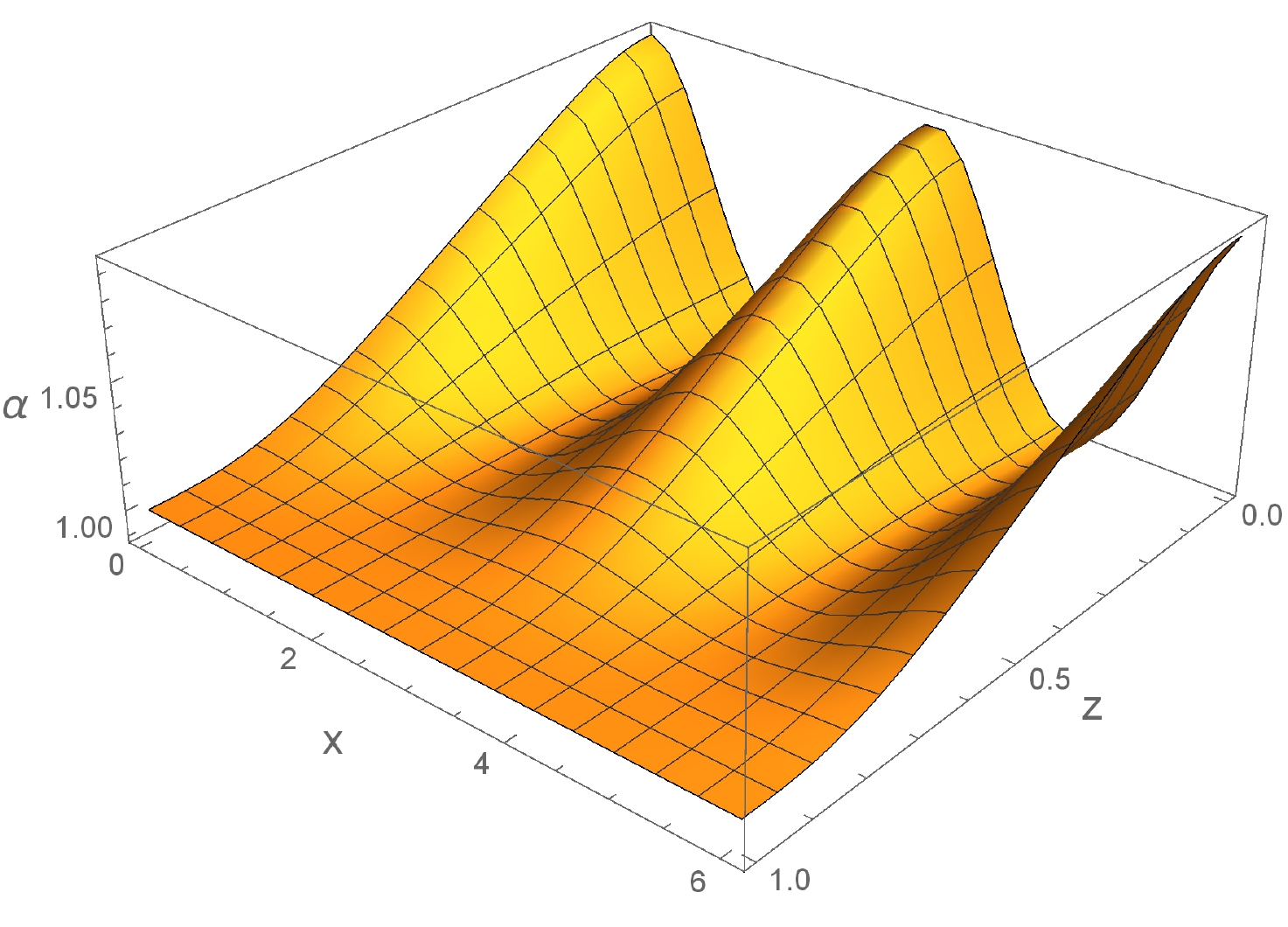}\qquad
\includegraphics[width=.45\textwidth]{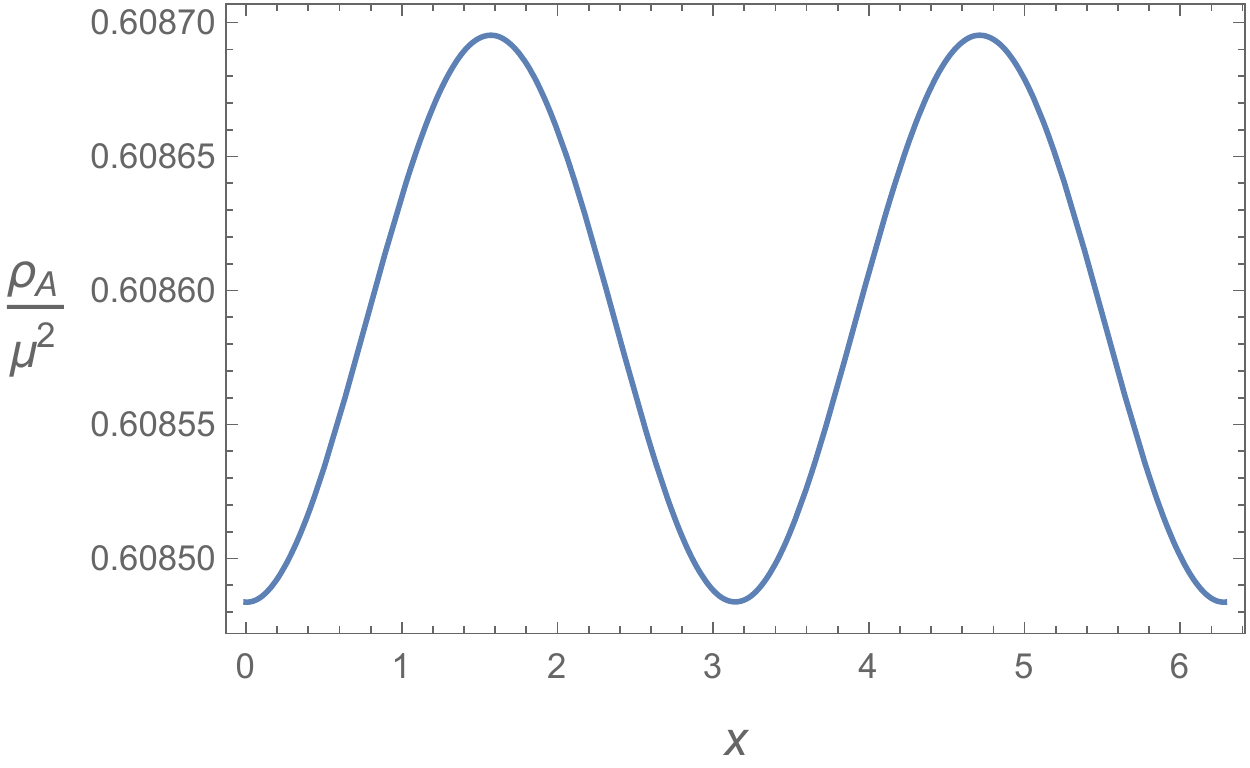}
\caption{Profile for the gauge field $\alpha$ (left panel) and charge density $\rho_A$ (right panel) for $T=0.01426$ and $k=1$,
for the parameters chosen in (\ref{couplingnumerics}) and $q_B=0$. }
\label{fig:solutionA}
\end{center}
\end{figure}

\begin{figure}[ht!]
\begin{center}
\includegraphics[width=.45\textwidth]{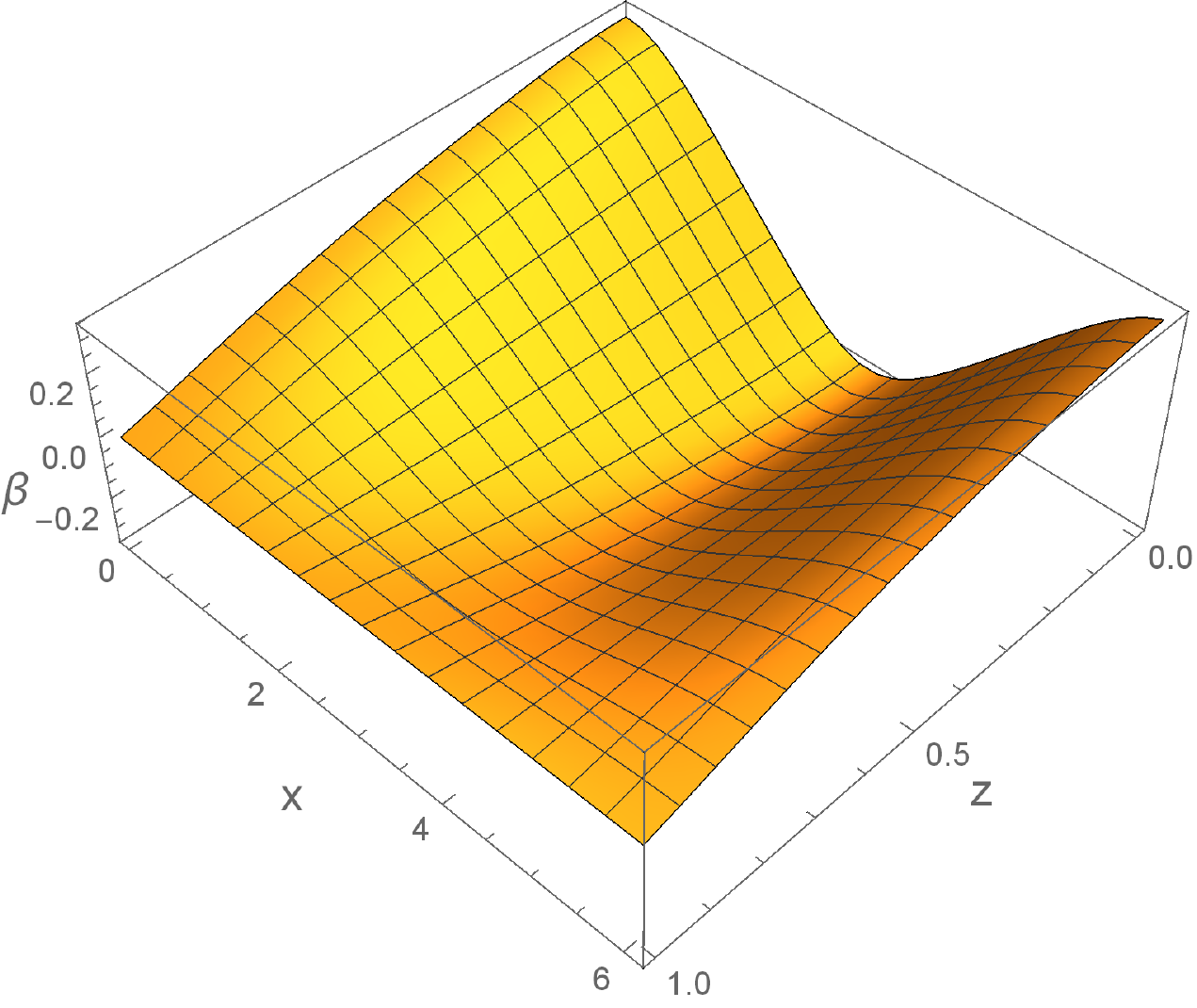}\qquad
\includegraphics[width=.45\textwidth]{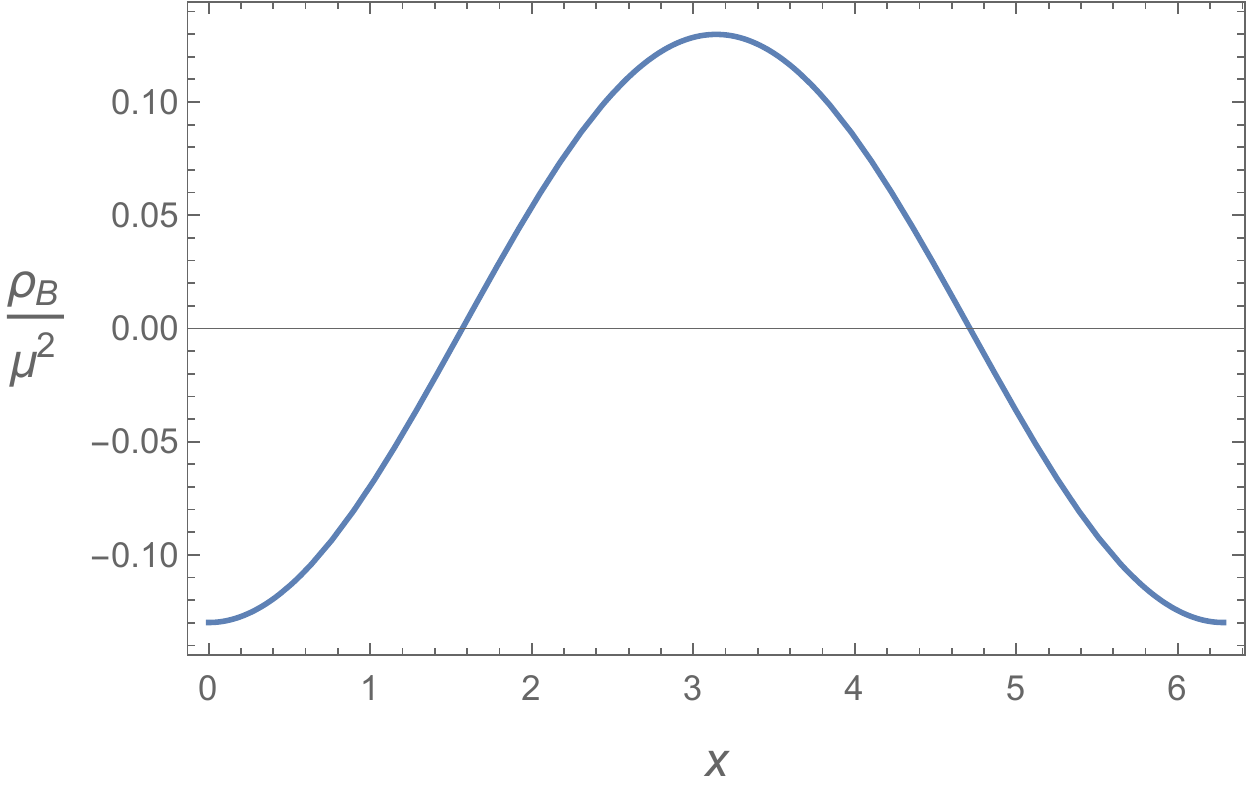}
\caption{Profile for the gauge field $\beta$ (left panel) and charge density $\rho_B$ (right panel) for $T=0.01426$ and $k=1$, for the parameters chosen in (\ref{couplingnumerics}) and $q_B=0$. Note that the average of $\rho_B$ is zero.}
\label{fig:solutionB}
\end{center}
\end{figure}

Representative profiles for the bulk fields for $q_B=0$ are shown in Figures~\ref{fig:solutionmetric} to~\ref{fig:solutionB}.
Note that one can clearly see that the spatial modulations are
imprinted on the horizon (at $z=0$), and decrease in overall magnitude as the UV is approached.
Indeed, the behavior of the strength of the striped oscillations, clearly visible from the figures, is in contrast to what is expected from explicit UV lattices of the type~\cite{Horowitz:2012ky,Donos:2014yya}.
In our model the striped feature is clearly strongest in the IR, and is therefore a relevant deformation of the UV field theory.

Figure~\ref{fig:solutionmetric} shows some of the functions appearing in the metric.
The left panel of Figure~\ref{fig:solutionphi} shows the dependence of the scalar field $\chi \propto \phi$ on both the holographic coordinate $z$ and the ``striped" spatial coordinate $x$, while in the right panel we plot the VEV of the scalar condensate as a function of $x$. We see from the latter that when $q_B=0$ the oscillations in the condensate $\langle O_\chi\rangle$ average out to zero, as required by a PDW state.

\begin{figure}[ht!]
\begin{center}
\includegraphics[width=0.65\textwidth]{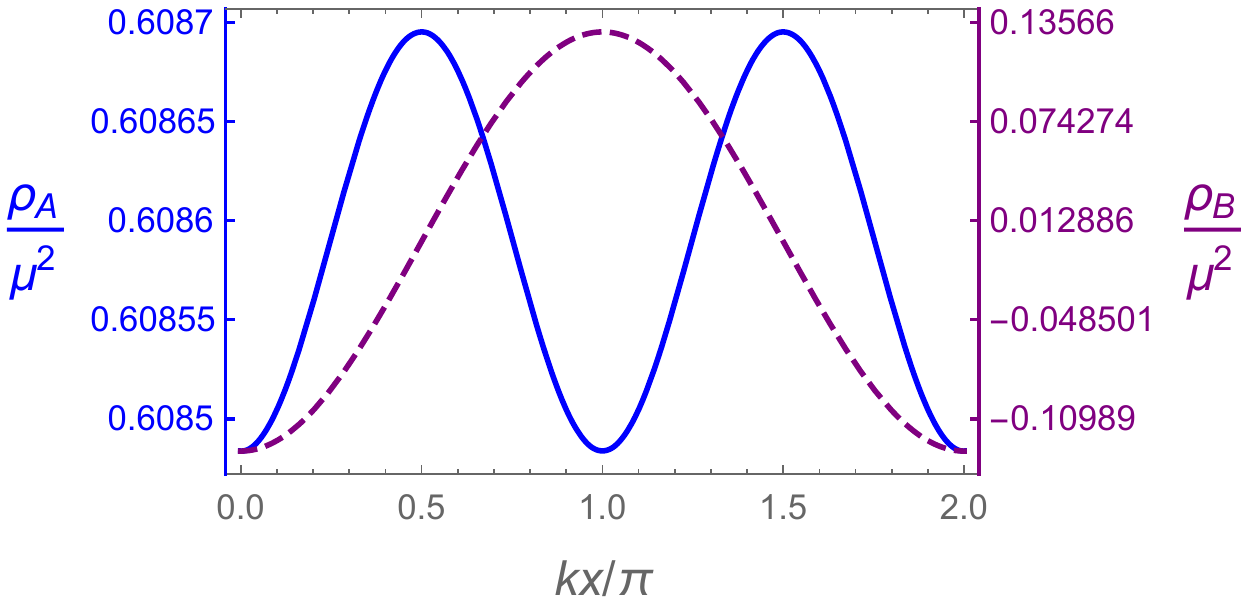}
\caption{The two charge densities for $q_B=0$, $T=0.01426$ and $k=1$. Here the frequency of $\rho_A$ is twice that of $\rho_B$ (and consequently of the scalar condensate).}
\label{fig:rhoAB-x}
\end{center}
\end{figure}

In Figure~\ref{fig:solutionA} we plot the profile for the gauge field $A_t \propto \alpha$ (left panel) and
for the corresponding charge density $\rho_A$ (right panel). The analogous plots for the second vector field are shown in Figure~\ref{fig:solutionB}.
We show the charge density profiles for the two vector fields as a function of the $x$ coordinate in Figure \ref{fig:rhoAB-x},
where they have been superimposed on top of each other.
It is clear from the figure that when $q_B=0$ the first vector field $A$ oscillates twice as fast as the second one, and also twice as fast as the scalar condensate, confirming the other crucial feature of the PDW.

\begin{figure}[ht!]
\begin{center}
\includegraphics[width=.45\textwidth]{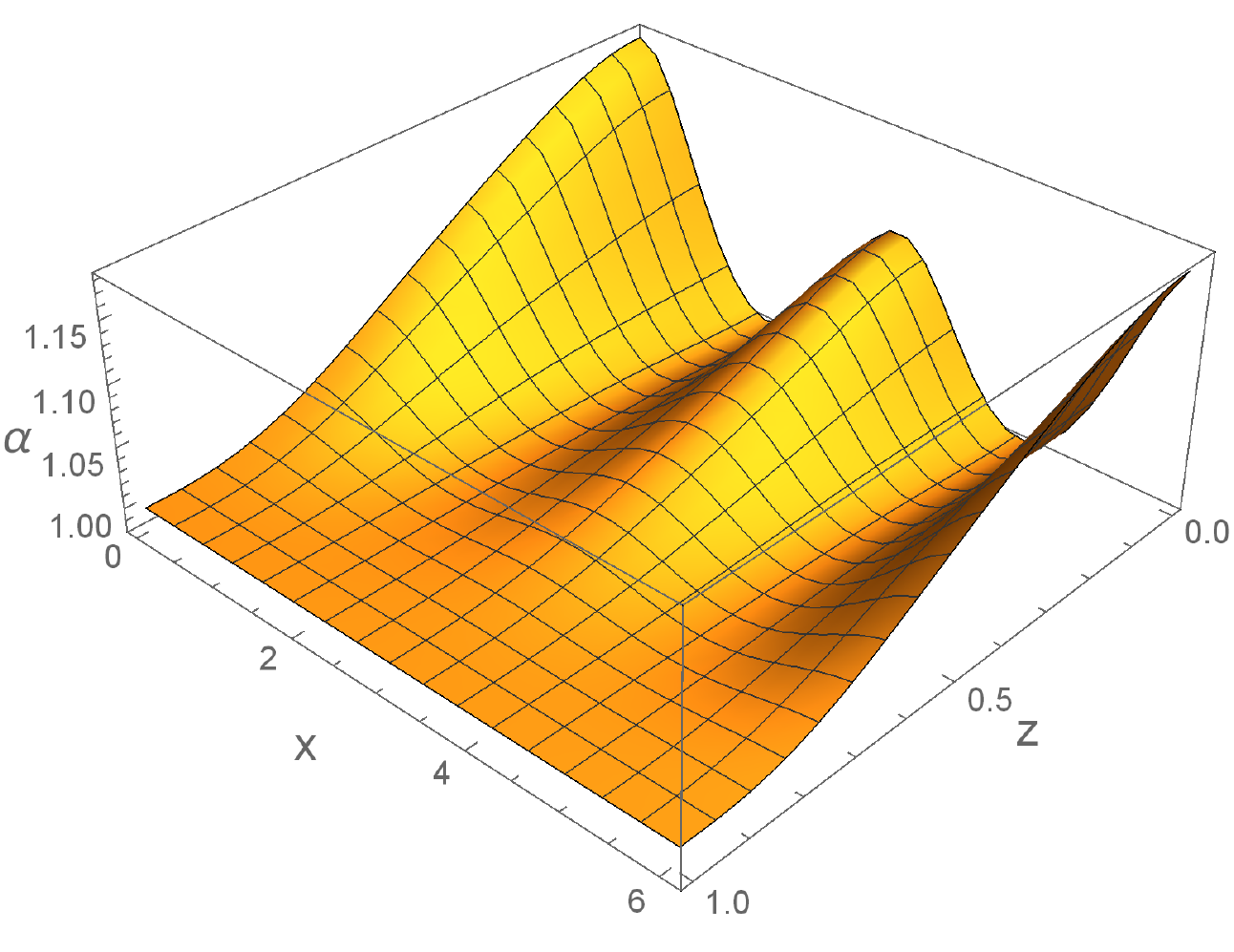}\qquad
\includegraphics[width=.45\textwidth]{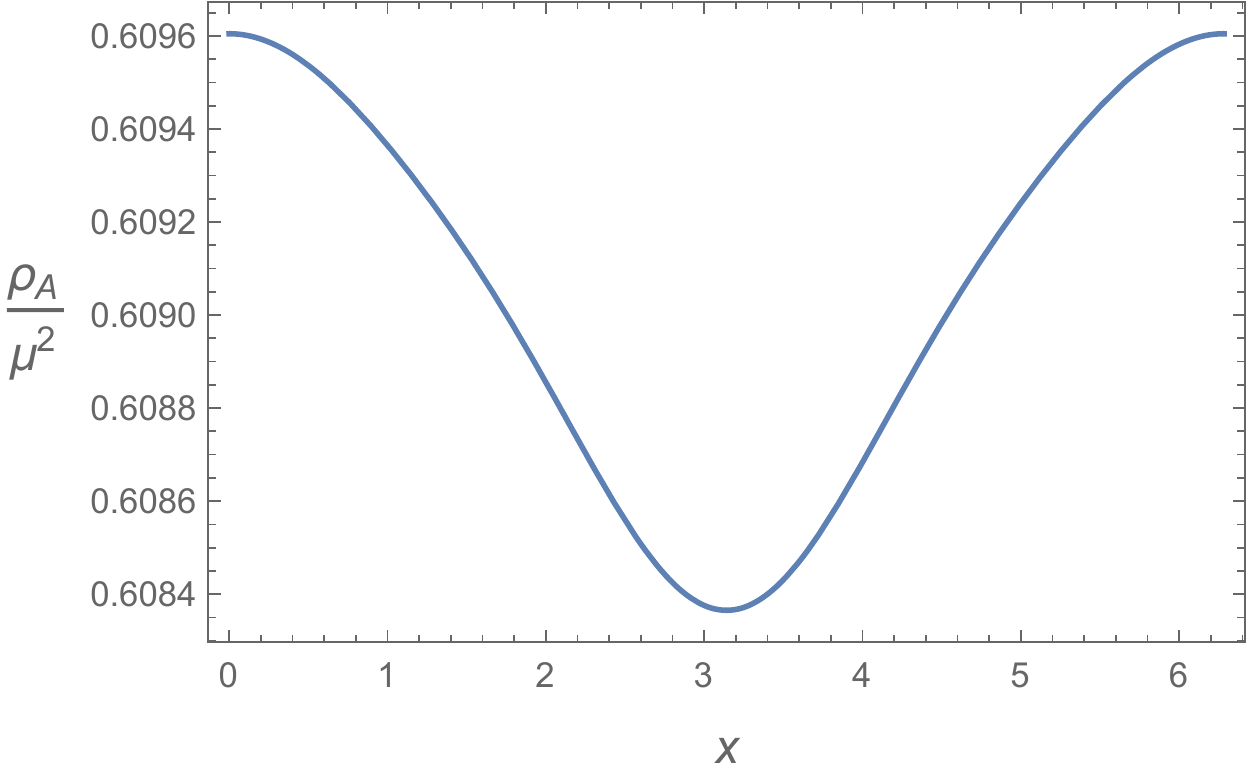}
\caption{Profile for the gauge field $\alpha$ (left panel) and charge density $\rho_A$ (right panel) for $T=0.01426$ and $k=1$, for the parameters chosen in (\ref{couplingnumerics}) and $q_B=0.5$.}
\label{fig:solutionAqB}
\end{center}
\end{figure}

\begin{figure}[ht!]
\begin{center}
\includegraphics[width=.45\textwidth]{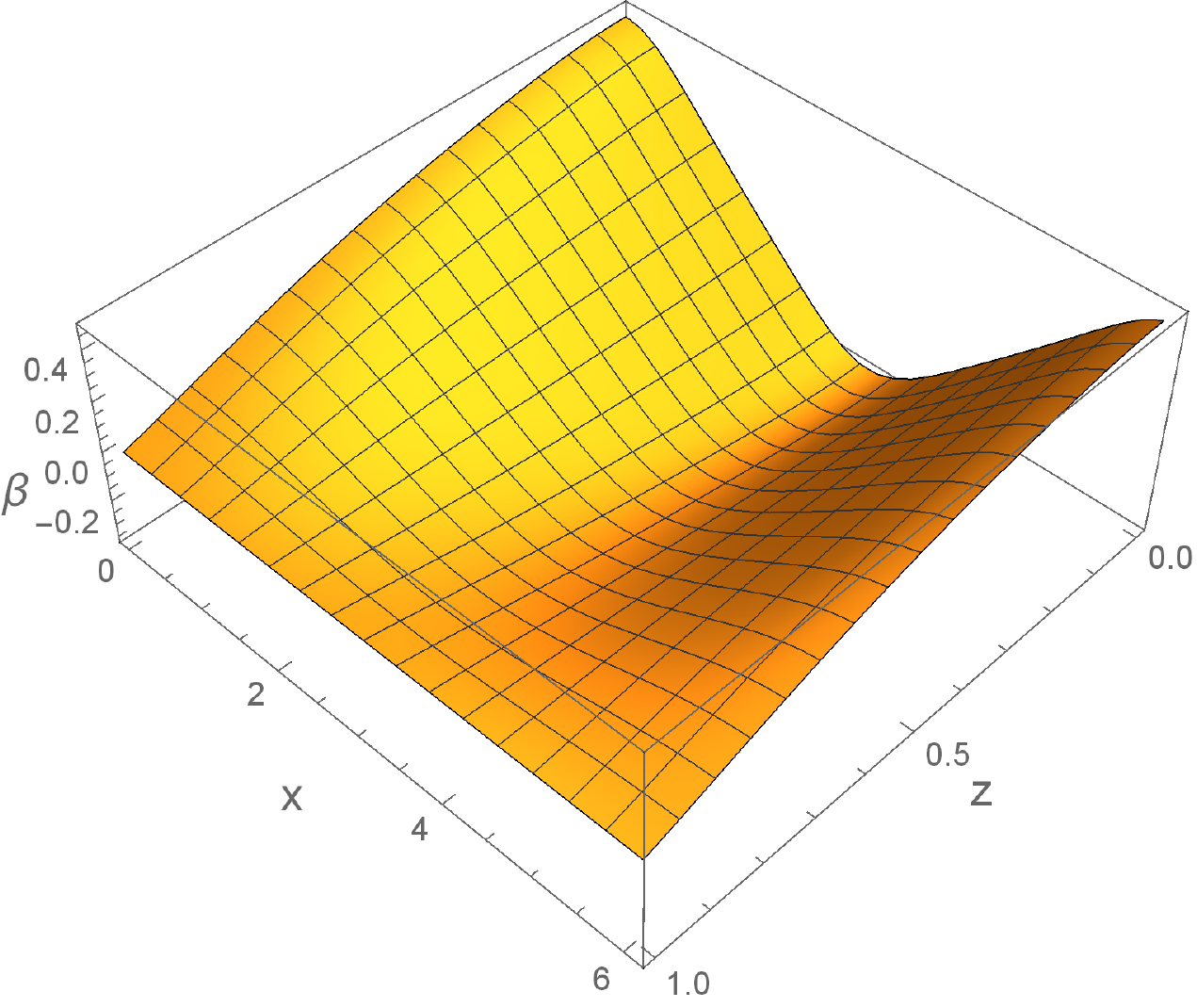}\qquad
\includegraphics[width=.45\textwidth]{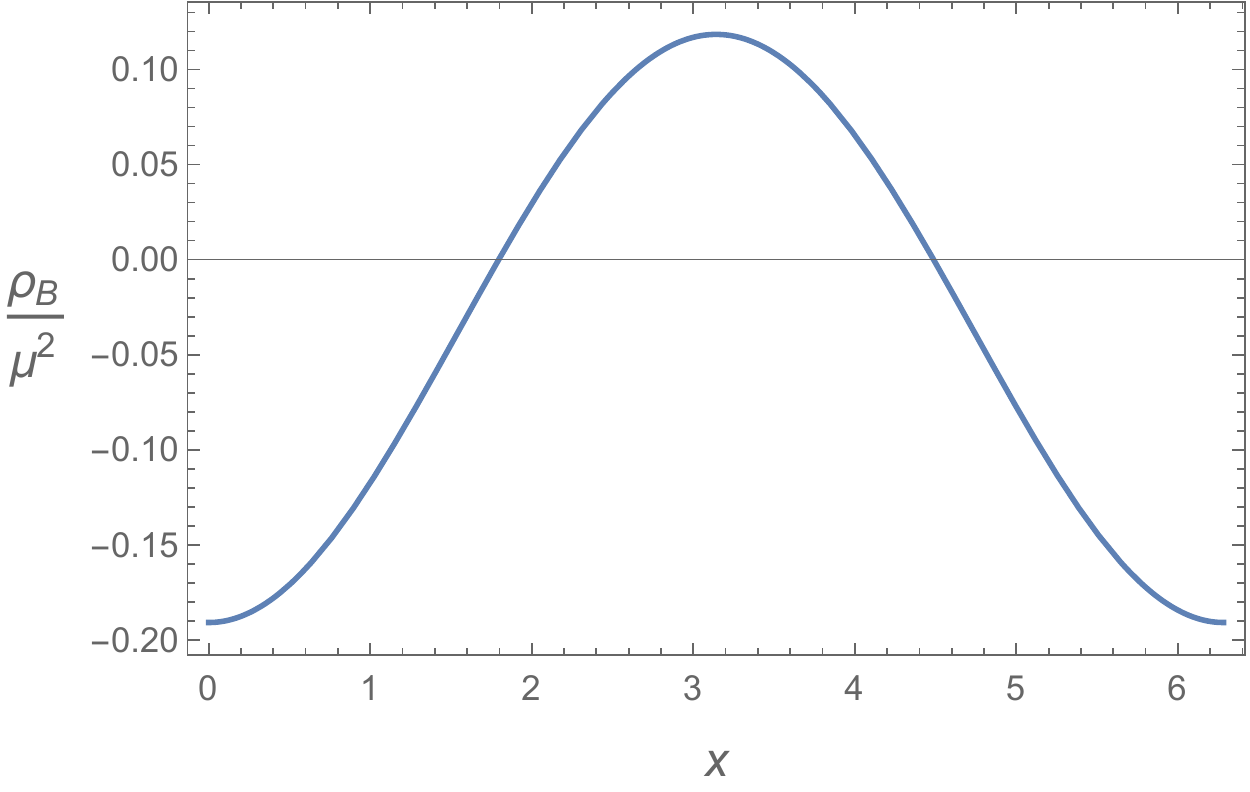}
\caption{Profile for the gauge field $\beta$ (left panel) and charge density $\rho_B$ (right panel) for $T=0.01426$ and $k=1$, for the parameters chosen in (\ref{couplingnumerics}) and $q_B=0.5$. In present case the average of $\rho_B$ is not zero.}
\label{fig:solutionBqB}
\end{center}
\end{figure}

The results are qualitatively different when the second charge is non-zero, $q_B \neq 0$.
Bulk configurations of the vector fields and their corresponding charge densities for $q_B=0.5$ are shown in Figures~\ref{fig:solutionAqB} and~\ref{fig:solutionBqB}.
It is clear that both vector fields share the same period, which is in sharp contrast to the $q_B=0$ case.
In this case one finds a homogeneous component to the scalar field condensate, which in turn implies that its oscillations no longer average out to zero.

\begin{figure}[ht!]
\begin{center}
\includegraphics[width=0.5\textwidth]{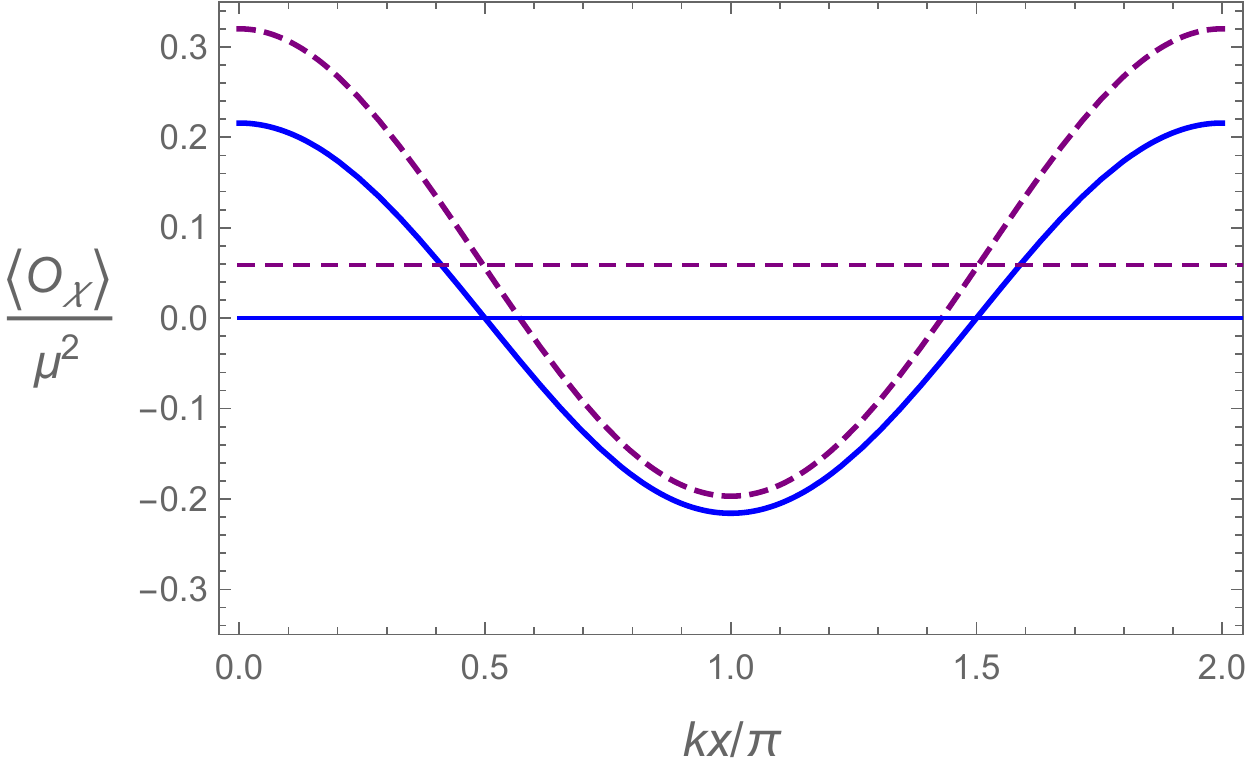}
\caption{The scalar condensate for $T=0.01426$ and $k=1$. The solid blue curve corresponds to $q_B=0$ and the dashed one is for $q_B=0.5$. Two horizontal lines denote their average values.}
\label{fig:condensatescomparison}
\end{center}
\end{figure}

\begin{figure}[ht!]
\begin{center}
\includegraphics[width=0.48\textwidth]{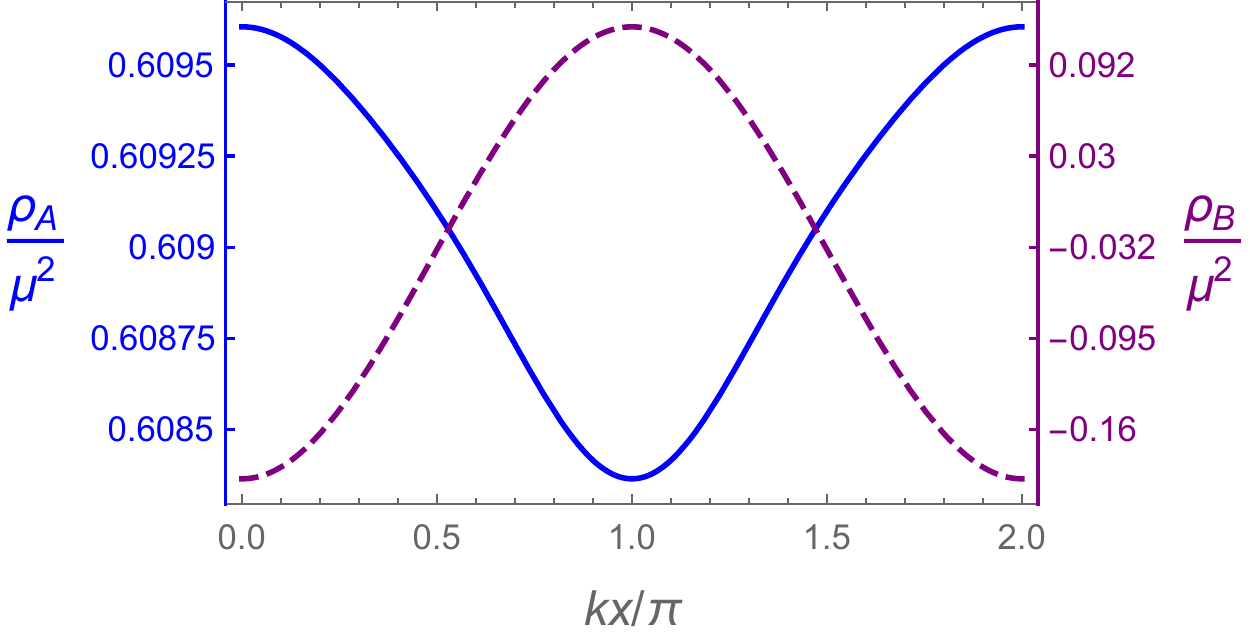}\quad
\includegraphics[width=0.48\textwidth]{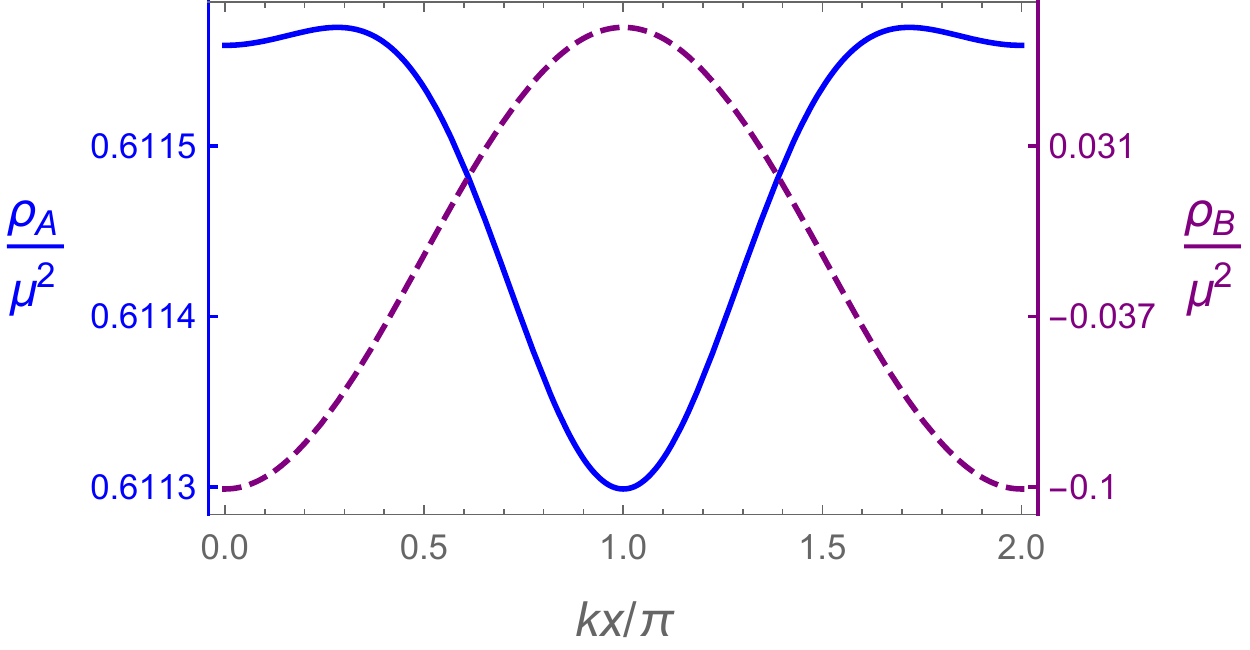}
\caption{The charge densities for $q_B=0.5$. We choose $T=0.01426$ (left panel) and $T=0.01571$ (right panel) with $k=1$.}
\label{fig:rhoAB-x-qB}
\end{center}
\end{figure}

The difference between the two cases, $q_B=0$ vs. $q_B\neq 0$, is shown in Figure~\ref{fig:condensatescomparison},
where we plot the corresponding profiles for $\langle O_\chi\rangle$ superimposed on each other. The dotted line describes the profile for $q_B =0.5$, with the average value of the oscillations given by the horizontal dotted line.
Finally, we plot the charge densities $\rho_A$ and $\rho_B$ when $q_B \neq 0$ in Figure~\ref{fig:rhoAB-x-qB}, for two different temperatures.
The crucial difference with the case shown in Figure~\ref{fig:rhoAB-x} is that the two densities now have the same period,
as already visible from the corresponding bulk fields in Figure~\ref{fig:solutionAqB} and Figure~\ref{fig:solutionBqB}.
Thus, we have reproduced the main features of a PDW ($q_B=0$) versus a CDW+SC ($q_B \neq 0$) state.


\section{Axions as Sources of Momentum Dissipation}
\label{AxionsSection}

The normal phase in our model, for which the gravitational background is the standard AdS-RN black hole,
is dual to a translationally invariant system with a finite charge density.
Therefore, it cannot relax momentum and would exhibit an infinite electric DC conductivity in the normal phase.
Also, in our model  translational invariance is broken spontaneously at low temperatures. Thus, even if the system were not superconducting, one would still expect to find generically an infinite
 conductivity, at least when the behavior of the Goldstone modes in the system is captured at large $N$ by the gravitational description.
Thus, to construct realistic models of dissipation in real materials, one needs to
 incorporate a mechanism for momentum relaxation.
A simple way in holography is to introduce massless scalar fields which are linear in one of the spatial coordinates~\cite{Andrade:2013gsa}, where the strength of momentum relaxation is identified with the associated proportionality constant.
Here we examine the role of such axionic scalars, on the details of the phase transition and on the background itself.
We should stress that the mechanism for generating striped order and the charge density modulations is the same as described in the previous sections, and is not due to the axions\,\footnote{Our paper provides a concrete holographic realization of the spontaneous formation of striped structures in two cases, with and without axions.
Alternatively, instead of using axions one could introduce a lattice via a spatially modulated chemical potential.
This was done recently in~\cite{Andrade:2017leb}, where the authors observed the commensurate lock-in between the stripes and the underlying lattice.}.

To this end, we would like to add two axions $(\psi_1,\psi_2)$ to our model~\eqref{actions},
\begin{equation}
\mathcal{L}_{m}\rightarrow \mathcal{L}_{m}-\frac{1}{2}(\partial_\mu\psi_1\partial^\mu\psi_1+\partial_\mu\psi_2\partial^\mu\psi_2).
\end{equation}
The normal phase  is now given by
\begin{equation}\label{RNadsaxion}
\begin{split}
&ds^2=\frac{1}{\tilde{f}(r)}dr^2- \tilde{f}(r)dt^2+\frac{r^2}{L^2}(dx^2+dy^2),\quad A_t=\mu\left(1-\frac{r}{r_h}\right),\\
&\tilde{f}(r)=\frac{r^2}{L^2}\left(1-\frac{r_h^3}{r^3}\right)+\frac{\mu^2 r_h^2}{4r^2}\left(1-\frac{r}{r_h}\right)-\frac{\tau^2 L^2}{2}\left(1-\frac{r_h}{r}\right),\\
&\psi_1=\tau\, x,\quad \psi_2=\tau\, y\,.
\end{split}
\end{equation}
with $r_h$ the horizon and $\mu$ the chemical potential. Notice that the axions depend on the spatial coordinates linearly, breaking translational invariance and giving rise to momentum relaxation, whose magnitude in this setup is going to be controlled by the parameter $\tau$.
The only role of the axions is to affect the blackening function and therefore the black brane temperature, which is now given by
\begin{equation}\label{temaxion}
T=\frac{r_h}{4\pi}\left[\frac{3}{L^2}-\frac{\mu^2}{4r_h^2}-\frac{L^2 \tau^2}{2 r_h^2}\right].
\end{equation}

The analysis of striped instabilities proceeds similarly to that of the standard AdS-RN background.
To take into account the axions one must turn on the following modes up to second order,
\begin{equation}\label{secdaxion}
\begin{split}
\delta \chi& =\varepsilon\, w(r)\cos(k\,x)+\varepsilon^2[\chi^{(1)}(r)+ \chi^{(2)}(r)\cos(2k\,x)]\,,\\
\delta B_t& =\varepsilon\, b_t(r)\cos(k\,x)+\varepsilon^2[b_{t}^{(1)}(r)+ b_{t}^{(2)}(r)\cos(2k\,x)]\,,\\
\delta \psi_1& = \varepsilon^2\, \psi^{(2)}(r)\sin(2k\,x)\,,\\
\delta A_t& = \varepsilon^2[a_{t}^{(1)}(r)+ a_{t}^{(2)}(r)\cos(2k\,x)]\,,\\
\delta g_{tt}& = \varepsilon^2[ h_{tt}^{(1)}(r)+h_{tt}^{(2)}(r)\cos(2k\,x)]\,,\\
\delta g_{xx}& =\varepsilon^2[h_{xx}^{(1)}(r)+ h_{xx}^{(2)}(r)\cos(2k\,x)]\,,\\
\delta g_{yy}& =\varepsilon^2[h_{yy}^{(1)}(r)+ h_{yy}^{(2)}(r)\cos(2k\,x)]\,.\\
\end{split}
\end{equation}
Substituting them into the equations of motion and expanding to linear order in $\varepsilon$ around the normal background~\eqref{RNadsaxion},
we obtain two coupled linear ordinary differential equations which are the same as those in (\ref{linearbt}), but with the original blackening function $f$ replaced by $ \tilde{f}$.
%
%
As usual, the critical temperature at the onset of striped order is determined by looking for zero mode solutions.
More precisely, we impose regularity conditions at the horizon $r=r_h$ and source free conditions at the AdS boundary $r\rightarrow \infty$.

\begin{figure}[ht!]
\begin{center}
\includegraphics[width=.5\textwidth]{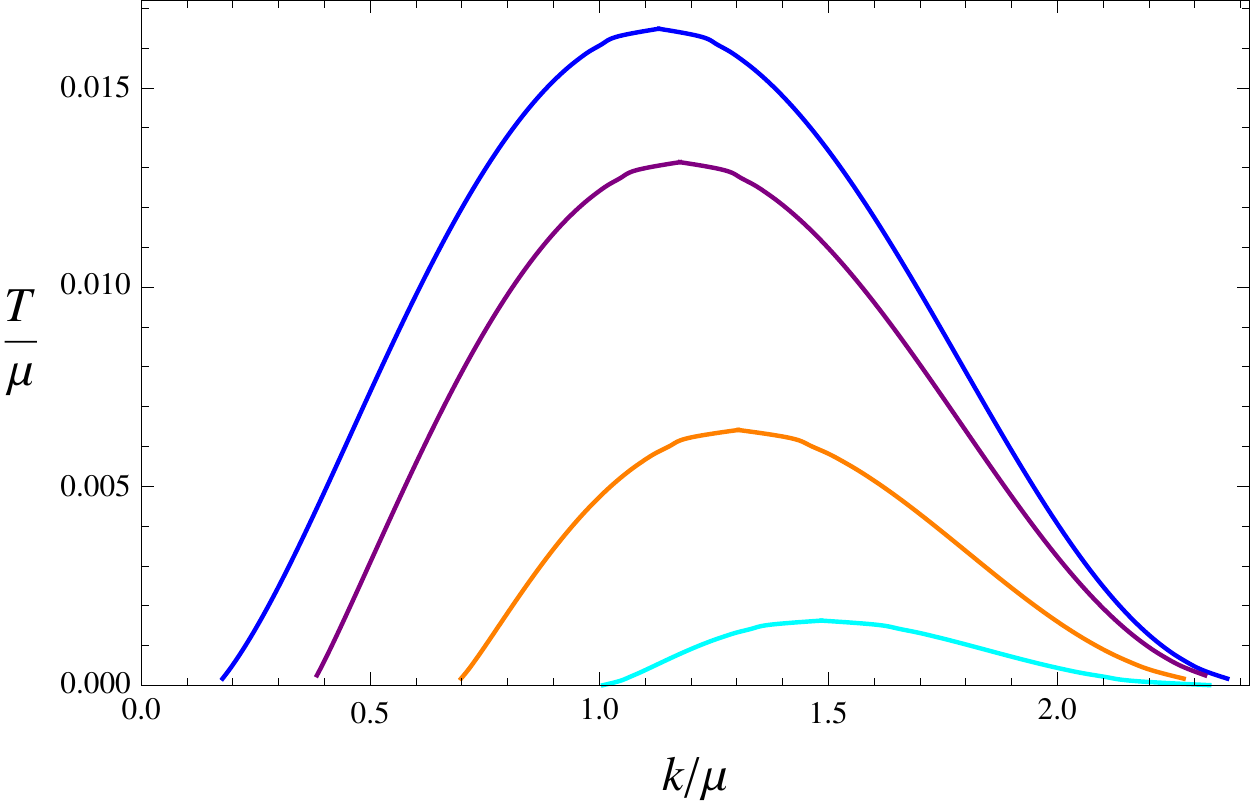}
\caption{Critical temperature versus wave number for the onset of striped instabilities in the presence of axions.
We have $m^2=-8, m_v^2=0, L=1/2, a=4, \kappa=q_A=1$, and from top to bottom $\tau/\mu=0.0, 0.5, 1.0, 1.5$.}
\label{fig:tckctau}
\end{center}
\end{figure}

The critical temperature will now depend on the strength of momentum dissipation $\tau$, and its dependence on wave number is shown in Figure~\ref{fig:tckctau}. We find bell curves similar to those in Figure~\ref{fig:tckccs}.
However, the presence of momentum dissipation suppresses the striped instabilities as compared to the case in which $\tau=0$ (the top curve in the figure).
Indeed, notice that as $\tau$ increases the magnitude of the peak decreases, and at the same time its location shifts to the right, \emph{i.e.} to larger values of $k$. We note that a similar feature in the behavior of $T_c$ was seen in the axionic model examined in~\cite{Andrade:2015iyf}.
While it would be interesting if this was a somewhat universal feature, it is not clear to us at this stage whether this is the case, and we suspect that it is instead model dependent.

We now construct the fully backreacted geometry that describes the striped phase. We choose the same couplings as~\eqref{couplingnumerics}.
Again, it is convenient to use the new coordinate~\eqref{rtoz}, in terms of which the background in the normal phase is still given by~\eqref{RNadsz} but with $F$ replaced by
\begin{equation}\label{RNadszaxion}
\begin{split}
&\tilde{F}(z)=z^2\left[2-z^2+(1-z^2)^2\left(1-\frac{L^4 \tau^2}{2 r_h^2}\right)-\frac{L^2 \mu^2}{4 r_h^2}(1-z^2)^3\right],
\end{split}
\end{equation}
the quantity explicitly impacted by the axions.
The horizon and boundary are still located, respectively, at $z=0$ and $z=1$, and the temperature is given by~\eqref{temaxion}.

\begin{figure}[h]
\begin{center}
\includegraphics[width=.49\textwidth]{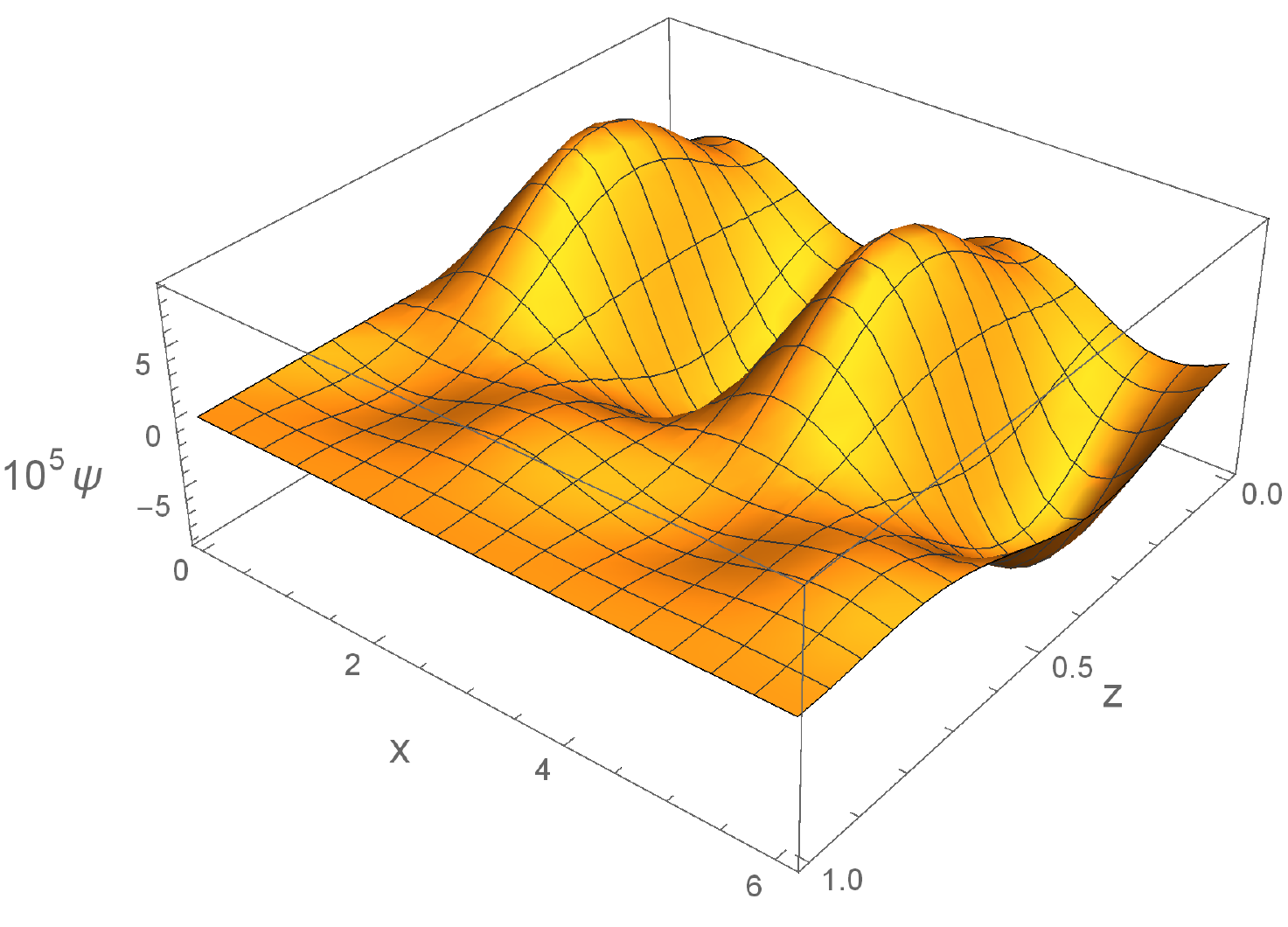}
\includegraphics[width=.49\textwidth]{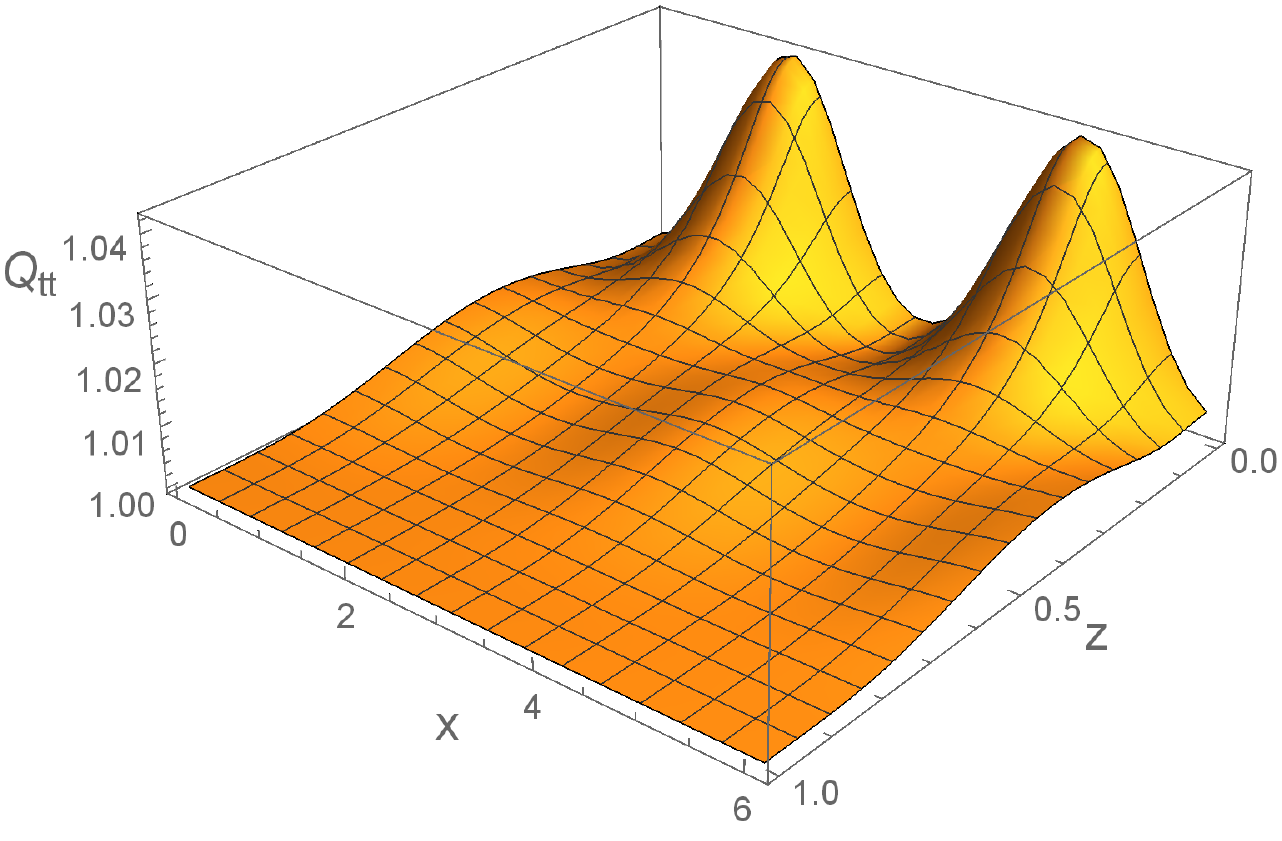}\\
\includegraphics[width=.49\textwidth]{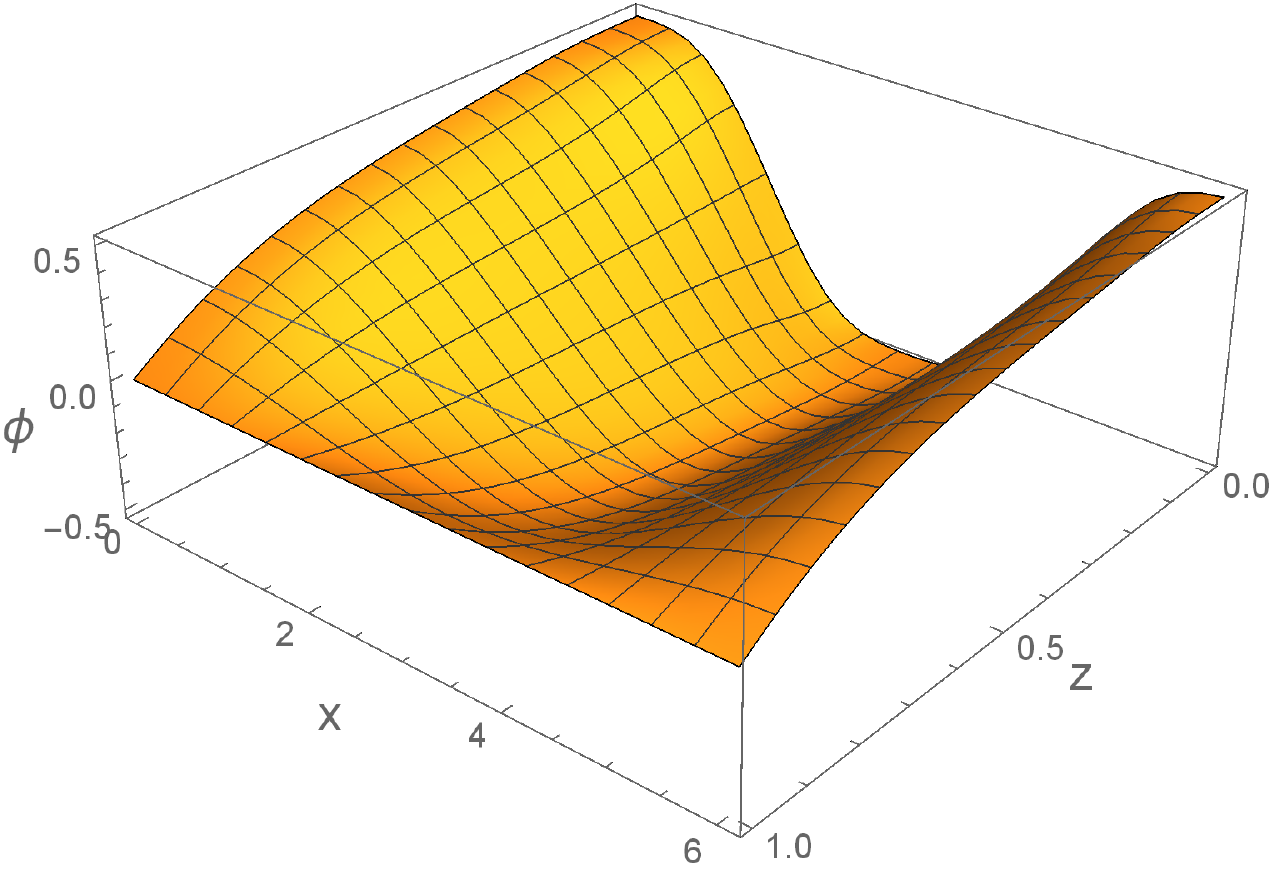}
\includegraphics[width=.49\textwidth]{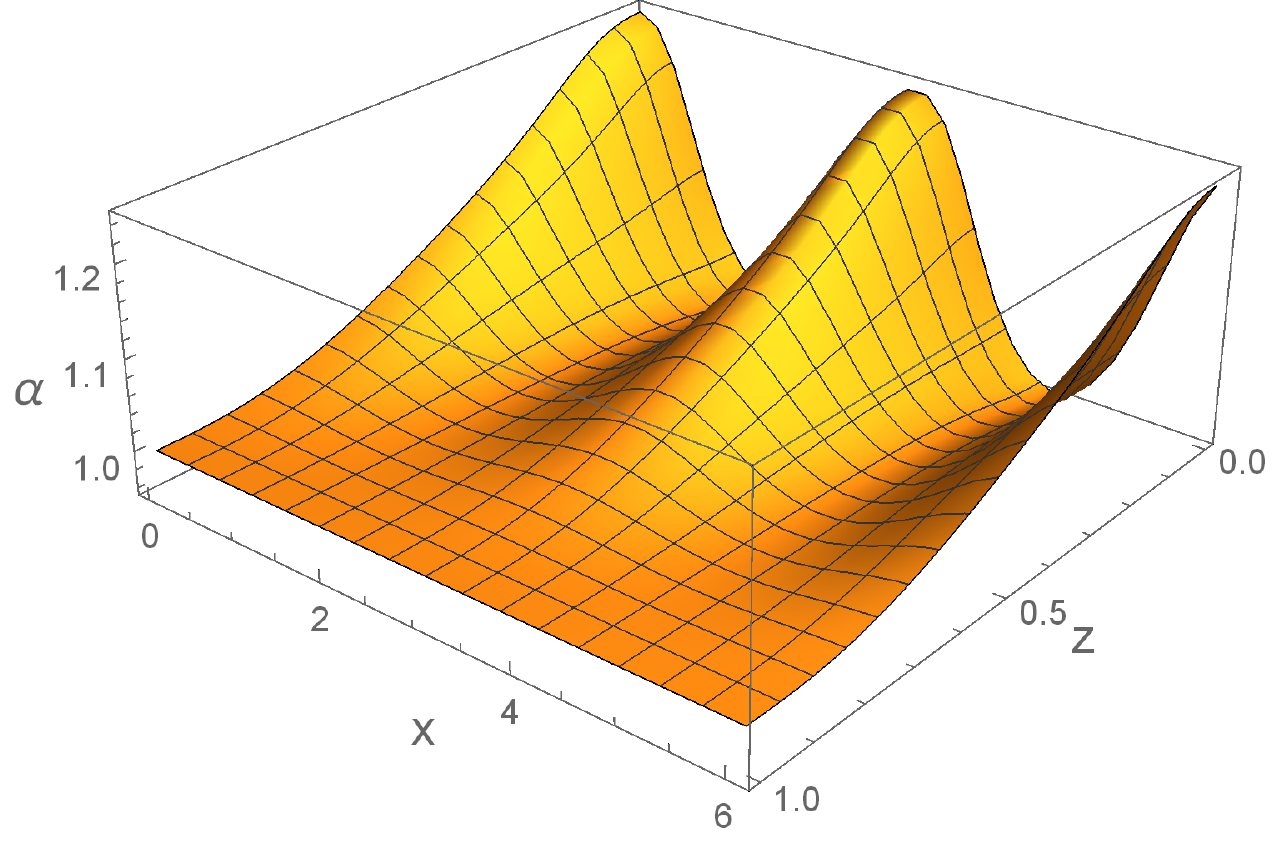}
\caption{Bulk solutions for $T=0.00737\mu$ and $k=1$ in the presence of axions.
The horizon is at $z=0$ and the conformal boundary at $z=1$. We have chosen the momentum relaxation strength  $\tau/\mu=0.5$ and the other parameters are the same as (\ref{couplingnumerics}).}
\label{fig:sol-axion}
\end{center}
\end{figure}

Our ansatz for the striped black brane geometry is
\begin{equation}\label{ansatzaxion}
\begin{split}
&ds^2=\frac{r_h^2}{L^2 (1-z^2)^2}\left[-\tilde{F}(z)Q_{tt} dt^2+\frac{4 z^2 L^4 Q_{zz}}{r_h^2 \tilde{F}(z)}dz^2+Q_{xx}(dx-2 z(1-z^2)^2Q_{xz}dz)^2+Q_{yy}dy^2\right],\\
&A_t=\mu\, z^2 \alpha,\quad B_t=z^2 \beta,\quad \chi=(1-z^2)\phi,\quad \psi_1=\tau\, x+\psi,\quad \psi_2=\tau\, y\,,
\end{split}
\end{equation}
where the nine functions $\mathcal{Q}=(\phi,\alpha,\beta, \psi, Q_{tt},Q_{zz},Q_{xx},Q_{yy}, Q_{xz})$ depend on $z$ and $x$.
The normal solution \eqref{RNadszaxion} is found by setting $\alpha=Q_{tt}=Q_{zz}=Q_{xx}=Q_{yy}=1$ and $\phi=\beta=\psi=Q_{xz}=0$, and once again we look for
solutions with a regular horizon at $z=0$, so that all the unknowns depend smoothly on $z^2$.
As we did for the case without axions, we impose one additional Dirichlet condition $Q_{tt}(0,x)=Q_{zz}(0,x)$ such that the temperature of the black brane~
is still given by~\eqref{temaxion}. In our numerics, we impose the following boundary condition at the conformal boundary $z=1$,
\begin{equation}\label{uvcondition}
\begin{split}
&Q_{tt}(1,x)=Q_{zz}(1,x)=Q_{xx}(1,x)=Q_{yy}(1,x)=\alpha(1,x)=1,\\
&\phi(1,x)=\beta(1,z)=Q_{xz}(1,x)=\psi(1,x)=0.
\end{split}
\end{equation}
In particular, notice that the last condition $\psi(1,x)=0$ means that the leading terms which are dual to the sources are just
$\psi_1=\tau\, x, \psi_2=\tau\, y$, the same as in the normal phase.
In the spatial direction $x$, we impose periodic boundary conditions.
The coupled PDEs are solved using the DeTurck method, and representative solutions for the bulk fields are given in Figure~\ref{fig:sol-axion}.
Their behavior qualitatively matches the perturbation analysis in~\eqref{secdaxion}.
While the phase structure is not impacted in any qualitative way by introducing $\psi_1$ and $\psi_2$, we expect to find interesting changes in the transport behavior in the system.

Indeed, we note that -- in the absence of a superconducting condensate-- when the translational symmetry breaking is spontaneous, the electrical conductivity should be divergent even along the direction of symmetry breaking~\cite{Delacretaz:2016ivq}. This can be explained by the presence of Goldstone modes in the system, which can be excited at no cost.
Without any explicit breaking of translational symmetry, the momentum relaxation rate $\Gamma$ and pinning frequency $\omega_o$ are both vanishing, so the DC conductivity is infinite.
Our model with axions provides a simple way to introduce a finite momentum relaxation rate $\Gamma$.
If the U(1) symmetry is preserved, so that one doesn't have a superconducting condensate,
then the axions would get rid of the divergence in the conductivity due to the
Goldstone modes.
However, when the U(1) symmetry is broken and the condensate forms, we expect to obtain an infinite DC conductivity, associated solely with the
superconducting excitation.
It will be interesting to see if one can see the pinning effect due to the CDW excitation from the behavior of the optical conductivity, in the presence of such axionic contributions.

\section{Final Remarks}
\label{FinalRemarks}

Our goal in this paper was to examine a gravitational model which would realize holographically certain features of a PDW, a phase in
which charge, superconducting and spin orders are intertwined with each other in a spatially modulated manner.
Specifically, we wanted to obtain a quantum system in which the scalar condensate and charge density are both striped, and are such that
the average value of the superconducting order parameter $\langle O_\chi\rangle$ vanishes (\emph{i.e.} there is no uniform component), and
the frequency of the charge density modulations is twice that of the condensate.
Our model indeed realizes both of these features -- two defining properties of PDW order -- by spontaneously breaking the global U(1) symmetry of the boundary theory, at the same time as translational invariance.
Thus, the resulting ordered phases are intertwined with each other and are understood to have a common origin.

We recall that our construction~\eqref{actions} is not of the form of the standard holographic superconductor, but instead falls within the generalized class of models advocated for in~\cite{Franco:2009yz}, in which the spontaneous breaking of the global U(1) symmetry happens via the St\"{u}ckelberg mechanism.
The local gauge invariance is now encoded in the transformation~\eqref{localgaugeinvariance},
and in these theories the two scalars $\chi$ (the field which condenses) and $\theta$ (which can be gauged fixed to zero)
are not necessarily associated with, respectively, the magnitude and phase of a complex scalar field.
In particular, in order to realize PDW order in our setup, $\chi$ must oscillate about zero, and is therefore allowed to be negative -- it cannot be directly identified with the magnitude of a complex scalar $\Psi = \chi e^{i \theta}$. In turn, this means that $\theta$ is not \emph{a priori} restricted to be periodic and
therefore the global U(1) symmetry that is being broken is non-compact (without imposing additional \emph{ad hoc} restrictions on the range of $\theta$).

Even though locally we expect to see the same features as in a theory with a compact U(1) symmetry (infinitesimally the two are the same), the global properties will be different,
and clearly one should be working with theories in which the symmetry is compact, as is usually the case.
We expect to be able to modify our model to ensure that the U(1) is indeed compact, and still obtain a PDW phase, but we anticipate that this requires
adding a more involved matter sector. Thus, we view our construction as only a first step towards realizing more sophisticated and realistic phase diagrams involving PDW order.
We have adopted this particular toy model because of its tractability, and take it as a proof of principle that the PDW features outlined above can indeed be realized. Guided by the intuition developed here, the model can then be refined and improved to be more relevant for phenomenology,
and to yield more realistic order parameters, along the lines of \emph{e.g.} \cite{JaefariFradkin2012}.
We postpone such efforts to future work.

Recall that the lack of a uniform component of $\langle O_\chi\rangle$ in our construction was associated with taking the charge $q_B$ to be vanishing.
When $q_B \neq 0$ on the other hand $\langle O_\chi\rangle$ contains a homogeneous, non-oscillating piece, and the CDW oscillations have the same period as those of the scalar condensate. Thus, we have co-existing SC and CDW orders, as opposed to a PDW.
While the density $\rho_A$ associated with the first vector field $A_\mu$ has the natural interpretation of a charge density (leading to CDW order),
the physical interpretation of the density $\rho_B$ of $B_\mu$ is less clear.
Naively, it obeys some of the properties of a SDW, \emph{i.e.} it is also striped and always oscillates at the same frequency as the scalar condensate.
However, to be able to claim that $\rho_B$ is really describing a density of spins (leading to SDW order), one must justify physically the
particular couplings and interactions chosen in our model, which are needed to realize these symmetry breaking patterns.
Perhaps more natural couplings can be introduced, which would allow for a more realistic identification between the degrees of freedom of our model and those in a
real high $T_c$ superconductor exhibiting these kinds of intertwined orders.
However, the second gauge field can also be thought of as a ``spectator'' field, whose only role is to seed the oscillations of the charge density $\rho_A$.
Indeed,  it is important to emphasize that in our model CDW order is  induced from the
oscillations of the scalar condensate and of $\rho_B$, as can be understood at least qualitatively by inspecting the perturbation analysis.

 An interesting question is that of the behavior of the conductivity in these systems.
As we discussed in Section \ref{AxionsSection}, the high temperature phase in our model is dual to a translationally invariant system, which lacks a mechanism
to dissipate momentum. Moreover, the low-temperature phase describes a superconducting condensate and therefore exhibits an infinite conductivity.
However, even if we turned off the scalar condensate and focused only on the CDW state of the broken symmetry phase, we would still expect the conductivity to be generically infinite.
This can be traced back to the fact that the translational symmetry in our model is broken spontaneously
-- the resulting Goldstone modes can in principle be excited at no cost, leading to an infinite conductivity.
To remedy this problem we have added axions to our model -- a simple way to generate momentum dissipation in the system -- and have examined the full backreaction.
While the axions don't affect in any qualitative way the PDW (or SC + CDW) features explored in this paper, we expect them to play a crucial role
in the transport properties of the system.
In particular, if we were to restrict our attention to the case in which the U(1) symmetry is preserved, \emph{i.e.} no superconducting condensate,
we would expect the axionic case to lead to a finite conductivity.
It would be interesting to explore the conductivity in this framework (as well as the pinning effect), and how it is impacted by the presence or absence of an explicit breaking of translational invariance.

Finally, we close with a few words on the nature of the ground state.
Models of the type we are considering are known to support hyperscaling violating geometries,
which can be obtained by suitable choices of scalar couplings and potential.
Such backgrounds are well-known to develop both striped~\cite{Cremonini:2013epa,Cremonini:2012ir,Iizuka:2013ag,Withers:2014sja,Gouteraux:2016arz}
and superfluid instabilities  (see~\cite{Cremonini:2016bqw} for general criteria).
It would be interesting to generalize our study of intertwined orders to this case.
We leave these questions to future work.

\newpage
\acknowledgments
\label{ACKNOWL}

We would like to thank Silviu Pufu for useful discussions.
The work of S.C. is supported in part by the National Science Foundation grant PHY-1620169. J.R. is partially supported by the American-Israeli Bi-National Science Foundation, the Israel Science Foundation Center of Excellence and the I-Core Program of the Planning and Budgeting Committee and The Israel Science Foundation ``The Quantum Universe."

\appendix
\renewcommand{\theequation}{\thesection.\arabic{equation}}
\addcontentsline{toc}{section}{Appendix}
\section*{Appendices}

\section{Perturbation Equations at Next-to-Leading Order \label{app:eoms}}

In this appendix we provide the detailed equations of motion at next-to-leading order in fluctuations for the theory~\eqref{actions} and~\eqref{coupling}. We turn on the following perturbations up to second order,
\begin{equation}\label{secondorder}
\begin{split}
\delta \chi& =\varepsilon\, w(r)\cos(k\,x)+\varepsilon^2[\chi^{(1)}(r)+ \chi^{(2)}(r)\cos(2k\,x)],\\
\delta B_t& =\varepsilon\, b_t(r)\cos(k\,x)+\varepsilon^2[b_{t}^{(1)}(r)+ b_{t}^{(2)}(r)\cos(2k\,x)],\\
\delta A_t& = \varepsilon^2[a_{t}^{(1)}(r)+ a_{t}^{(2)}(r)\cos(2k\,x)],\\
\delta g_{tt}& = \varepsilon^2[ h_{tt}^{(1)}(r)+h_{tt}^{(2)}(r)\cos(2k\,x)],\\
\delta g_{xx}& =\varepsilon^2[h_{xx}^{(1)}(r)+ h_{xx}^{(2)}(r)\cos(2k\,x)],\\
\delta g_{yy}& =\varepsilon^2[h_{yy}^{(1)}(r)+ h_{yy}^{(2)}(r)\cos(2k\,x)].\\
\end{split}
\end{equation}
Plugging all modes~\eqref{secondorder} into the equations of motion and expanding around the AdS-RN background~\eqref{RNads},
we obtain four sets of decoupled equations of motion at $\mathcal{O}(\varepsilon^2)$, for the following perturbations:
\begin{itemize}
  \item (I) The homogeneous components of $\chi$ and $B_t$: ($\chi^{(1)}, b_{t}^{(1)}$)
  \item (II) The non-homogeneous components of $\chi$ and $B_t$: ($\chi^{(2)}, b_{t}^{(2)}$)
  \item (III) The homogeneous components of the metric and $A_t$: ($a_{t}^{(1)}, h_{tt}^{(1)}, h_{xx}^{(1)}, h_{yy}^{(1)}$)
   \item (IV) The non-homogeneous components of the metric and $A_t$: ($a_{t}^{(2)}, h_{tt}^{(2)}, h_{xx}^{(2)}, h_{yy}^{(2)}$).
\end{itemize}
All these modes are in general sourced by the linear order zero modes $(w, b_t)$, which can be obtained by solving~\eqref{linearbt}.

\subsection{Set (I)}
This set describes the homogeneous contribution to the condensation $\langle O_\chi\rangle$ dual to $\chi$ and to the charge density $\rho_B$ associated with $B_t$.
The equations of motion are given by
\begin{eqnarray}
\chi''^{(1)}+\left(\frac{f'}{f}+\frac{2}{r}\right)\chi'^{(1)}-\frac{1}{f}\left[m^2-\frac{a \mu^2 r_h^2}{2 r^4}-\frac{\kappa q_A^2 \mu^2}{f}(1-\frac{r_h}{r})^2\right]\chi^{(1)} \nonumber \\
+\frac{c\mu r_h}{r^2 f} b_{t}'^{(1)}=-q_A q_B\frac{\kappa\mu}{f^2}\left(1-\frac{r_h}{r}\right) b_t w\,,\\
b_{t}''^{(1)} +\frac{2}{r} b_{t}'^{(1)}+\frac{c \mu r_h}{r^2}\chi'^{(1)}-\frac{m_v^2}{f}b_{t}^{(1)}=  q_A q_B\frac{\kappa\mu}{2f}\left(1-\frac{r_h}{r}\right)w^2\,.
\end{eqnarray}
Notice that the source terms are proportional to $q_A q_B$.
In particular, when $q_A q_B\neq 0$, the modes $\chi^{(1)}, b_{t}^{(1)}$ have to be turned on, and therefore the scalar condensate $\langle O_\chi\rangle$ and charge density $\rho_B$ dual to $B_t$ acquire uniform values.
On the other hand, if $q_A q_B=0$, we can consistently set $\chi^{(1)}=b_{t}^{(1)}=0$.
In this case $\langle O_\chi\rangle$ and $\rho_B$ would have no homogeneous components.

\subsection{Set (II)}
This set gives the non-homogeneous contribution to the condensation $\langle O_\chi\rangle$ dual to $\chi$ and charge density $\rho_B$ associated with $B_t$.
The associated equations of motion are given by
\begin{eqnarray}
\chi''^{(2)}+\left(\frac{f'}{f}+\frac{2}{r}\right)\chi'^{(2)}-\frac{1}{f}\left[m^2+\frac{4 k^2 L^2}{r^2}-\frac{a \mu^2 r_h^2}{2 r^4}-\frac{\kappa q_A^2 \mu^2}{f}(1-\frac{r_h}{r})^2\right]\chi^{(1)} \nonumber \\
+\frac{c\mu r_h}{r^2 f} b_{t}'^{(2)}=-q_A q_B\frac{\kappa\mu}{f^2}\left(1-\frac{r_h}{r}\right) b_t w\,,\\
b_{t}''^{(2)} +\frac{2}{r} b_{t}'^{(2)}+\frac{c \mu r_h}{r^2}\chi'^{(2)}-\frac{1}{f}\left(m_v^2+\frac{4 k^2 L^2}{r^2}\right)b_{t}^{(2)}=  q_A q_B\frac{\kappa\mu}{2f}\left(1-\frac{r_h}{r}\right)w^2\,,
\end{eqnarray}
with the sources still proportional $q_A q_B$.

If $q_A q_B\neq 0$, $\chi^{(2)}$ and $b_{t}^{(2)}$ have to be turned on, and therefore the scalar condensate $\langle O_\chi\rangle$ and density $\rho_B$
acquire periodic modes with wave number $2 k$.
On the other hand, when $q_A q_B=0$ we can consistently set $\chi^{(2)}=b_{t}^{(2)}=0$ and $\langle O_\chi\rangle$ and $\rho_B$ have no periodic modes with wave number $2 k$.

\subsection{Set (III)}
This set involves the homogeneous components of the metric and $A_t$, \emph{i.e.} ($a_{t}^{(1)}, h_{tt}^{(1)}, h_{xx}^{(1)}, h_{yy}^{(1)}$).
The corresponding equations of motion are given by:
\begin{eqnarray}
a_t''^{(1)}+\frac{2}{r}a_t'^{(1)}+\frac{r_h \mu}{2 r^2 f}h_{tt}'^{(1)}+\frac{L^2 r_h \mu}{2 r^4}\left(h_{xx}'^{(1)}+h_{yy}'^{(1)}\right)-\frac{r_h\mu f'}{2 r^2 f^2}h_{tt}^{(1)}\nonumber\\-\frac{L^2 r_h\mu}{r^5}(h_{xx}^{(1)}+h_{xx}^{(1)})=\mathcal{C}_1^{(1)},
\end{eqnarray}
\begin{eqnarray}\label{2ndtt1}
h_{xx}''^{(1)}+h_{yy}''^{(1)}+\frac{1}{2}\left(\frac{f'}{f}-\frac{2}{r}\right)(h_{xx}'^{(1)}+h_{yy}'^{(1)})-\frac{2}{L^2 f^2}\left(f+r f'-\frac{3 r^2}{L^2}\right)h_{tt}^{(1)}\nonumber\\+\frac{r_h\mu}{L^2 f}a_{t}'^{(1)}-\frac{f'}{r f}(h_{xx}^{(1)}+h_{yy}^{(1)})=\mathcal{C}_2^{(1)},
\end{eqnarray}
\begin{eqnarray}
h_{xx}''^{(1)}-\frac{r^2}{L^2 f}h_{tt}''^{(1)}+\left(\frac{f'}{f}-\frac{2}{r}\right)h_{xx}'^{(1)}+\frac{r(r f'-2f)}{2 L^2 f^2} h_{tt}'^{(1)}-\frac{r_h\mu}{L^2 f}a_{t}'^{(1)}\nonumber\\+\frac{2(f-rf')}{r^2 f}h_{xx}^{(1)}-\frac{L^2 r^2 f'^2+2rf(L^2 f'-6r)}{2 L^4 f^3}h_{tt}^{(1)}=\mathcal{C}_3^{(1)},
\end{eqnarray}
\begin{eqnarray}
h_{yy}''^{(1)}-\frac{r^2}{L^2 f}h_{tt}''^{(1)}+\left(\frac{f'}{f}-\frac{2}{r}\right)h_{yy}'^{(1)}+\frac{r(r f'-2f)}{2 L^2 f^2} h_{tt}'^{(1)}-\frac{r_h\mu}{L^2 f}a_{t}'^{(1)}\nonumber\\+\frac{2(f-rf')}{r^2 f}h_{yy}^{(1)}-\frac{L^2 r^2 f'^2+2rf(L^2 f'-6r)}{2 L^4 f^3}h_{tt}^{(1)}=\mathcal{C}_4^{(1)},
\end{eqnarray}
\begin{eqnarray}\label{2ndrr1}
h_{tt}'^{(1)}-\frac{L^2(rf'+2f)}{4 r^2}(h_{xx}'^{(1)}+h_{yy}'^{(1)})-\frac{r_h\mu}{2r}a_{t}'^{(1)}+\frac{L^2(r f'+2f)}{2 r^3}(h_{xx}^{(1)}+h_{yy}^{(1)})\nonumber\\-\frac{4 r^3 f'+r_h^2\mu^2}{4 r^3 f}h_{tt}^{(1)}=\mathcal{C}_5^{(1)}.
\end{eqnarray}
Here the five source terms $\mathcal{C}_i^{(1)} (i=1,...,5)$ depend on the leading order modes $(w, b_t)$.
We haven't explicitly written down their form for simplicity, as they are not essential to our arguments.
Note that ~\eqref{2ndrr1} is a constraint equation, constraining~\emph{e.g.} $h_{tt}^{(1)}$. Moreover, the $tt$ component of Einstein's equation~\eqref{2ndtt1} is implied by the remaining equations.
Thus, what we have can be reduced to four independent differential equations: three second order equations  for $(a_{t}^{(1)}, h_{xx}^{(1)}, h_{yy}^{(1)})$ and one first order equation for $h_{tt}^{(1)}$.

\subsection{Set (IV)}
This case comprises six equations of motion for the four modes ($a_{t}^{(2)}, h_{tt}^{(2)}, h_{xx}^{(2)}, h_{yy}^{(2)}$),
\begin{eqnarray}
a_t''^{(2)}+\frac{2}{r}a_t'^{(2)}+\frac{r_h \mu}{2 r^2 f}h_{tt}'^{(2)}+\frac{L^2 r_h \mu}{2 r^4}\left(h_{xx}'^{(2)}+h_{yy}'^{(2)}\right)-\frac{r_h\mu f'}{2 r^2 f^2}h_{tt}^{(2)}\nonumber\\-\frac{L^2 r_h\mu}{r^5}(h_{xx}^{(2)}+h_{xx}^{(2)})-\frac{4 L^2 k^2}{r^2 f}a_t^{(2)}=\mathcal{C}_1^{(2)},
\end{eqnarray}
\begin{eqnarray}\label{2ndtt2}
h_{xx}''^{(2)}+h_{yy}''^{(2)}+\frac{1}{2}\left(\frac{f'}{f}-\frac{2}{r}\right)(h_{xx}'^{(2)}+h_{yy}'^{(2)})-\frac{2}{L^2 f^2}\left(f+r f'-\frac{3 r^2}{L^2}\right)h_{tt}^{(2)}\nonumber\\+\frac{r_h\mu}{L^2 f}a_{t}'^{(2)}-\frac{f'}{r f}(h_{xx}^{(2)}+h_{yy}^{(2)})-\frac{4 L^2 k^2}{r^2 f}h_{yy}^{(2)}=\mathcal{C}_2^{(2)},
\end{eqnarray}
\begin{eqnarray}
h_{xx}''^{(2)}-\frac{r^2}{L^2 f}h_{tt}''^{(2)}+\left(\frac{f'}{f}-\frac{2}{r}\right)h_{xx}'^{(2)}+\frac{r(r f'-2f)}{2 L^2 f^2} h_{tt}'^{(2)}-\frac{r_h\mu}{L^2 f}a_{t}'^{(1)}\nonumber\\+\frac{2(f-rf')}{r^2 f}h_{xx}^{(2)}-\frac{L^2 r^2 f'^2+2rf(L^2 f'-6r)}{2 L^4 f^3}h_{tt}^{(2)}+\frac{4 k^2}{f^2}h_{tt}^{(2)}=\mathcal{C}_3^{(2)},
\end{eqnarray}
\begin{eqnarray}
h_{yy}''^{(2)}-\frac{r^2}{L^2 f}h_{tt}''^{(2)}+\left(\frac{f'}{f}-\frac{2}{r}\right)h_{yy}'^{(2)}+\frac{r(r f'-2f)}{2 L^2 f^2} h_{tt}'^{(2)}-\frac{r_h\mu}{L^2 f}a_{t}'^{(2)}\nonumber\\+\frac{2(f-rf')}{r^2 f}h_{yy}^{(2)}-\frac{L^2 r^2 f'^2+2rf(L^2 f'-6r)}{2 L^4 f^3}h_{tt}^{(2)}=\mathcal{C}_4^{(2)},
\end{eqnarray}
\begin{eqnarray}\label{2ndrr2}
h_{tt}'^{(2)}-\frac{L^2(rf'+2f)}{4 r^2}(h_{xx}'^{(2)}+h_{yy}'^{(2)})-\frac{r_h\mu}{2r}a_{t}'^{(2)}+\frac{L^2(r f'+2f)}{2 r^3}(h_{xx}^{(2)}+h_{yy}^{(2)})\nonumber\\-\frac{4 r^3 f'+r_h^2\mu^2}{4 r^3 f}h_{tt}^{(2)}+\frac{2 L^4 k^2}{r^3}h_{yy}^{(2)}-\frac{2 L^2 k^2}{r f}h_{tt}^{(2)}=\mathcal{C}_5^{(2)},
\end{eqnarray}
\begin{eqnarray}\label{2ndrx2}
h_{tt}'^{(2)}-\frac{L^2 f}{r^2}h_{yy}'^{(2)}-\frac{1}{2}\left(\frac{f'}{f}+\frac{2}{r}\right)h_{tt}^{(2)}+\frac{2L^2 f}{r^3}h_{yy}^{(2)}+\frac{r_h\mu}{r^2}a_{t}^{(2)}=\mathcal{C}_6^{(2)}.
\end{eqnarray}
As in the previous case, the six source terms $\mathcal{C}_i^{(2)} (i=1,...,6)$ depend on the two leading order perturbations $(w, b_t)$.
The precise form of the source terms is not important for our arguments and is not included for simplicity.
Again, the $tt$ component of Einstein's equation~\eqref{2ndtt2} can be obtained from the remaining equations. Using the $rr$ and $rx$ components of~\eqref{2ndrr2} and~\eqref{2ndrx2}, we find that $h_{tt}^{(2)}$ satisfies an algebraic equation.
Finally, we obtain three independent second order equations for $(a_{t}^{(2)}, h_{xx}^{(2)}, h_{yy}^{(2)})$ and one algebraic equation for $h_{tt}^{(2)}$.
Since $A_t$ corresponds to the charge density $\rho_A$ of the dual field theory, the mode $a_{t}^{(2)}$ means $\rho_A$ is spatially modulated with wave number $2 k$.

\section{Thermodynamics in Grand Canonical Ensemble \label{therm}}

Recall that in order to obtain a renormalized action, we must supplement it with appropriate boundary terms.
For the model we are considering in Section~\ref{backreaction}, such terms take the form
\begin{equation}\label{fullaction}
S_\partial=\frac{1}{2\kappa_N^2}\int_{z\rightarrow 1} dx^3\sqrt{-\gamma}\left(2K-\frac{4}{L}-\frac{1}{2L}\chi^2\right),
\end{equation}
where $\gamma_{\mu\nu}$ is the induced metric at the AdS boundary and $K_{\mu\nu}$ is the extrinsic curvature defined by the outward pointing normal vector to the boundary $n^\mu$.
According to the holographic dictionary, the stress-energy tensor of the dual field theory is
\begin{equation}
T_{\mu\nu}=\frac{1}{\kappa_N^2}\lim_{z\rightarrow 1}\frac{r_h}{(1-z^2)L}\left(K \gamma_{\mu\nu}-K_{\mu\nu}-\frac{2}{L}\gamma_{\mu\nu}-\frac{1}{4 L}\chi^2 \gamma_{\mu\nu}\right).
\end{equation}
Notice that we are interested in the one point functions with respect to the metric
\begin{equation}
ds^2=-dt^2+dx^2+dy^2.
\end{equation}
Substituting the expansion~\eqref{uvexpand}, we obtain the following components of $T_{\mu\nu}$,
\begin{equation}\label{stress}
\begin{split}
2 \kappa_N^2 T_{tt}=\frac{r_h^3}{L^4} \left(2+\frac{L^2}{2}\frac{\mu^2}{r_h^2}+3(q_{xx}+q_{yy})\right),\\
2 \kappa_N^2 T_{xx}=\frac{r_h^3}{L^4} \left(1+\frac{L^2}{4}\frac{\mu^2}{r_h^2}+3(q_{tt}+q_{yy})\right),\\
2 \kappa_N^2 T_{yy}=\frac{r_h^3}{L^4} \left(1+\frac{L^2}{4}\frac{\mu^2}{r_h^2}+3(q_{tt}+q_{xx})\right),
\end{split}
\end{equation}
with all other components vanishing. Moreover, using the relation~\eqref{uvrelation}, one can explicitly check that the stress-energy tensor is traceless and conserved,
\begin{equation}
{T^\mu}_\mu=0,\quad \partial_\nu T^{\nu\mu}=0,
\end{equation}
which is consistent with the expectation that it describes a CFT.

The expectation values of the currents can be obtained by the variation of the on-shell action with respect to the gauge fields,
\begin{equation}
\left<J^\mu_A\right>=\frac{1}{2\kappa_N^2}\lim_{z\rightarrow 1}\sqrt{-\gamma} n_\nu (Z_A A^{\mu\nu}+Z_{AB} B^{\mu\nu}),
\end{equation}
\begin{equation}
\left<J^\mu_B\right>=\frac{1}{2\kappa_N^2}\lim_{z\rightarrow 1}\sqrt{-\gamma} n_\nu (Z_B B^{\mu\nu}+Z_{AB} A^{\mu\nu}).
\end{equation}
Plugging in~\eqref{uvexpand}, we obtain
\begin{equation}
\rho_A=\left<J^t_A\right>=\frac{1}{2\kappa_N^2}\frac{r_h\mu}{L^2}(1-\rho_a(x)),\quad \rho_B=\left<J^t_B\right>=-\frac{1}{2\kappa_N^2}\frac{r_h^2}{L^2}\rho_b(x).
\end{equation}

Working in the grand canonical ensemble, the local free energy density $\Omega(x)$ is given by
\begin{equation}\label{freemu}
\Omega(x)=T_{tt}(x)-T\,s(x)-\mu\,\rho_A(x),
\end{equation}
where the thermal entropy density $s$ is given by
\begin{equation}
s(x)=\frac{2\pi}{\kappa_N^2}\frac{r_h^2}{L^2}\sqrt{Q_{xx}(0,x) Q_{yy}(0,x)}.
\end{equation}
For the normal solution~\eqref{RNadsz}, we then have
\begin{equation}
T_{tt}=\frac{r_h^3}{2 \kappa_N^2 L^4} \left(2+\frac{L^2}{2}\frac{\mu^2}{r_h^2}\right),\quad s(x)=\frac{2\pi}{\kappa_N^2}\frac{r_h^2}{L^2}, \quad \rho_A=\frac{1}{2 \kappa_N^2 } \frac{r_h \mu}{L^2},
\end{equation}
and the corresponding free energy density is given by
\begin{equation}
\Omega_{RN}=-\frac{1}{2 \kappa_N^2}\frac{r_h^3}{L^4}\left(1+\frac{L^2}{4}\frac{\mu^2}{r_h^2}\right).
\end{equation}
Notice that the UV expansion for $\chi$ in the $r$ coordinate is in general given by
\begin{equation}
\chi=\frac{\chi_s}{r}+\frac{\chi_v}{r^2}+\mathcal{O}(r^{-3}),
\end{equation}
where $\chi_s$ is identified as the source of the dual scalar operator ${O}_\chi$.
Therefore, the one point function for ${O}_\chi$ is computed by
\begin{equation}
\left<{O}_\chi\right>=\frac{1}{2\kappa_N^2}\lim_{z\rightarrow 1}\frac{1-z^2}{r_h}\sqrt{-\gamma}\,(-n_\nu \nabla^\nu\chi-\frac{1}{L}\chi)=\frac{1}{2 \kappa_N^2}\frac{r_h^2}{L^4}\phi_v(x).
\end{equation}
In the grand canonical ensemble the two free thermodynamic variables are the chemical potential $\mu$ and temperature $T$.
The system~\eqref{ansatzbh} has the following scaling symmetry,
\begin{equation}
(r_h,\mu, \beta)\rightarrow c\, (r_h,\mu,\beta),\quad (t, x, y)\rightarrow (t,x,y)/c,
\end{equation}
where $c$ is a constant, and
$(z, L, \phi,\alpha,Q_{tt},Q_{zz},Q_{xx},Q_{yy}, Q_{xz})$ are unchanged.
To fix the above scaling symmetry, we use the dimensionless quantities
\begin{equation}
\frac{\left<{O}_\chi\right>}{\mu^2},\quad \frac{\Omega}{\mu^3},\quad \frac{T_{tt}}{\mu^3},\quad \frac{\rho_A}{\mu^2},\quad \frac{\rho_B}{\mu^2},\quad \frac{s}{\mu^2}, \quad \frac{T}{\mu}.
\end{equation}
As one can check, they only depend on the dimensionless combination $r_h/\mu$, or equivalently $T/\mu$.

\section{Accuracy of Numerics}\label{app:test}

\begin{figure}[ht!]
\begin{center}
\includegraphics[width=.55\textwidth]{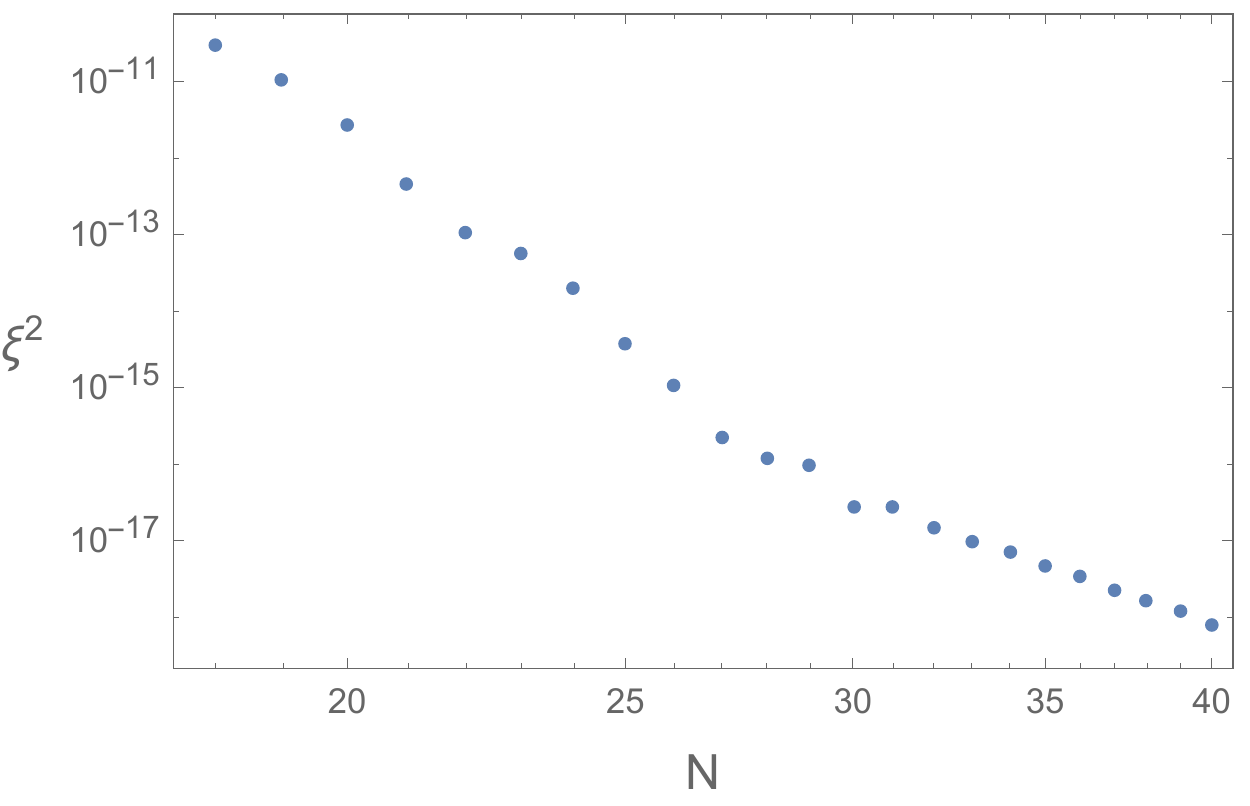}
\caption{$\xi^2$ as a function of $N_z=N_x=N$ for $T=0.01426$ and $k=1$ with $\mu=1$. We consider the same parameters as in Figure~\ref{fig:Conden-FreeEn}.}
\label{fig:xi2}
\end{center}
\end{figure}

In this appendix we check the quality of our numerics.
We start by checking the convergence of $\xi^2=\xi_\mu \xi^\mu$,
where the DeTurck vector field $\xi$ has been defined in~\eqref{eqxi}. The number of grid points in the $z$ and $x$ directions are $N_z$ and $N_x$, respectively.
As one can see in Figure~\ref{fig:xi2}, the maximum value of $\xi^2$ converges towards zero so that indeed we have a solution to Einstein's equations.
We also find an approximatively exponential rate of convergence with the number of grid points, which is expected by the pseudospectral collocation method.

\begin{figure}[ht!]
\begin{center}
\includegraphics[width=.55\textwidth]{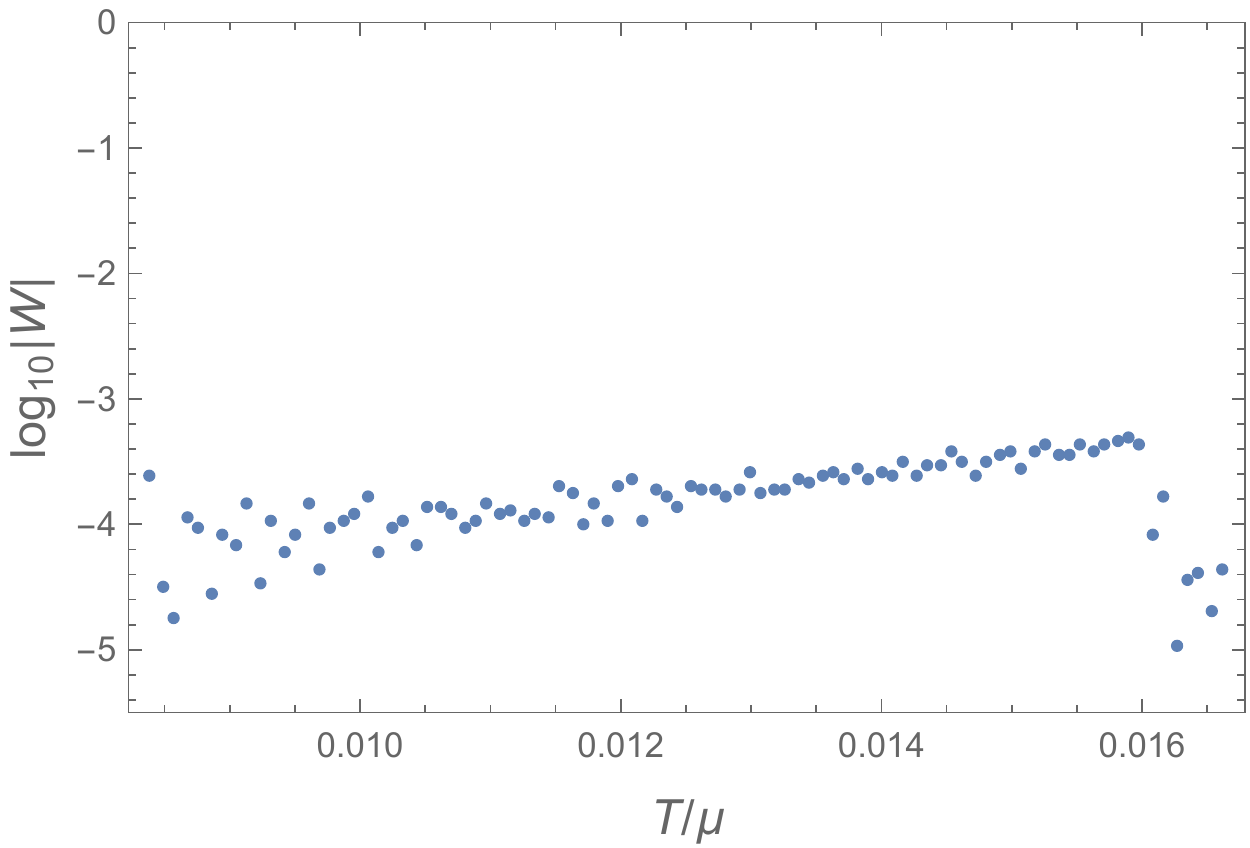}
\caption{Numerical check of the first law $W$ for a set of solutions corresponding to a branch with fixed $k/\mu=1$. We choose the same parameters as Figure~\ref{fig:Conden-FreeEn}.}
\label{fig:logWT}
\end{center}
\end{figure}

Next, we perform a check of the first law of thermodynamics.
Note that we are working in the grand canonical ensemble with the free energy given by~\eqref{freemu}. Together with the first law,
\begin{eqnarray}
d  \, T_{tt}(x)=T \,d s(x)+\mu\, d\rho(x)\,,
\end{eqnarray}
one finds that
\begin{eqnarray}
W=\frac{d (\bar{\Omega}/\mu^3)}{d (T/\mu)}+\frac{\bar{s}}{\mu^2}=0\,,
\end{eqnarray}
evaluated on a fixed $k/\mu$ branch of the striped solutions. As shown in Figure~\ref{fig:logWT}, this condition is satisfied along the set of solutions we are considering in Figure~\ref{fig:Conden-FreeEn}.


\begin{thebibliography}{100}

\def\arXiv#1{\href{http://arxiv.org/abs/#1}{arXiv:#1}}
\def\arXiv#1#2{\href{http://arxiv.org/abs/#1}{arXiv:#1}}
\def\arXivid#1#2{\href{http://arxiv.org/abs/#1/#2}{#1/#2}}
\def\doi#1#2{\href{http://doi.org/#1}{#2}}


\bibitem{Cremonini:2016rbd}
  S.~Cremonini, L.~Li and J.~Ren,
  {\em Holographic Pair and Charge Density Waves},
  \doi{10.1103/PhysRevD.95.041901}{Phys.\ Rev.\ D {\bf 95} (2017) 041901}
  [\arXiv{1612.04385}{[hep-th]}].

\bibitem{EmeryKivelsonTranquada}
  V.~J.~Emery, S.~A.~Kivelson, and J.~M.~Tranquada,
  {\em Stripe phases in high-temperature superconductors},
  \doi{10.1073/pnas.96.16.8814}{Proc. Natl. Acad. Sc 96 (1999) 8814}
  [\arXivid{cond-mat}{9907228}].

\bibitem{Berg:2009dga}
  E.~Berg, E.~Fradkin, S.~A.~Kivelson and J.~M.~Tranquada,
  {\em Striped superconductors: how spin, charge and superconducting orders intertwine in the cuprates},
  \doi{10.1088/1367-2630/11/11/115004}{New J.\ Phys.\  {\bf 11} (2009) 115004}
  [\arXiv{0901.4826}{[cond-mat.supr-con]}].


\bibitem{Colloquium}
  E.~Fradkin, S.~A.~Kivelson and J.~M.~Tranquada,
  {\em Colloquium: Theory of intertwined orders in high temperature superconductors},
  \doi{	10.1103/RevModPhys.87.457}{Rev. Mod. Phys. {\bf 87} (2015) 457}
  [\arXiv{1407.4480}{[cond-mat.supr-con]}].


\bibitem{HimedaPRL}
  A.~Himeda, T.~Kato, and M.~Ogata,
  {\em Stripe States with Spatially Oscillating $d$-Wave Superconductivity in the Two-Dimentional $t-t^\prime-J$ Model},
  \doi{10.1103/PhysRevLett.88.117001}{Phys. Rev. Lett. {\bf 88} (2002) 117001}.

\bibitem{BergPRL}
  E.~Berg, E.~Fradkin, E.~A.~Kim, S.~Kivelson, V.~Oganesyan, J.~M.~Tranquada, and S.~Zhang,
  {\em Dynamical Layer Decoupling in a Stripe-Ordered High-$T_c$ Superconductor},
  \doi{10.1103/PhysRevLett.99.127003}{Phys. Rev. Lett. {\bf 99} (2007) 127003}
  [\arXiv{0704.1240}{[cond-mat.str-el]}].

\bibitem{Gubser:2008px}
  S.~S.~Gubser,
  {\em Breaking an Abelian gauge symmetry near a black hole horizon},
  \doi{10.1103/PhysRevD.78.065034}{Phys.\ Rev.\ D {\bf 78} (2008) 065034}
  [\arXiv{0801.2977}{[hep-th]}].

\bibitem{Hartnoll:2008vx}
  S.~A.~Hartnoll, C.~P.~Herzog and G.~T.~Horowitz,
  {\em Building a Holographic Superconductor},
  \doi{10.1103/PhysRevLett.101.031601}{Phys.\ Rev.\ Lett.\  {\bf 101} (2008) 031601}
  [\arXiv{0803.3295}{[hep-th]}].

\bibitem{Franco:2009yz}
  S.~Franco, A.~Garcia-Garcia and D.~Rodriguez-Gomez,
  {\em A General class of holographic superconductors},
  \doi{10.1007/JHEP04(2010)092}{JHEP {\bf 1004} (2010) 092}
  [\arXiv{0906.1214}{[hep-th]}].


\bibitem{Gubser:2009qm}
  S.~S.~Gubser, C.~P.~Herzog, S.~S.~Pufu and T.~Tesileanu,
  {\em Superconductors from Superstrings},
  \doi{10.1103/PhysRevLett.103.141601}{Phys.\ Rev.\ Lett.\  {\bf 103} (2009) 141601}
  [\arXiv{0907.3510}{[hep-th]}].


\bibitem{Aprile:2009ai}
  F.~Aprile and J.~G.~Russo,
  {\em Models of Holographic superconductivity},
  \doi{10.1103/PhysRevD.81.026009}{Phys.\ Rev.\ D {\bf 81} (2010) 026009}
  [\arXiv{0912.0480}{[hep-th]}].

\bibitem{Aprile:2010yb}
  F.~Aprile, S.~Franco, D.~Rodriguez-Gomez and J.~G.~Russo,
  {\em Phenomenological Models of Holographic Superconductors and Hall currents},
  \doi{10.1007/JHEP05(2010)102}{JHEP {\bf 1005} (2010) 102}
  [\arXiv{1003.4487}{[hep-th]}].

\bibitem{Liu:2010ka}
  Y.~Liu and Y.~W.~Sun,
  {\em Holographic Superconductors from Einstein-Maxwell-Dilaton Gravity},
  \doi{10.1007/JHEP07(2010)099}{JHEP {\bf 1007} (2010) 099}
  [\arXiv{1006.2726}{[hep-th]}].

\bibitem{Peng:2011gh}
  Y.~Peng, Q.~Pan and B.~Wang,
  {\em Various types of phase transitions in the AdS soliton background},
  \doi{10.1016/j.physletb.2011.04.025}{Phys.\ Lett.\ B {\bf 699} (2011) 383}
  [\arXiv{1104.2478}{[hep-th]}].

\bibitem{Cai:2012es}
  R.~G.~Cai, S.~He, L.~Li and L.~F.~Li,
  {\em Entanglement Entropy and Wilson Loop in St\'{u}ckelberg Holographic Insulator/Superconductor Model},
  \doi{10.1007/JHEP10(2012)107}{JHEP {\bf 1210} (2012) 107}
  [\arXiv{1209.1019}{[hep-th]}].


\bibitem{Gouteraux:2012yr} 
  B.~Gouteraux and E.~Kiritsis,
  ``Quantum critical lines in holographic phases with (un)broken symmetry,''
  \doi{10.1007/JHEP04(2013)053}{JHEP {\bf 1304}, (2013) 053}
  [\arXiv{1212.2625}{[hep-th]}].

\bibitem{Kiritsis:2015oxa} 
  E.~Kiritsis and J.~Ren,
  {\em ``On Holographic Insulators and Supersolids},
  \doi{10.1007/JHEP09(2015)168}{JHEP {\bf 1509}, (2015) 168}
  [\arXiv{1503.03481}{[hep-th]}].





\bibitem{Flauger:2010tv}
  R.~Flauger, E.~Pajer and S.~Papanikolaou,
  {\em A Striped Holographic Superconductor},
  \doi{10.1103/PhysRevD.83.064009}{Phys.\ Rev.\ D {\bf 83} (2011) 064009}
  [\arXiv{1010.1775}{[hep-th]}].



\bibitem{Hutasoit:2011rd}
  J.~A.~Hutasoit, S.~Ganguli, G.~Siopsis and J.~Therrien,
  {\em Strongly Coupled Striped Superconductor with Large Modulation},
  \doi{10.1007/JHEP02(2012)086}{JHEP {\bf 1202} (2012) 086}
  [\arXiv{1110.4632}{[cond-mat.str-el]}].

\bibitem{Hutasoit:2012ib}
  J.~A.~Hutasoit, G.~Siopsis and J.~Therrien,
  {\em Conductivity of Strongly Coupled Striped Superconductor},
  \doi{10.1007/JHEP01(2014)132}{JHEP {\bf 1401} (2014) 132}
  [\arXiv{1208.2964}{[hep-th]}].


\bibitem{Erdmenger:2013zaa}
  J.~Erdmenger, X.~H.~Ge and D.~W.~Pang,
  {\em Striped phases in the holographic insulator/superconductor transition},
  \doi{10.1007/JHEP11(2013)027}{JHEP {\bf 1311} (2013) 027}
  [\arXiv{1307.4609}{[hep-th]}].

\bibitem{Kuang:2013jma}
  X.~M.~Kuang, B.~Wang and X.~H.~Ge,
  {\em Observing the inhomogeneity in the holographic models of superconductors},
  \doi{10.1142/S0217732314500709}{Mod.\ Phys.\ Lett.\ A {\bf 29} (2014) 1450070}
  [\arXiv{1307.5932}{[hep-th]}].

\bibitem{Arean:2013mta}
  D.~Arean, A.~Farahi, L.~A.~Pando Zayas, I.~S.~Landea and A.~Scardicchio,
  {\em Holographic superconductor with disorder},
  \doi{10.1103/PhysRevD.89.106003}{Phys.\ Rev.\ D {\bf 89} (2014) 106003}
  [\arXiv{1308.1920}{[hep-th]}].





\bibitem{Horowitz:2013jaa}
  G.~T.~Horowitz and J.~E.~Santos,
  {\em General Relativity and the Cuprates},
  \doi{10.1007/JHEP06(2013)087}{JHEP {\bf 1306} (2013) 087}
  [\arXiv{1302.6586}{[hep-th]}].


\bibitem{Donos:2012yu}
  A.~Donos, J.~P.~Gauntlett, J.~Sonner and B.~Withers,
  {\em Competing orders in M-theory: superfluids, stripes and metamagnetism},
  \doi{10.1007/JHEP03(2013)108}{JHEP {\bf 1303} (2013) 108}
  [\arXiv{1212.0871}{[hep-th]}].

\bibitem{Cremonini:2014gia}
  S.~Cremonini, Y.~Pang, C.~N.~Pope and J.~Rong,
  {\em Superfluid and metamagnetic phase transitions in $\omega$-deformed gauged supergravity},
  \doi{10.1007/JHEP04(2015)074}{JHEP {\bf 1504} (2015) 074}
  [\arXiv{1411.0010}{[hep-th]}].


\bibitem{Donos:2011bh}
  A.~Donos and J.~P.~Gauntlett,
  {\em Holographic striped phases},
  \doi{10.1007/JHEP08(2011)140}{JHEP {\bf 1108} (2011) 140}
  [\arXiv{1106.2004}{[hep-th]}].

\bibitem{Donos:2013gda}
  A.~Donos and J.~P.~Gauntlett,
  {\em Holographic charge density waves},
  \doi{10.1103/PhysRevD.87.126008}{Phys.\ Rev.\ D {\bf 87} (2013) 126008}
  [\arXiv{1303.4398}{[hep-th]}].


\bibitem{Ling:2014saa}
  Y.~Ling, C.~Niu, J.~Wu, Z.~Xian and H.~b.~Zhang,
  {\em Metal-insulator Transition by Holographic Charge Density Waves},
  \doi{10.1103/PhysRevLett.113.091602}{Phys.\ Rev.\ Lett.\  {\bf 113} (2014) 091602}
  [\arXiv{1404.0777}{[hep-th]}].

\bibitem{Kiritsis:2015hoa}
  E.~Kiritsis and L.~Li,
  {\em Holographic Competition of Phases and Superconductivity},
  \doi{10.1007/JHEP01(2016)147}{JHEP {\bf 1601} (2016) 147}
  [\arXiv{1510.00020}{[cond-mat.str-el]}].

\bibitem{Donos:2016hsd}
  A.~Donos and C.~Pantelidou,
  {\em Holographic Magnetisation Density Waves},
  \doi{10.1007/JHEP10(2016)038}{JHEP {\bf 1610} (2016) 038}
  [\arXiv{1607.01807}{[hep-th]}].


\bibitem{Donos:2011ff}
  A.~Donos and J.~P.~Gauntlett,
  {\em Holographic helical superconductors},
  \doi{10.1007/JHEP12(2011)091}{JHEP {\bf 1112} (2011) 091}
  [\arXiv{1109.3866}{[hep-th]}].

\bibitem{Donos:2012gg}
  A.~Donos and J.~P.~Gauntlett,
  {\em Helical superconducting black holes},
  \doi{10.1103/PhysRevLett.108.211601}{Phys.\ Rev.\ Lett.\  {\bf 108} (2012) 211601}
  [\arXiv{1203.0533}{[hep-th]}].



\bibitem{Headrick:2009pv}
  M.~Headrick, S.~Kitchen and T.~Wiseman,
  {\em A New approach to static numerical relativity, and its application to Kaluza-Klein black holes},
  \doi{10.1088/0264-9381/27/3/035002}{Class.\ Quant.\ Grav.\  {\bf 27} (2010) 035002}
  [\arXiv{0905.1822}{[gr-qc]}].



\bibitem{Horowitz:2012ky}
  G.~T.~Horowitz, J.~E.~Santos and D.~Tong,
  {\em Optical Conductivity with Holographic Lattices},
  \doi{10.1007/JHEP07(2012)168}{JHEP {\bf 1207} (2012) 168}
  [\arXiv{1204.0519}{[hep-th]}].

\bibitem{Donos:2014yya}
  A.~Donos and J.~P.~Gauntlett,
  {\em The thermoelectric properties of inhomogeneous holographic lattices},
  \doi{10.1007/JHEP01(2015)035}{JHEP {\bf 1501} (2015) 035}
  [\arXiv{1409.6875}{[hep-th]}].



\bibitem{Andrade:2013gsa}
  T.~Andrade and B.~Withers,
  {\em A simple holographic model of momentum relaxation},
  \doi{10.1007/JHEP05(2014)101}{JHEP {\bf 1405} (2014) 101}
  [\arXiv{1311.5157}{[hep-th]}].


\bibitem{Andrade:2017leb}
  T.~Andrade and A.~Krikun,
  {\em Commensurate lock-in in holographic non-homogeneous lattices},
  \doi{10.1007/JHEP03(2017)168}{JHEP {\bf 1703} (2017) 168}
  [\arXiv{1701.04625}{[hep-th]}].




 \bibitem{Andrade:2015iyf}
  T.~Andrade and A.~Krikun,
  {\em Commensurability effects in holographic homogeneous lattices},
  \doi{10.1007/JHEP05(2016)039}{JHEP {\bf 1605} (2016) 039}
  [\arXiv{1512.02465}{[hep-th]}].

\bibitem{Delacretaz:2016ivq}
  L.~V.~Delacretaz, B.~Gouteraux, S.~A.~Hartnoll and A.~Karlsson,
  {\em Bad Metals from Fluctuating Density Waves},
  \arXiv{1612.04381}{[cond-mat.str-el]}.


\bibitem{JaefariFradkin2012}
A.~Jaefari and E.~Fradkin.
{\em Pair-Density-Wave Superconducting Order in Two-Leg Ladders},
\doi{10.1103/PhysRevB.85.035104}{Phys. Rev. B {\bf 85} (2012) 035104}
[\arXiv{1111.6320}{[cond-mat.str-el]}].



\bibitem{Cremonini:2013epa}
  S.~Cremonini,
  {\em Spatially Modulated Instabilities for Scaling Solutions at Finite Charge Density},
  \doi{10.1103/PhysRevD.95.026007}{Phys.\ Rev.\ D {\bf 95} (2017) 026007}
  [\arXiv{1310.3279}{[hep-th]}].

\bibitem{Cremonini:2012ir}
  S.~Cremonini and A.~Sinkovics,
  {\em Spatially Modulated Instabilities of Geometries with Hyperscaling Violation},
  \doi{10.1007/JHEP01(2014)099}{JHEP {\bf 1401} (2014) 099}
  [\arXiv{1212.4172}{[hep-th]}].

\bibitem{Iizuka:2013ag}
  N.~Iizuka and K.~Maeda,
  {\em Stripe Instabilities of Geometries with Hyperscaling Violation},
  \doi{10.1103/PhysRevD.87.126006}{Phys.\ Rev.\ D {\bf 87} (2013) 126006}
  [\arXiv{1301.5677}{[hep-th]}].

\bibitem{Withers:2014sja}
  B.~Withers,
  {\em Holographic Checkerboards},
  \doi{10.1007/JHEP09(2014)102}{JHEP {\bf 1409} (2014) 102}
  [\arXiv{1407.1085}{[hep-th]}].

\bibitem{Gouteraux:2016arz}
  B.~Gouteraux and V.~L.~Martin,
  {\em Spectral weight and spatially modulated instabilities in holographic superfluids},
  \doi{10.1007/JHEP05(2017)005}{JHEP {\bf 1705} (2017) 005}
  [\arXiv{1612.03466}{[hep-th]}].

\bibitem{Cremonini:2016bqw}
  S.~Cremonini and L.~Li,
  {\em Criteria For Superfluid Instabilities of Geometries with Hyperscaling Violation},
  \doi{10.1007/JHEP11(2016)137}{JHEP {\bf 1611} (2016) 137}
  [\arXiv{1606.02745}{[hep-th]}].


\end{thebibliography}
\end{document}